%% file: a_main.tex
\documentclass[twoside,11pt]{article}

\usepackage{jmlr2e}
\input{extra/acronyms}
\usepackage[inline]{enumitem} %
\usepackage{amsmath}
\usepackage{amssymb}
\usepackage{mathtools}
\usepackage{xcolor}
\usepackage{microtype}
\usepackage[caption=false]{subfig}
\usepackage{comment}
\usepackage{extra/algorithmic}
\usepackage{extra/algorithm}
\usepackage{tikz} %
\usepackage{hyperref}
\usepackage{multirow}

\usepackage{makecell}
\usepackage{array}

\usepackage{tabularx}

\usepackage{lastpage}
\jmlrheading{24}{2023}{1-\pageref{LastPage}}{9/22; Revised5/23}{8/23}{22-0994}{Arrasy Rahman, Ignacio Carlucho, Niklas H\"opner and Stefano V. Albrecht}
\ShortHeadings{A General Learning Framework for Open Ad Hoc Teamwork}{Rahman, Carlucho, H\"opner and Albrecht}

\firstpageno{1}

\begin{document}

\title{A General Learning Framework for Open Ad Hoc Teamwork Using Graph-based Policy Learning}

\author{\name Arrasy Rahman \email arrasy.rahman@ed.ac.uk \\
       \addr School of Informatics \\
       University of Edinburgh \\
       Edinburgh, UK
       \AND
       \name Ignacio Carlucho \email ignacio.carlucho@ed.ac.uk \\
       \addr School of Informatics \\
       University of Edinburgh \\
       Edinburgh, UK
       \AND
       \name Niklas H\"opner \email n.r.hopner@uva.nl \\
        \addr Institute of Informatics \\
       University of Amsterdam \\
       Amsterdam, Netherlands
       \AND
       \name Stefano V.\ Albrecht \email s.albrecht@ed.ac.uk \\
       \addr School of Informatics \\
       University of Edinburgh \\
       Edinburgh, UK}

\editor{Laurent Orseau}

\maketitle

\begin{abstract}%

Open ad hoc teamwork is the problem of training a single agent to efficiently collaborate with an unknown group of teammates whose composition may change over time. 
A variable team composition creates challenges for the agent, such as the requirement to adapt to new team dynamics and dealing with changing state vector sizes. These challenges are aggravated in real-world applications in which the controlled agent only has a partial view of the environment. %
In this work, we develop a class of solutions for open ad hoc teamwork under full and partial observability. 
We start by developing a solution for the fully observable case that leverages graph neural network architectures to obtain an optimal policy based on reinforcement learning. We then extend this solution to partially observable scenarios by proposing different methodologies that maintain belief estimates over the latent environment states and team composition. These belief estimates are combined with our solution for the fully observable case to compute an agent's optimal policy under partial observability in open ad hoc teamwork. Empirical results demonstrate that our solution can learn efficient policies in open ad hoc teamwork in fully and partially observable cases. Further analysis demonstrates that our methods' success is a result of effectively learning the effects of teammates' actions while also inferring the inherent state of the environment under partial observability.\footnote{We provide an implementation of the algorithms and the environments in~\url{https://github.com/uoe-agents/PO-GPL}}

\end{abstract}


\input{sections/introduction}

\input{sections/related}

\input{sections/objective}

\input{sections/methods_1}

\input{sections/results_1}

\input{sections/methods_2}
\input{sections/results_2}

\input{sections/conclusion}

\newpage
\acks{This research received financial support from the US Office of Naval Research (ONR) via grant N00014-20-1-2390, and the Google Cloud Research Credits program award.
}

\input{sections/appendix}

\vskip 0.2in
\bibliography{sample}

\end{document}

%% file: extra/acronyms.tex
\usepackage[nolist,nohyperlinks]{acronym}
\newacro{SBG}[SBG]{stochastic Bayesian game}
    \newacroindefinite{SBG}{an}{a}
    \newacroplural{SBG}[SBGs]{stochastic Bayesian games}
\newacro{MCTS}[MCTS]{Monte Carlo tree search}
    \newacroindefinite{MCTS}{an}{a}
    \newacroplural{MCTS}[MCTSs]{Monte Carlo tree searches}
\newacro{UCT}[UCT]{upper confidence tree}
    \newacroindefinite{UCT}{a}{an}
    \newacroplural{UCT}[UCTs]{upper confidence trees}
\newacro{RL}[RL]{reinforcement learning}
    \newacroindefinite{RL}{an}{a}
\newacro{MARL}[MARL]{multi-agent reinforcement learning}
    \newacroindefinite{MARL}{a}{a}
\newacro{NN}[NN]{neural network}
    \newacroindefinite{NN}{an}{a}
    \newacroplural{NN}[NNs]{neural networks}
\newacro{GNN}[GNN]{graph neural network}
    \newacroindefinite{GNN}{a}{a}
    \newacroplural{GNN}[GNNs]{graph neural networks}
\newacro{CNN}[CNN]{convolutional neural network}
    \newacroindefinite{CNN}{a}{a}
    \newacroplural{CNN}[CNNs]{convolutional neural networks}
\newacro{VAE}[VAE]{variational autoencoder}
    \newacroindefinite{VAE}{a}{a}
    \newacroplural{VAE}[VAEs]{variational autoencoders}
\newacro{OSBG}[OSBG]{open stochastic Bayesian game}
    \newacroindefinite{OSBG}{a}{a}
\newacro{PO-OSBG}[PO-OSBG]{partially observable open stochastic Bayesian game}
    \newacroindefinite{PO-OSBG}{a}{a}
\newacro{SBG}[SBG]{stochastic Bayesian game}
    \newacroindefinite{OSBG}{a}{a}
\newacro{CG}[CG]{coordination graph}
    \newacroindefinite{CG}{a}{a}
\newacro{ELBO}[ELBO]{evidence lower bound}
    \newacroindefinite{ELBO}{a}{a}
\newacro{GPL}[GPL]{Graph-based Policy Learning}
    \newacroindefinite{GPL}{a}{a}
\newacro{RFM}[RFM]{relational forward model}
    \newacroindefinite{RFM}{a}{a}
\newacro{POMDP}[POMDP]{partially observable Markov decision process}
    \newacroindefinite{POMDP}{a}{a}
    \newacroplural{POMDP}[POMDPs]{partially observable Markov decision processes}
\newacro{RNN}[RNN]{recurrent neural network}
    \newacroindefinite{RNN}{a}{a}
\newacro{SMC}[SMC]{Sequential Monte Carlo}
    \newacroindefinite{SMC}{an}{a}
\newacro{RMSE}[RMSE]{room mean squared error}
    \newacroindefinite{RMSE}{an}{a} 
\newacro{LSTM}[LSTM]{long short term memory}

\newacro{LBF}[LBF]{Level-Based Foraging}
    \newacroindefinite{LBF}{a}{a}
\newacro{PCN}[PCN]{Penalized Cooperative Navigation}
    \newacroindefinite{PCN}{an}{a} 

%% file: sections/introduction.tex
\section{Introduction}
\label{sec:introduction}

Current research in multi-agent systems has demonstrated how teams of agents can be co-trained to learn policies to solve a number of problems, such as smart grid management \citep{app10196900}, navigating human-shared environments  \citep{BOLDRER2022103979}, multi-robot warehouse management~\citep{krnjaic2023scalable}, and real-time strategy games \citep{ZHOU2021115707}. In many real-world applications, a controlled agent may be required to collaborate in diverse teams without the possibility of previous joint training. This problem is commonly referred to as \textit{ad hoc teamwork}~\citep{Stone2010-adhoc}. The objective of ad hoc teamwork is to train a single agent, which we refer to as the \textit{learner}, that can successfully collaborate ``on the fly'' with a group of teammates with unknown policies. Previous research on ad hoc teamwork has focused on the application of agent modelling \citep{ALA12-Barrett, albrecht_autonomous_2018} or communication techniques \citep{Reuth2020,surveyadhoc}. However, most prior ad hoc teamwork approaches are based on assumptions which may not hold in real-world applications.

One of these assumptions is that the number of other agents in the team is fixed. Real-world scenarios, such as autonomous driving and robotics rescue tasks, may require the learner to interact with a changing number of agents between timesteps. 
In open ad hoc teamwork, teammates may enter or leave the environment without prior notification. We refer to the variable team size nature as \textit{open teams} or environmental openness \citep{eck2020scalable}. The open nature of the team presents a set of additional challenges to the learner, which increases the difficulty of the ad hoc teamwork task. We call the problem of designing an ad hoc teamwork learner in open teams the \textit{open ad hoc teamwork problem}.

There are two main challenges in solving open ad hoc teamwork. The first challenge that needs to be addressed is the change in the size of the observation vector resulting from teammates entering or leaving the environment, which prevents many function approximation models that assume fixed input sizes from being directly applicable to estimating the learner's optimal policy. Second, the unknown team composition resulting from teammates joining or leaving the environment requires the learner to rapidly adapt its policy to effectively collaborate with its teammates. For instance, the learner may need to adopt different roles when dealing with teams that consist of distinct collections of teammate behavioural policies~\citep{surveyadhoc}. 

Another commonly violated assumption in real-world ad hoc teamwork problems is related to state information availability. In many problems, agents only have access to observations containing partial information of the state.
The learner then has to infer the latent environment state based only on a sequence of partial observations.
The combination of partial observability and open ad hoc teamwork provides new challenges to the learner, which now has to model the effects of environment openness while also inferring the latent state of the environment. Since the actions taken by other agents can affect the learner's returns, the learner also needs to maintain a model of all existing teammates' actions, even for unobserved teammates. Previous works have addressed the ad hoc teamwork problem under partial observability~\citep{gu2021online,ribeiro2022assisting}, but have not considered team openness. 

In this work, we investigate approaches for solving the challenges introduced by open ad hoc teamwork under full and partial observability. 
First, we present different algorithms that solve the fully observable open ad hoc teamwork problem. 
Our algorithms are based on three main components that a learner requires for effective ad hoc collaboration with teams of variable sizes. These three components are respectively used for teammate type inference, action prediction, and joint action value modelling. The output of these three components can be combined together to estimate a learner's optimal policy for open ad hoc teamwork. To deal with environment openness, we implement these components as \ac{GNN} architectures, which have been demonstrated as effective function approximation models for input data with variable sizes~\citep{jiang2019graph, huang2020one}. We call our proposed learning framework \acfi{GPL} and we demonstrate GPL's ability to train a learner's optimal policy in fully observable open ad hoc teamwork problems. 

Our results show that a GPL-based learner achieves significantly higher returns in open ad hoc teamwork than learners that use value-based reinforcement learning and multi-agent reinforcement learning algorithms for training. Furthermore, a GPL-based learner also achieves significantly higher performance compared to baseline methods when evaluated under an open process it has not experienced during training. Our experiments demonstrate that GPL's significantly higher performance results from its usage of GNNs and joint action value modelling, which enables GPL to learn the effects of other teammates' actions towards the learner. Through additional experiments we empirically demonstrate that learning the effects of teammates' actions via the joint action value model enables a learner to acquire useful behaviour from teammates.

We then address the open ad hoc teamwork problem under partial observability by extending GPL with belief inference methods that estimate the latent environment state. We evaluate different belief inference methods which allow the learner to maintain representations of important latent variables for decision making, such as the environment state, teammates' existence, as well as teammates' joint actions. The belief inference methods proposed in this work are inspired by latent variable inference methods such as \ac{SMC} methods~\citep{doucet2001introduction}, autoencoders~\citep{rumelhart1985learning}, and variational autoencoders~\citep{kingma2013auto}. We enable the proposed belief inference methods to handle data resulting from variable team sizes by using GNN-based models and graph generation techniques. Additionally, our extension to partial observability utilises the representations from the belief inference model as inputs to the different modules in GPL to estimate the learner's optimal policy.

We evaluate the performance of the different proposed belief inference models when combined with GPL to solve partially observable open ad hoc teamwork problems. 
Additionally, we compare the combination of GPL with the proposed belief inference models against different single-agent RL baselines~\cite{schulman2017proximal, IglICML2018}. Our results show that autoencoder-based belief inference models achieve significantly higher returns than the other proposed methods and baselines in the different evaluated environments. Further investigation into the information encoded by the belief inference models demonstrates that using autoencoder-based architectures yields representations that more accurately encode the latent environment state and teammates' joint actions. This improved representational quality enables the learner to achieve higher returns when using the resulting representations for decision-making under partial observability. We also investigate the proposed belief inference models' ability to encode the existence of teammates not perceived in the partial observation. Our results demonstrate that SMC-based methods are significantly more effective at encoding unobserved teammates' existence. However, these methods do not translate to higher returns during decision-making, since SMC-based representations are unable to accurately represent the latent environment state and teammates' joint actions. On the other hand, our generalisation results show that autoencoder-based architectures are able to generalise better to different numbers of teammates, as well as to previously unseen teammates during training.

The remainder of this paper introduces our proposed methods and details our experiments in open ad hoc teamwork. In Section~\ref{sec:related} we discuss related works, followed by Section~\ref{sec:formulation} which formalises the learning problem in open ad hoc teamwork. Section~\ref{sec:GPLGeneralOverview} introduces GPL as a method for solving the open ad hoc teamwork problem under full observability. We then describe our fully observable open ad hoc teamwork experiments and analyse the results from GPL in Section~\ref{sec:results_1}. Section~\ref{sec:POGPL} then presents different methodologies to solve the open ad hoc teamwork problem under partial observability, which is followed by the description of our experiments under this setting alongside the analysis of its associated results in Section~\ref{sec:results_2}. Finally, we summarise our findings in this work and provide pointers to potential future work in Section~\ref{sec:conclusions}.\footnote{Parts of this work have previously been published by \citet{rahman2021towards}. This new version includes a new formal model for open ad hoc teamwork under partial observability alongside new algorithms and experimental results that are specifically designed to deal with partial observability.}

%% file: sections/related.tex
\section{Related Work}
\label{sec:related}

\textit{Ad Hoc Teamwork.} Ad hoc teamwork is the problem of training a single agent to perform optimally in a team of unknown teammates \citep{Stone2010-adhoc}. 
Three main assumptions characterise ad hoc teamwork \citep{surveyadhoc}. First is the lack of prior coordination between agents. This means that the learner should be able to cooperate with the team on-the-fly, without the opportunity to rely on previously agreed collaboration strategies. The second assumption is the fact that the learner has no control over its teammates. Lastly, teammates are assumed to be collaborative. That is, all agents in the team have a common goal and are able to take actions that will benefit the team. However, teammates might have additional objectives, that may vary per teammate, and even have different rewards.
Early works in ad hoc teamwork operated under the assumption that the teammate's behaviour was known to the learner~\citep{AAMAS10-adhoc,Agmon2012}. 
Other approaches have relaxed this restrictive assumption from these early works and assumed teammate behaviour to be unknown during the interaction. These approaches ~\citep{AlbrechtReasoning2017,BARRETT2017132} proposed methods that infer teammates' policy based on their displayed behaviour and utilise it for decision making.
Such methods typically utilise the concept of \emph{types}, which encapsulates the important information determining an agent's behaviour. In type-based methods, each teammate's behaviour is assigned a particular hypothesised type~\citep{albrecht2016belief}. Then during the interaction, new unseen teammates are assigned one of the previously hypothesised types~\citep{ravula_ad_2019, Reuth2020}. One issue with these methods is that during training they require diverse teammates, to allow the ad hoc agent to learn policies that are useful for collaborating with novel partners. Current research has thus focused on how to generate diverse teammates for ad hoc teamwork training~\citep{rahman2023generating, rahman2023minimum}.
Another avenue of research within ad hoc teamwork focuses on how communication can be leveraged inside the team to improve the overall performance \citep{barrett_communicating_2014, Reuth2021IAAI}. 
However, all these prior works assume that the teams are closed, meaning that the number of teammates and their types are fixed during episodes.  
A very limited number of works consider the case of open ad hoc teamwork, which poses an even more difficult problem \citep{chandrasekaran_individual_2016}. 

Recent works have addressed the ad hoc teamwork problem under partial observability. \citet{gu2021online} presented ODITS, a reinforcement learning-based approach, which utilises an information-based regulariser to estimate proxy representations based solely on the learner's observations. Recently, \citet{ribeiro2022assisting} presented a Bayesian prediction algorithm for addressing partial observability in ad hoc teamwork. However, these works only considered closed teams. Unlike previous works, our proposal is the first to address the open ad hoc teamwork problem under partial observability.

\textit{Zero-Shot Coordination.} A related problem to ad hoc teamwork is that of zero-shot coordination (ZSC) \citep{huOtherplayZeroshotCoordination2020}. In ZSC agents are paired together and need to coordinate on-the-fly. However, in ZSC reward functions are assumed to be the same across the different agents, 
while our definition of ad hoc teamwork assumes that the learner has no specific knowledge about the rewards of other agents~\citep{surveyadhoc}. Initial works in ZSC proposed an updated version of self-play, called other-play, that aims to break symmetries in MDPs to improve coordination. Other works have focused on how to generate teammates that have diverse policies \citep{TrajectoryDiversityZSC, rahman2023generating}.

\textit{Belief States.} Many approaches to find optimal policies for a \acp{POMDP} are based on computing a probability distribution, or a belief state, regarding the actual state of the learner \citep{Izadi2005UsingRF, albrecht2016causality,AlbrechtExploitingCausality2017}. Other works have also suggested the use of \acp{RNN} to deal with the uncertainty in the observations \citep{Wierstra2007, hausknecht2015deep}. However, these models are incapable of modelling the learner's uncertainty of the latent state due to only modelling the latent state as a single representation. 
A different approach was taken by \citet{NIPS2008_42e77b63}, which proposed the use of a particle filter (or sequential Monte-Carlo) \citep{Arulampalam2002} for estimating the belief state in \acp{POMDP}. 
More recently,~\citet{le2018auto} solved issues of belief update using variational autoencoders trained by optimising \ac{ELBO}. These improvements were later leveraged for estimating belief states in single agent \acp{POMDP} \citep{IglICML2018, SWB2021}.

\textit{Interactive POMDPs (I-POMDPs).} Interactive POMDPs are an extension of POMDPs to the multi-agent domain~\citep{IPOMDP}. It achieves this by including agent models in the state space. As such, I-POMDPs are related to \acp{SBG} \citep{albrecht2016belief}, however, I-POMDPs have mostly focused on the nested belief which makes their solutions complex. \citet{chandrasekaran_individual_2016} leveraged instead I-POMPD-Lite \citep{hoang2013interactive}, a simplified version of I-POMDPs that assumes the behaviour of other agents follows a nested MDP, for solving the planning problem in open multi-agent systems. Our problem formulation is instead based on \acp{SBG}, which utilises the joint action to model the effects of teammates in the observations of the agent~\citep{albrecht2016belief}.

\textit{Graph Neural Networks (GNNs).} GNNs are a newly proposed type of neural network that can work with graph-structured data \citep{Wu2019ACS}.  %
\acp{GNN} have been used for solving a diverse domain of problems, such as chemical reaction prediction \citep{Kien2019},  generative models \citep{Graphite2019} or traffic prediction \citep{GMAN2020}. 
In multi-agent settings, \acp{GNN} have been used for modelling other agents' behaviour \citep{tacchetti_relational_2018}, although only in closed environments. 
In \ac{MARL}, \acp{GNN} \citep{BoehmerDCG,naderializadeh2021graph} have been used to factorise value functions~\citep{GuestrinCG2002} as \acp{CG}. 
Deep \acp{CG} have also been used in ad hoc teamwork \citep{rahman2021towards}, although under full observability.

\textit{Multi-agent Reinforcement Learning (MARL).} MARL explores the application of reinforcement learning to jointly train a collection of agents, whose goal is to maximise their individual returns in each other's presence~\citep{marl-book}. Previous MARL approaches have focused on ideas such as counterfactual credit assignment~\citep{foerster2018a}, joint action value factorisation~\citep{sunehag2017value, rashid2018qmix, BoehmerDCG}, and learning communication protocols between agents~\citep{foerster2016learning, jiang2019graph}. MARL differs from ad hoc teamwork in two aspects. First, MARL approaches assume control over a set of agents during training and execution, whereas ad hoc teamwork only assumes control over the learner. Second, MARL assumes that an agent will only interact with other agents encountered during joint training, while ad hoc teamwork methods neither assumes knowledge nor control over the teammates that are encountered during evaluation. As a result of these differences, MARL has previously been demonstrated to yield poor performances when dealing with teammate policies not encountered during training~\citep{vezhnevetsOPtionsREsponsesGrounding2020, huOtherplayZeroshotCoordination2020}, which we also show in our experiments.

%% file: sections/objective.tex
\section{Problem Formulation}
\label{sec:formulation}

A formal model for ad hoc teamwork must achieve two requirements. First, the model must formalise the interaction between agents and the effects these interactions have towards the information perceived by the \textit{learner}. Second, these models must represent the absence of knowledge regarding teammates' decision-making process. 

The \acf{SBG} model~\citep{albrecht2016belief} fulfils the aforementioned requirements by combining the Stochastic Game~\citep{shapleyStochasticGames1953} and Bayesian Game model~\citep{harsanyiGamesIncompleteInformation1967}. Using the formalism defined in Stochastic Games, \ac{SBG} formally models the effects of the agents' joint actions on the learner's observed states and rewards. At the same time, \ac{SBG} adopts the concept of \textit{types} from Bayesian Games to encapsulate the set of unknown information regarding teammates' decision making process. 

While \acp{SBG} adequately formalise ad hoc teamwork where team sizes are fixed, they are insufficient for \emph{open} ad hoc teamwork due to their inability to formalise the changing number of teammates. 
To address the limitations of the SBG model as a formal definition of open ad hoc teamwork, we introduce the \ac{OSBG} model.
OSBG extends the \ac{SBG} model to open environments by adding components that formalise the changing number of teammates.
Additionally, we present an extension of \acp{OSBG}, that formalises the ad hoc teamwork problem under partial observability, the \ac{PO-OSBG} model.  
We introduce the \ac{OSBG} model in Section~\ref{sec:OSBG}, and in  Section~\ref{subsec:learningFullObs} we specify the learner's learning objective when solving an open ad hoc teamwork problem under the proposed \ac{OSBG}. We then introduce the \ac{PO-OSBG} extension in Section \ref{sec:POOSBG}, and present the learning objectives when solving open ad hoc teamwork problems under partial observability in Section~\ref{sec:LearningObj}.

\subsection{Open Stochastic Bayesian Games (OSBG)}
\label{sec:OSBG}
In this section, we define the \acf{OSBG} model, which is an extension of \ac{SBG} that formalises the open ad hoc teamwork problem.\footnote{This model definition appeared originally in \citet{rahman2021towards}.} We define \ac{OSBG} as follows: 

\begin{definition}
\label{def:OSBGDef}
\normalfont
An \ac{OSBG}, is a 7-tuple containing the following components:
\begin{itemize}
    \item $S:$ The finite state space.
    \item $A:$ The finite set of possible actions for each agent\footnote{$A$ is assumed to be shared between agents for simplicity. This can be extended to cases where agents' action spaces are different by assuming that $A$ is the union of all agents' individual action spaces.}.
    \item $\Theta:$ The finite set of types that can be assumed by teammates.
    \item $N:$ The finite set of possible agents, $N = \{1,2,...,n\}$. 
    \item $\gamma:$ The discount rate. 
\end{itemize}
Before defining the remaining components of an OSBG, we first introduce notations regarding the agent type assignment and action selection under a variable number of agents. Note that a valid action selection and type assignment only allows an agent to be associated with a single type and action. Assuming $\mathcal{P}(S)$, $a^{i}$, and  $\theta^{i}$ denote the power set of set $S$, action selected by agent $i$, and type assigned to agent $i$, respectively, we define notations for valid agent type and action assignments as follows:
\begin{itemize}
    \item $\boldsymbol{A_{N}} = \{a|a\in\mathcal{P}(N \times A), \forall{(i,a^{i}),(j,a^{j})} \in a: i=j \Rightarrow a^{i}=a^{j}\}$ denotes the \textit{joint agent-action space}, which is the set of all possible joint action selections under a variable number of agents. The predicates that define the membership of $a$ to  $\boldsymbol{A_{N}}$ ensure that each agent can only select a single action in a valid joint action selection. Its elements, $a\in \boldsymbol{A_{N}}$,  are referred as \textit{joint agent-actions}. 
    \item $\boldsymbol{\Theta_{N}}=\{\theta|\theta\in\mathcal{P}(N \times \Theta), \forall{(i,\theta^{i}),(j,\theta^{j})}\in\theta: i=j\Rightarrow\theta^{i}=\theta^{j}\}$ is the \textit{joint agent-type space}, which denotes the set of all possible assignments of types under a variable number of agents. Similar to $\boldsymbol{A_{N}}$, the predicates in the membership conditions of $\boldsymbol{\Theta_{N}}$ ensures that each agent can only be assigned a single type. Its elements, $\theta \in \boldsymbol{\Theta_{N}}$, are then referred to as the \emph{joint agent-type}. 
    Each type $\theta^{i}$ in the joint agent-type $\theta$ represents a set of information or action selection mechanisms underlying a different existing agent's behaviour. Note that during action selection, teammates' types are unknown to the learner, since we do not assume knowledge over teammates' types in ad hoc teamwork.
\end{itemize}
With the defined notation of $\boldsymbol{A_{N}}$ and $\boldsymbol{\Theta_{N}}$ and with $\Delta(X)$ denoting the set of all probability distributions over a random variable $X$, the remaining components of an OSBG are defined as follows:
\begin{itemize}
    \item $R:S\times \boldsymbol{A_{N}} \mapsto \mathbb{R}$, which is the reward function that determines the rewards received by the learner, given the actions of all agents in the environment. 
    \item $P: S \times \boldsymbol{\Theta_{N}} \times \boldsymbol{A_{N}} \mapsto \Delta(S \times \boldsymbol{\Theta_{N}})$, which is the transition function which determines the next state and joint agent-types encountered by the learner, given the current state, joint agent-types and joint agent-actions.
\end{itemize}

The interaction between a learner and its teammates in an \ac{OSBG} starts from an initial state, $s_{0} \in S$. To model changing numbers of agents in the open environment, different subsets of agents are sampled from $N$ to model the set of existing agents in the environment at each timestep. At the beginning of the episode, the initial set of teammates, $N_0\subseteq{N}$ is sampled and assigned the joint agent-type $\theta_{0}\in\boldsymbol{\Theta_{N}}$. The initial state ($s_{0}$), set of teammates ($N_0$), and joint agent-type ($\theta_{0}$), are sampled from the initial distribution $P_{0} \in \Delta(S \times \boldsymbol{\Theta_{N}})$. As an example, if the task would be to play a game of football then each type $\theta^{i}$ will correspond to  a different policy that controls a teammate to play a specific position (e.g winger, defender, or striker) with different skill levels.

At each timestep, the interaction between the learner and its teammates undergoes two distinct processes. First, teammates select their respective actions according to the observed state of the environment $s_{t}$ at time  $t$. Each teammate selects its action based on its current policy, $\pi : S\times{N}\times\Theta\mapsto\Delta(A)$, conditioned on the state, the existing set of agents, and its assigned type.  In the football example, each teammate will select their action based on their own sequence of observations. Meanwhile, the learner chooses its actions based on its sequence of previously observed states and executed actions, $H_{t} = \{s_{\leq{t}}, a_{<t}\}$, without knowing teammates' types or actions, which formalises the lack of knowledge regarding teammates' decision-making process assumed by ad hoc teamwork problems. Unlike its teammates, the learner chooses its actions based on $H_{t}$ because it has no knowledge of its teammates' types and must infer it through their observed behaviour throughout the interaction.

The second step occurs as a result of the execution of joint actions chosen by agents at the first step. Following the execution of the agents' joint action, the learner receives a scalar reward, $r_{t}$, which is determined by the reward function $R:S\times \boldsymbol{A_{N}} \mapsto \mathbb{R}$. The environment state, the set of existing teammates, and the joint agent-type all change following the transition function defined by $P: S \times \boldsymbol{\Theta_{N}} \times \boldsymbol{A_{N}} \mapsto \Delta(S \times \boldsymbol{\Theta_{N}})$. Aside from determining the next state observed by the learner, the transition function $P$ also models the way teammates may enter or leave the environment. This is done by determining the set of teammates encountered by the learner at the next timestep and alongside their respective types.
\end{definition}

\subsection{Learning Objective Under Full Observability}
\label{subsec:learningFullObs}

Solving an OSBG requires the learner, denoted by $i$, to find an optimal policy, $\pi^{i,*}$, which selects the optimal action based on the learner's previously experienced environment states, observed teammates, and executed actions, $H_{t} = \{s_{\leq{t}}, a_{<t}\}$. We define the optimal policy as follows:

\begin{definition}
\normalfont Let the joint actions and the joint policy of teammates at time $t$ be denoted by $a^{-i}_{t}$ and $\boldsymbol{\pi}^{-i}_{t}$, respectively. Given $0\leq\gamma<1$, and the learner's previous experience, $H = \{s_{\leq{t}}, a_{<t}\}$, we define the action-value of a policy $\pi^{i}$, as: 
\begin{equation}
    \label{def:LearningObjOSBG}
    \bar{Q}_{\pi^{i}}(H,a^{i}) = \mathbb{E}_{\substack{\theta^{-i}_{t}\sim{p(.|H_{t})}, \text{ } a^{i}_{T} \sim \pi^{i}, \\ a^{-i}_{T}\sim{\pi}^{-i}_{T}(\cdot|s_{T},\theta^{-i}_{T}), \\(s_{T+1},\theta^{-i}_{T+1})\sim P(.,.|s_{T},\theta^{-i}_{T}, a_{T})} \\}\Bigg[\sum_{T=t}^{\infty} \gamma^{T-t}R(s_{T}, a_{T})\bigg| \, H_{t} = H, a^{i}_{t}=a^{i}\Bigg].
\end{equation}
A learner's policy, $\pi^{i,*}$, is then optimal if:
\begin{equation}
    \bar{Q}_{\pi^{i,*}}(H,a^{i}) \geq \bar{Q}_{\pi^{i}}(H,a^{i}),
\end{equation}
\noindent for all possible $\pi^{i}$, $H$, and $a^{i}$. Given $\bar{Q}_{\pi^{i,*}}(H,a^{i}_{t})$, a learner's optimal policy is to greedily choose actions with the highest state-action value. Note that to remove any ambiguity in the text we use the bar notation $\bar{Q}_{\pi^{i}}$ to denote the action-value of the learner. 
\end{definition}

\subsection{Partially Observable Open Stochastic Bayesian Games (PO-OSBG)}
\label{sec:POOSBG}

In this section, we define the \ac{PO-OSBG} model, which is an extension of OSBG to problems with partial observability. PO-OSBG extends OSBG by introducing components which model the observations received by the learner during interaction with its teammates. We define the PO-OSBG model as follows:

\begin{definition}
\normalfont A PO-OSBG is an 9-tuple, consisting of the following components $(N, S, A, \allowbreak \Theta, \allowbreak R, \allowbreak P, \Omega, O, \gamma)$. In a PO-OSBG, $N,S,A,$ $\Theta$, $R,$ $P$, and $\gamma$ are defined exactly as their respective counterparts in an OSBG. The remaining components of a PO-OSBG which model the information received by the learner under partial observability are defined as follows: 
\begin{itemize}
    \item $\Omega$: The learner's set of possible finite observations.
    \item $O: S\times{N} \mapsto \Delta(\Omega$), which is the observation function that determines the distribution of observations received by the learner given the current state of the environment and the learner's set of teammates. 
\end{itemize}
\end{definition}

In a PO-OSBG, the interaction between a learner and its teammates is similar to their interactions in OSBGs. The main difference is that the learner will not perceive the subsequent state information after all agents execute their respective actions. Instead, at each timestep the learner receives an observation sampled from the distribution outputted by the observation function $O: S\times{N} \mapsto \Delta(\Omega)$, which is determined by the state and set of teammates that exist in the environment at $t$. This absence of state information forces the learner $i$ to choose actions based only on the sequence of observations and its actions until the present time, $H_{t} = \{o_{\leq{t}}, a^{i}_{<t}\}$. On the other hand, teammates receive their own observations of the environment according to their own observation function. We assume that teammates' observation functions are part of their type, and as a consequence unknown to the learner. This means that the PO-OSBG formulation defines the observation function only for the learner. The main reason behind this design is because many scenarios have learners that have different perception capabilities than its teammates. For instance, robots created from different factories may be equipped with different sets of sensors. Since teammates' observation functions are unknown and important for their decision making process, the unknown observation function of teammates are instead encapsulated as part of their types under the PO-OSBG model.

\subsection{Learning Objective Under Partial Observability}
\label{sec:LearningObj}
Solving a \ac{PO-OSBG} amounts to finding an optimal policy for action selection based on the sequence of the learner's previous observations and executed actions, $\pi^{i,*}( \cdot | \, o_{\leq{t}},a^{i}_{<t})$. We define the optimal policy as follows:
\begin{definition}
\normalfont Let the unknown joint actions and the joint policy of teammates at time $t$ be denoted by $a^{-i}_{t}$ and $\boldsymbol{\pi}^{-i}_{t}$, respectively. Given $0\leq\gamma<1$, and $H_{t} = \{o_{\leq{t}}, a^{i}_{<t}\}$, the action-value of a policy $\pi^{i}$, is defined as: 
\begin{equation}
    \label{def:LearningObjPOOSBG}
    \bar{Q}_{\pi^{i}}(H_{t},a_t^{i}) = \mathbb{E}_{\substack{(s_{t}, \theta_{t}^{-i})\sim{p(.,.|H_{t})},\text{ } a^{i}_{T} \sim \pi^{i}, \\ a^{-i}_{T}\sim\pi^{-i}_{T}(\cdot|s_{T},\theta^{-i}_{T}),\\ (s_{T+1},\theta^{-i}_{T+1})\sim P(.,.|s_{T},\theta^{-i}_{T}, a_{T}), \\ o_{T+1}\sim{O(.|s_{T+1})}}}\Bigg[\sum_{T=t}^{\infty} \gamma^{T-t}R(s_{T}, a_{T})\bigg| \, H_{t} , a^{i}_{t}\Bigg].
\end{equation}
A learner's policy, $\pi^{i,*}$, is then optimal if:
\begin{equation}
    \bar{Q}_{\pi^{i,*}}(H_t,a_t^{i}) \geq \bar{Q}_{\pi^{i}}(H_t,a_t^{i}),
\end{equation}
\noindent for all possible $\pi^{i}$, $H_t$, and $a^{i}$. The learner's optimal policy is to then greedily choose actions with the highest state-action value for any given $H_{t}$ experienced by the learner. 

\end{definition}

%% file: sections/methods_1.tex
\section{Open Ad Hoc Teamwork in Fully Observable Environments}
\label{sec:GPLGeneralOverview}

We present here a general learning framework designed to achieve optimal decision-making in open ad hoc teamwork. We first provide a general overview regarding the role of the three main components that constitute our proposed method \acfi{GPL}. We then describe details of the models designed for each component, starting with the type inference component in Section~\ref{sec:TypeInference}, the joint action value modelling component in Section~\ref{sec:JointActionValModelling}, and the agent modelling component in Section~\ref{sec:GPLAgentModelling}. Finally, Section~\ref{sec:GPLActionSelection} details the learner's action selection process. and Section~\ref{sec:GPLLearningObjective} outlines the learning objective for training GPL's neural network-based components.

\begin{figure*}[t]
    \centering
    \includegraphics[width=\textwidth, trim={0.5cm 11.5cm 0.cm 0.35cm},clip]{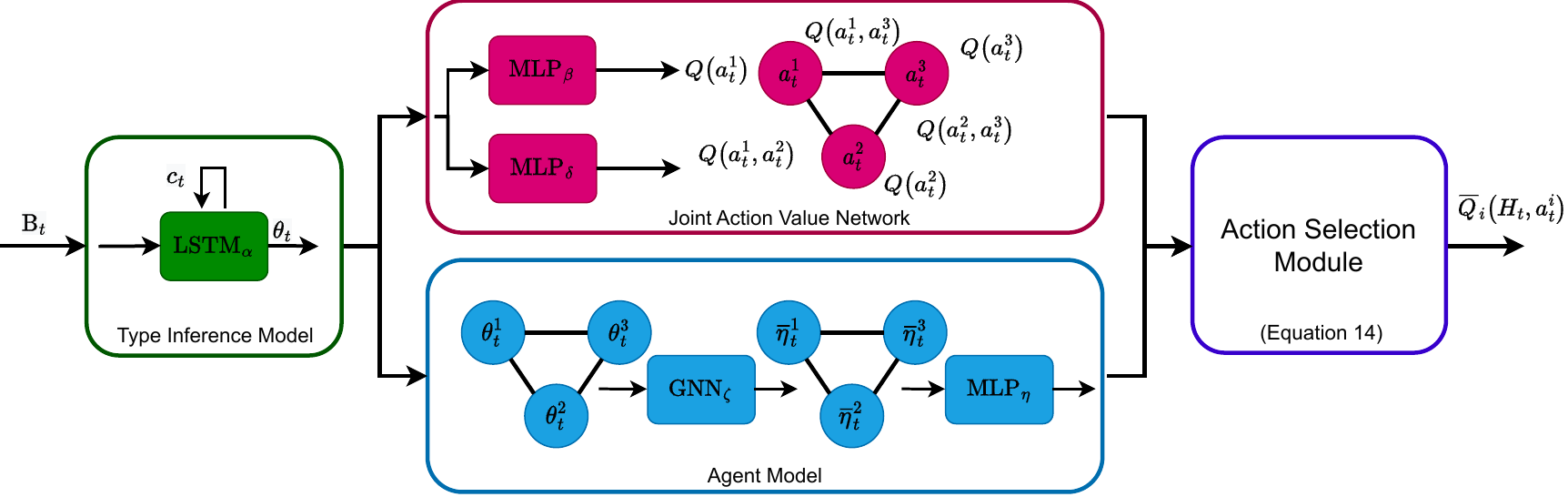}
    \caption{\textnormal{Overview of GPL.} GPL receives a collection of input vectors, $B_{t}$, containing state features of currently existing agents in the current state. The type inference model, parameterised by $\alpha$, then produces a type vector $\theta_t$ based on $B_{t}$. The type vectors from the type inference model are then inputted to the joint-action value network and the agent model parameterised by $(\beta, \delta)$ and $(\zeta,\eta)$ respectively. The joint action value network outputs individual $Q_{\beta}(a^j_t)$, and pairwise value function estimates $Q_{\delta}(a^j_t,a^k_t)$. Meanwhile, the agent model network outputs the likelihood of taking action $a$ for each agent $q_{\zeta,\eta}(a^j,s_t)$. The outputs from the joint action value network and the agent model are then combined in the action selection module, to compute the learner's action-value function ($\bar{Q}_{\pi^{i}}$), following Equation~\eqref{SingleActionValue}. The obtained $\bar{Q}_{\pi^{i}}$ is then used to select the learner's optimal action.}
    \label{Fig:GPL_overview}
\end{figure*}

\subsection{Overview}

 Our \ac{GPL} framework is based around three main components: a \textit{type inference model}, a \textit{joint action value model}, and an \textit{agent model}, whose role in decision making is explained in the following sections. These components are trained to estimate the learner's optimal policy using the experience collected by the learner while interacting with its teammates. The main components and interactions between them are illustrated in Figure~\ref{Fig:GPL_overview}.

The \textit{type inference model} is needed to infer teammates' unknown types in open ad hoc teamwork. Knowing teammates' types is crucial for decision-making since it gives information regarding teammates' selected actions, which can have a strong impact on the returns achieved by the learner. In the absence of knowledge regarding teammates' inherent types, the type inference model infers each teammate's type based solely on the sequence of past state features observed by the learner.

A \textit{joint action value model} predicts the learner's returns following existing agents' joint actions. Modelling the joint action value is crucial to solving open ad hoc teamwork for two reasons. First, the agents' joint actions influence the learner's current and future rewards, following the reward and transition function definition in OSBGs. Second, the joint action value estimation provides better credit assignment than value-based RL methods that directly model the learner's action-value function, such as Q-Learning~\citep{watkins1992q}. 
Joint action value estimation prevents the agent from assigning too much credit to its action when it has minimal impact on the observed rewards, which has been demonstrated to be crucial for credit assignment in previous works related to MARL~\citep{lowe2017multi, foerster2018a}. In this work, we extend its applicability to ad hoc teamwork by incorporating a joint action value model as part of GPL. Section~\ref{Sec:GPLJointActValAnalysis} shows that the joint action model contributes towards improved credit assignment, which enables the learner to identify and learn useful behaviour from well-performing teammates. 

Given the joint agent-action $a$, the joint-action value of a learner's policy, $\pi^{i}$, is defined as follows:
\begin{equation}
Q_{\pi^{i}}(H, a) = \mathbb{E}_{\substack{\theta^{-i}_{t} \sim p(.|H), a^{i}_{T} \sim \pi^{i}, \\ a^{-i}_{T}\sim{\pi}^{-i}_{T}(\cdot|s_{T},\theta^{-i}_{T}),\\ (s_{T+1},\theta^{-i}_{T+1})\sim P(.,.|s_{T},\theta^{-i}_{T}, a_{T})}}\Bigg[\sum_{T=t}^{\infty} \gamma^{T-t}R(s_{T}, a_{T})\bigg| \, H_{t} = H, a_{t}=a\Bigg].
\label{Eq:JointActionReturns}
\end{equation}
In contrast to Equation~\eqref{def:LearningObjOSBG}, this joint-action value denotes the learner's expected returns after joint agent-action $a$ is executed after a history of previously observed states and actions $H$, assuming other agents follow $\pi^{-i}$, while the learner follows $\pi^{i}$. 
This value is directly influenced by the set of existing teammates and their respective types, which determines the joint action selected by the joint teammate policy $\pi^{-i}$. GPL accounts for the influence of existing teammate types to $Q_{\pi^{i}}(H, a)$ by incorporating the inferred teammate types when estimating this value. 

It is not possible to use Equation~\eqref{Eq:JointActionReturns} directly to decide the learner's optimal action. Using the joint action value model to decide the learner's optimal action requires knowledge about actions that will be selected by teammates, a fact that is unknown to the learner when deciding its own actions. Nonetheless, we can still use the joint action value estimate for decision-making by exploiting the following equation:
\begin{equation}
    \bar{Q}_{\pi^{i}}(H_{t},a_{t}^{i}) = \mathbb{E}_{a_{t}^{-i}\sim{\pi}^{-i}(\cdot|s_{t},\theta^{-i}_{t})}\bigg[Q_{\pi^{i}}(H_{t},a_{t})\bigg|{}a^{i}=a_{t}^{i}\bigg],
    \label{Eq:Marginalisation}
\end{equation}
which expresses the learner's action value function in terms of $Q_{\pi^{i}}(H_{t}, a)$. Equation~\eqref{Eq:Marginalisation} dictates that the learner's action value is the expected value of $Q_{\pi^{i}}(H_{t},a)$ under the distribution of teammates' actions. In problems with discrete possible actions, $\bar{Q}_{\pi^{i}}(H_{t},a_{t}^{i})$ may therefore be computed by evaluating $Q_{\pi^{i}}(H_{t}, a)$ for all possible joint actions and computing their weighted average according to teammates' joint action probability.

Equation~\eqref{Eq:Marginalisation}, highlights the importance of the \textit{agent model}, which is the third component of GPL. The agent model's role is to estimate teammates' joint action likelihood $\pi^{-i}(a^{-i}_{t}|s_{t},\theta^{-i}_{t})$. Estimating the likelihood gives the learner predictions regarding which actions will be selected by the teammates, which enables the learner to use its joint action value model to compute $\bar{Q}_{\pi^{i}}(H_{t},a_{t}^{i})$. 
Note that the learner's prediction regarding teammates' actions is based on the environment state and the teammates' inferred types. The use of state and inferred teammate types for action prediction follows from OSBG's formulation of teammate's decision-making process as outlined in Section~\ref{def:OSBGDef}. Definition~\ref{sec:GPLAgentModelling} further details a model which predicts teammates' actions based on the state and inferred teammates' types.

In the following subsections, we will describe the models implemented for each of the aforementioned components. We start by describing the type inference model in Section~\ref{sec:TypeInference}, followed by a description of the joint action value model in Section~\ref{sec:JointActionValModelling}, and of the agent model in Section~\ref{sec:GPLAgentModelling}. We then explain how the output of these models are combined for action selection in Section~\ref{sec:GPLActionSelection}. Section~\ref{sec:GPLLearningObjective} concludes our method description by outlining the learning objective for training the defined models based on the learner's experience interacting with teammates. Additionally, we present in Algorithm \ref{alg:SimplifiedGPL} a simplified pseudocode of GPL. A more detailed pseudocode can be found in Appendix~\ref{sec:GPLPseudocode}.  

\begin{algorithm*}[t]
\caption{Simplified GPL}\label{alg:SimplifiedGPL}
\begin{algorithmic}[1]
\STATE \textbf{Input:} Number of training steps $T$  
\STATE Initialise network parameters
\STATE Get initial observation $s_t$ from environment
\FOR{$t=1$ to $T$}
    \STATE $B_t$ = \textbf{Preprocess} ($s_t$) 
    \STATE  $\theta_{t}$ = \textbf{Type\_inference\_network} $(B_t, c_{t-1}, \theta_{t-1})$
    \STATE $\pi^{-i}(\cdot|H_{t})$  = \textbf{Agent\_model} $(\theta_{t})$
    \STATE $ Q^{j}(a_t^{j}|H_{t})$, $Q^{j,k}(a_{t}^{j},a_{t}^{k}|H_{t})$ =  \textbf{Joint\_action\_value\_model} $(\theta_{t})$
    \STATE  $\bar{Q}(H_{t}, a^{i}_{t})$ = \textbf{Action\_selection} $(\pi^{-i}(\cdot|H_{t}), \,  Q^{j}(a_t^{j}|H_{t}), \, Q^{j,k}(a_{t}^{j},a_{t}^{k}|H_{t}))$
    \STATE Sample action according to the learning algorithm being used, \\\begin{center}
   $a^{i}_{t} \sim \begin{cases}
    \text{$\epsilon$-greedy}(\epsilon, \bar{Q}(H_t,\cdot)),& \text{if Q-Learning}\\
    p_{\text{SPI}}(\tau,\bar{Q}(H_t,\cdot))             & \text{if SPI}
    \end{cases}
    $
    \end{center}
    \STATE Execute action $a_t^i$, get $r_t, s'$ and $a_t^{-i}$
    \STATE Accumulate parameter gradients for update
\IF{$t$ mod $t_{update}$ } 
\STATE Update networks following Eq.\eqref{ValueLoss}-\eqref{Boltzmann}
\ENDIF
\ENDFOR \label{euclidendwhile}
\end{algorithmic}
\end{algorithm*}

\subsection{Type Inference}
\label{sec:TypeInference}

There are three challenges in designing type inference models for open ad hoc teamwork. 
First, the model must accurately predict the teammates' types, even when interacting with previously unseen teammates. Second, the model must learn without ground truth knowledge regarding the current types of other members of the team. 
Third, the model should be able to handle inputs of different sizes, since the number of agents can vary between timesteps. 
In many real-world applications of ad hoc teamwork, the first two challenges result from the absence of any type-related information pertaining to teammates. For instance, obtaining ground truth types and knowledge of the wide-ranging type space of human agents can be difficult in applications of ad hoc teamwork to human-robot interaction.
The third challenge is directly related to the problem of openness in ad hoc teamwork.

We address the aforementioned challenges by representing teammate types as continuous vectors.
To compute such vectors we utilise \acp{RNN}, which are trained to produce similar type vectors for teammates that have similar behaviour to each other. The RNN infers the current teammates' type based on the input vector and the RNN hidden states. In this way, the current inferred type not only depends on the current observed state but on the sequence of observed behaviour for each teammate. 
As a result, the type inference model can be trained without ground truth teammate types while also generalising well against unseen teammates if the learner has previously interacted with teammates displaying similar behaviour.

The type inference model is implemented as a \ac{LSTM} network~\citep{SchmidhuberLSTM} with parameters are denoted by $\alpha$. Assuming that $\theta_{t}$ and $c_{t}$ are the hidden and cell states of the \ac{LSTM} at timestep $t$, the LSTM updates the type vectors following this expression:
\begin{equation}
    \label{Eq:GPLTypeInference}
    c_{t}, \theta_{t} = \text{LSTM}_{\alpha}(B_t, c_{t-1}, \theta_{t-1}),
\end{equation}
where $B_t$ is an input batch that contains information about the current state of the system. This LSTM-based type update is illustrated on the left side of Figure~\ref{Fig:GPL_overview}.

After the update process, additional steps are required to ensure only type vectors of existing agents are used in GPL's optimal action value estimation. Between subsequent timesteps, the type inference model removes the type vectors of teammates that are no longer in the environment due to environment openness. At the same time, type vectors of newly-arrived teammates are added. More details on how state information is preprocessed and post-processed are given in Appendix~\ref{sec:GPLInputPreprocessing}. 

The parameters $\alpha$ of the type inference network are optimised by back-propagating the losses from the agent model and the joint action value network. In practice, we utilise two separate networks, one that feeds the Joint Action Value model and one that feeds the Agent Model. This is done to avoid the losses from each model interfering during training. More details on how these losses are computed are given in Section~\ref{sec:GPLLearningObjective} and in Appendix~\ref{sec:GPLInputPreprocessing}.

\subsection{Joint Action Value Model}
\label{sec:JointActionValModelling}

A joint action value model for open ad hoc teamwork must address three challenges. First, the model must be capable of handling inputs of variable sizes resulting from environment openness.  Second, it must facilitate efficient computation of the learner's action value function based on Equation~\eqref{Eq:Marginalisation}. Third, the model must also estimate the effects of teammates' actions towards the learner's returns. 

One way to fulfil the aforementioned requirements is to represent the joint action value model as a fully connected \acf{CG}~\citep{guestrin2002coordinated}. CGs facilitate the factorisation of joint action value functions into singular and pairwise utility terms, which we demonstrate in Section~\ref{sec:GPLActionSelection} to have enabled a more efficient action value computation process. Implementation of CG models can also be based on GNNs~\citep{BoehmerDCG}, which are designed to handle inputs of variable sizes. Finally, CG's joint action value factorisation also enables modelling the effects of teammates' individual and pairwise actions on the learner's returns, as demonstrated in Section~\ref{sec:results_1}.

Given a history of past states and actions from the learner, $H_{t}$, and a set of existing agents $N_{t}$, a fully connected CG factorises the learner's joint action value into the sum of singular utility terms, $Q_{\pi^{i}}^{j}(a^{j}_{t}|H_{t})$,  and pairwise utility terms, $Q_{\pi^{i}}^{j,k}(a_{t}^{j},a_{t}^{k}|H_{t})$. The joint action value factorisation for a fully connected CG follows this Equation:
\begin{equation}
    \label{JointActionValueComputation}
    Q_{\pi^{i}}(H_{t}, a_{t}) = \sum_{j\in{N_{t}}} Q^{j}_{\pi^{i}}(a^{j}_{t}|H_{t}) + \sum_{\substack{j,k\in{N_{t}}\\j\neq{k}}}Q^{j,k}_{\pi^{i}}(a_{t}^{j},a_{t}^{k}|H_{t}).
\end{equation}
In terms of the contributions towards the learner's returns, $Q^{j}_{\pi^{i}}(a^{j}_{t}|H_{t})$ can be viewed as the contribution of agent $j$'s action $a^{j}$, while $Q^{j,k}_{\pi}(a^{j},a^{k}|H_{t})$ is the contribution of agents $j$ and $k$ jointly choosing $a^{j}$ and $a^{k}$ respectively.

To enable generalisation across different input $H_{t}$, $Q^{j}_{\pi_{i}}(a^{j}_{t}|H_{t})$ and $Q^{j,k}_{\pi_{i}}(a_{t}^{j},a_{t}^{k}|H_{t})$ are implemented as multilayer perceptrons (MLPs) parameterised by $\beta$ and $\delta$ respectively. For two reasons, both models that compute the singular and pairwise utilities receive input solely consisting of agents' type representations outputted by the type inference network instead of $H_{t}$. First, the output of the type inference network contains information regarding the unknown teammate types,
Second, it also contains important information on $s_{t}$ since $s_{t}$ is used as input for the type inference model.
This way of calculating the joint action depends heavily on the type inference network. So if the type inference is unable to adequately classify the teammates' types, we expect the joint value to produce poor estimates. This could be the case in teams that have very rapid changes in composition. However, our results have shown that the type network is able to produce good estimates, at least for the evaluated environments.

We define the types as $\theta^{i}_{t}$ and $\theta^{j}_{t}$, where $\theta^{i}_{t}$ is the type vector associated to the learner and $\theta^{j}_{t}$ is the type vector of agent $j$. These type vectors are provided as input to $\text{MLP}_{\beta}$ and $\text{MLP}_{\delta}$, which allows the estimation of agents' individual and pairwise action contributions towards the learner's returns. Given the types vectors as input, $\text{MLP}_{\beta}$ outputs a vector with a length of $|A|$ that estimates $Q^{j}_{\beta, \alpha}(a^{j}|s_{t})$ for each possible actions of $j$ following:
\begin{equation}
    Q^{j}_{\beta, \alpha}(a^{j}|H_{t}) = \text{MLP}_{\beta}(\theta^{j}_{t},\theta^{i}_{t})(a^{j}).
\end{equation}

Instead of outputting the pairwise utility for the $|A|\times|A|$ possible pairwise actions of agent $j$ and $k$, $\text{MLP}_{\delta}$ outputs an $K\times{|A|}$ matrix ($K\ll{|A|}$) given its type vector inputs. $\text{MLP}_{\delta}$ computes its output matrix solely based on the type vectors, following the same reasoning as $\text{MLP}_{\beta}$. Assuming a low-rank factorisation of the pairwise utility terms, the output of $\text{MLP}_{\delta}$ is used to compute $Q^{j,k}_{\delta,\alpha}(a_{t}^{j},a_{t}^{k}|H_{t})$ with the following equation: 
\begin{equation}
    Q^{j,k}_{\delta,\alpha}(a_{t}^{j},a_{t}^{k}|H_{t}) =  (\text{MLP}_{\delta}(\theta^{j}_{t}, \theta^{i}_{t})^{\mathsf{T}}  \text{MLP}_{\delta}(\theta^{k}_{t},\theta^{i}_{t}))(a^{j}_{t},a^{k}_{t}).
\end{equation}
We expect that this way of computing the joint-action value will work well on small teams, and this is confirmed by our results in Section~\ref{sec:results_1}. In fact, previous work from~\citet{zhou2019factorized} demonstrated that low-rank factorisation enables scalable pairwise utility computation even under thousands of possible pairwise actions. However, evaluating on teams of such dimensions is out of the scope of this work. 

Finally, note that we use the same parameters for $\text{MLP}_{\beta}$ and $\text{MLP}_{\delta}$ to model each teammate or pair of teammates, to encourage knowledge reuse for utility term computation. We show the importance of knowledge reuse via parameter sharing to GPL's performance in Section~\ref{Sec:GPLJointActValAnalysis}.

\subsection{Agent Modelling}
\label{sec:GPLAgentModelling}

Due to the openness related to ad hoc teamwork, the agent model has to efficiently predict the joint action probability distribution, $\pi^{-i}(.|s_{t}, \theta_{t}^{-i})$, of a variable number of teammates. 
In order to deal with this issue, we implement the agent model as a GNN, and utilise as input the inferred types from the type inference model.
This GNN-based agent model facilitates an efficient computation of the learner's optimal action value estimate, as we will see in Section~\ref{sec:GPLActionSelection}. 

While the agent model assumes that the teammates choose their actions independently, it models the potential effect that teammates have on each other by using the Relational Forward Model (RFM) architecture~\citep{DBLP:conf/iclr/TacchettiSMZKRG19}. As in other GNN models, RFM contains message passing operations, which enables improved reasoning regarding the relations between nodes in a graph. Furthermore, RFM has been demonstrated to be accurate in predicting the likelihood of agents' next actions ~\citep{DBLP:conf/iclr/TacchettiSMZKRG19}. 

The RFM-based agent model only receives agents' types as input. The type of all existing agents, $\theta_{t}$, is then treated as node input to compute a fixed-length embedding, $\bar{n}$, for each agent $j$, as: 
\begin{equation}
\bar{n}_{j} = (\text{GNN}_{\zeta}(\theta_{t}))_{j}.
\end{equation}

This embedding is used together with the actions taken by agent $j$ to obtain a likelihood estimate: 
\begin{equation}
q_{\zeta,\eta, \alpha}(a^{j}|s) = \textrm{Softmax}(\text{MLP}_{\eta}(\bar{n}_{j}))(a^{j}).
\end{equation}

Each individual estimate is then combined for each agent to obtain the likelihood of taking action $a^{-i}$ at state $s$:  
\begin{equation}\label{GNNequations}
\pi^{-i}(.|s_{t}, \theta_{t}^{-i}) \approx q_{\zeta,\eta,\alpha}(.|s_{t}, \theta_{t}^{-i}) = \prod_{j\in{-i}} q_{\zeta,\eta,\alpha}(a^{j}|s),
\end{equation}

As we will see in the following section, the obtained likelihood and the joint action value modelling can be utilised for computing the optimal policy for the learner. 
It is important to note that, while teammates can leave and enter the environment at any time, the observed teammates have fixed policies. Types can indeed change as agents leave and enter the environment, but when an agent enters the environment its type remains the same. Having teammates that change their behaviour over time (i.e. during an episode), might require more complex agent modelling techniques~\citep{albrecht_autonomous_2018, xie2021learning}. However, we do not make any further assumptions regarding how other agents choose their actions.

\subsection{Action Selection}
\label{sec:GPLActionSelection}

Computing the exact value of Equation~\eqref{Eq:Marginalisation} for action selection can be challenging in many practical applications. For instance, a team of $k$ agents which may choose from $n$ possible actions requires the evaluation of $n^{k}$ joint-action terms. This exponential increase in the number of terms makes the evaluation of Equation~\eqref{Eq:Marginalisation} unfeasible for large teams. 

A way to reduce the computational complexity of evaluating Equation~\eqref{Eq:Marginalisation} is to factorise $Q_{\pi^{i}}(s_{t}, a_{t})$ and $\pi^{-i}(a^{-i}_{t}|s_{t},\theta^{-i}_{t})$ into simpler terms. For example, we can use Equation~\eqref{JointActionValueComputation} and Equation~\eqref{GNNequations} to define an action-value function that is factorised into smaller terms.
Substituting the joint-action value and agent models from Equation~\eqref{JointActionValueComputation} and Equation~\eqref{GNNequations} into  Equation~\eqref{Eq:Marginalisation} results in an action value function with the following expression:
\begin{align}\label{SingleActionValue}
\begin{split}
\bar{Q}(H_{t}, a^{i}_{t}) & = Q^{i}_{\beta, \alpha}(a^{i}_{t}|H_{t})  \\& + \sum_{ \mathclap{\substack{a^{j}\in{A_{j}},  j\neq{i}}}}\big(Q^{j}_{\beta, \alpha}(a^{j}|H_{t})+Q^{i,j}_{\delta, \alpha}(a^{i}_{t}, a^{j}|H_{t})\big)q_{\zeta,\eta, \alpha}(a^{j}|s_{t}) \\
&+ \sum_{\mathclap{\substack{a^{j}\in{A_{j}},a^{k}\in{A_{k}}, j\neq{i}, k\neq{i}}}}Q^{j,k}_{\delta, \alpha}(a^{j}, a^{k}|H_{t})q_{\zeta,\eta, \alpha}(a^{j}|s_{t})q_{\zeta,\eta, \alpha}(a^{k}|s_{t}).
\end{split}
\end{align}

Unlike Equation~\eqref{Eq:Marginalisation}, Equation~\eqref{SingleActionValue} is defined in terms of singular and pairwise action terms. This limits the number of computed terms to only increase quadratically as the team size increases. Furthermore, the computation of the singular and pairwise terms in Equation~\eqref{SingleActionValue} can be efficiently done in parallel with existing GNN libraries~\citep{wang2019dgl}.

\subsection{Learning Objective}
\label{sec:GPLLearningObjective}

Optimising GPL's models requires interaction experiences that are collected by the learner. We assume that the learner collects these experiences according to an $\epsilon$-greedy action selection policy with its action value computation method as described in Section~\ref{sec:GPLActionSelection}. Given a batch of interaction experiences $D = \{(H^{n}_{t}, a^{n}_{t}, r^{n}_{t}, H^{n}_{t+1})\}_{n=1}^{|D|}$, the agent modelling network is trained to estimate $\pi(a^{-i}_{t}|s_{t},\theta^{-i}_{t})$ through supervised learning by minimising the negative log likelihood loss defined below:
\begin{align}\label{ActionModelLoss}
\begin{split}
    L_{\zeta, \eta, \alpha}(D) &= \sum_{(H_{t}, a_{t}, r_{t}, H_{t+1})\in{D}}\left(-\sum_{j\in{-i}}\text{log}(q_{\zeta,\eta, \alpha}(a^{j}_{t}|s_{t}))\right).
\end{split}
\end{align}
Also, the collected data set is used to update GPL's joint-action value network using value-based reinforcement learning. Unlike standard value-based deep reinforcement learning approaches~\citep{mnih2015human}, we use the joint action value as the predicted value. The loss function for the joint action value network is defined as:
\begin{equation}
    L_{\beta,\delta,\alpha}(D) = \sum_{(H_{t}, a_{t}, r_{t}, H_{t+1})\in{D}}\left(\dfrac{1}{2}\left(Q_{\beta,\delta,\alpha}\left(H_{t}, a_{t}\right)-y\left(r_{t}, H_{t+1}\right)\right)^2\right),
    \label{ValueLoss}
\end{equation}
with $y(r_{t}, H_{t+1})$ being a target value which depends on the algorithm being used. We train GPL with Q-Learning (GPL-Q) ~\citep{watkins1992q} and Soft-Policy Iteration (GPL-SPI)~\citep{haarnoja2018soft}, which produces a greedy and stochastic policy, respectively. The target value computations of GPL-Q and GPL-SPI are defined as the following:
\begin{align}
    y_{\mathrm{QL}}\left(r_{t}, H_{t+1}\right) &= r_{t} + \gamma \text{max}_{a^{i}}\bar{Q}\left(H_{t+1},a^{i}\right),\\
    y_{\mathrm{SPI}}\left(r_{t}, H_{t+1}\right) &= r_{t} + \gamma \sum_{a^{i}}p_{\mathrm{SPI}}(a^{i}|H_{t+1})\bar{Q}\left(H_{t+1},a^{i}\right).
    \label{TargetValues}
\end{align}
GPL-SPI's target values in Equation~\eqref{TargetValues} assume that the learner's policy selects actions using the following expression:
\begin{equation}\label{Boltzmann}
    p_{\mathrm{SPI}}(a^{i}_{t}|H_{t}) \propto \mathrm{exp}\left(\dfrac{\bar{Q}(H_{t},a^{i})}{\tau}\right),
\end{equation}
with $\tau$ being the temperature parameter.

Finally, the optimisation of the type inference model is carried out with the losses defined in Eq.~\eqref{ActionModelLoss} and~\eqref{ValueLoss}. These losses are back-propagated through the type inference network. This allows us to train the type inference network without knowledge of other agents' types. More details about the inputs and outputs of the type inference are given in the Appendix~\ref{sec:GPLInputPreprocessing}. 

One important aspect of the training is how the data is collected to obtain the buffer $D$. In practice the learner could collect all the data in an experience replay buffer, and then sample transitions to optimise the model~\citep{mnih2015human}. However, we utilise a synchronous data collection mechanism, based on the asynchronous Q-learning~\citep{mnih2016asynchronous}. Instead of using an experience replay buffer, these types of methods collect data from several environments in parallel. The data collected this way is decorrelated, and avoids having to use a large experience replay buffer.

%% file: sections/results_1.tex
\section{Fully Observable Open Ad Hoc Teamwork Experiments}
\label{sec:results_1}

This section describes our experiments, which demonstrates how the methods introduced in Section~\ref{sec:GPLGeneralOverview} can solve the open ad hoc teamwork problem under full observability.   
We start this section by describing the open environments utilised in our experiments (Section~\ref{Sec:Environments}), as well as the different baseline algorithms (Section~\ref{Sec:FullObservedAlgorithms}). We then present a performance comparison between different versions of GPL and the proposed baselines when solving open ad hoc teamwork problems. Finally, we provide a comprehensive analysis of the joint action-value function of GPL, and discuss why this is the main reason why GPL outperforms the proposed baselines.  

\subsection{Experimental Setup}
\label{sec:exp_setup}
This section outlines the setup of our open ad hoc teamwork experiments. Section~\ref{Sec:Environments} provides an overview of the environments utilised in our experiments. We then describe how we induce openness in Section~\ref{Sec:EnvironmentOpennessSpecification}. Section~\ref{Sec:TeammateTypesFullObs} provides a description of the different teammates' types utilised for our simulations. Finally, we give an overview of the different evaluated algorithms in Section~\ref{Sec:FullObservedAlgorithms}.

\subsubsection{Environments}
\label{Sec:Environments}
Assuming that the environment's state is always fully observed, we describe three environments for our open ad hoc teamwork experiments: 

\begin{figure}[t]
\centering
 \subfloat[Level-based Foraging]{%
      \includegraphics[height=0.25\textwidth, trim={1cm 12.5cm 4cm 2.5cm},clip]{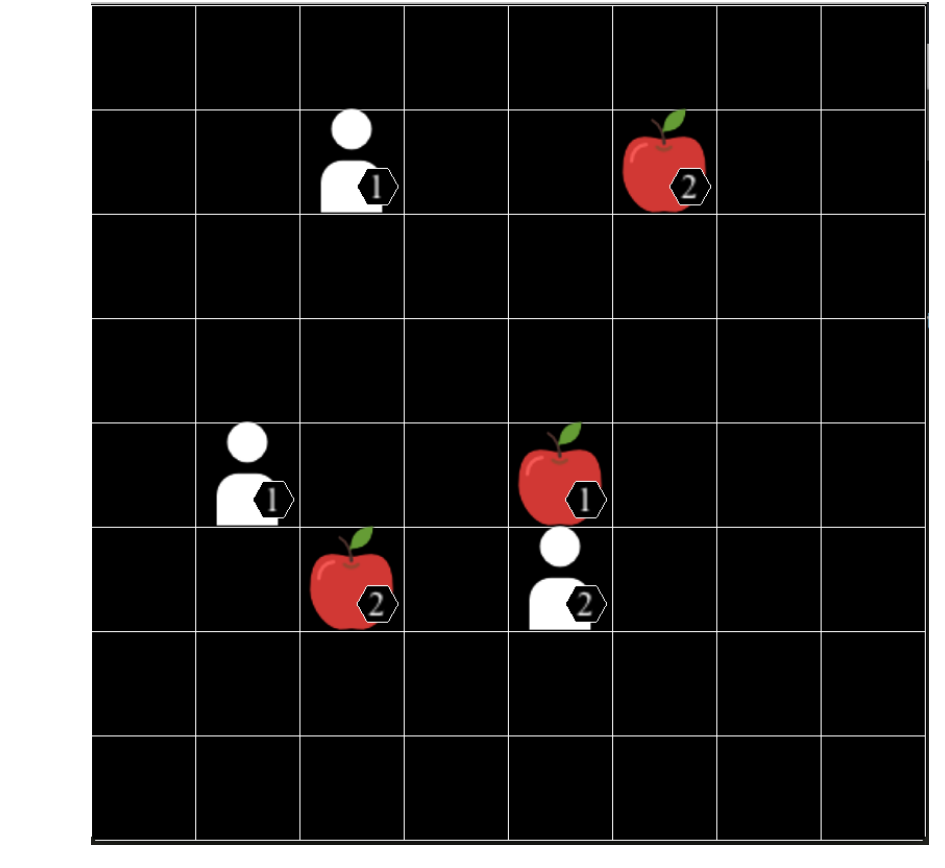}
    }  \hfil
    \subfloat[Wolfpack]{%
      \includegraphics[height=0.25\textwidth, trim={1cm 13.5cm 4cm 2.5cm},clip]{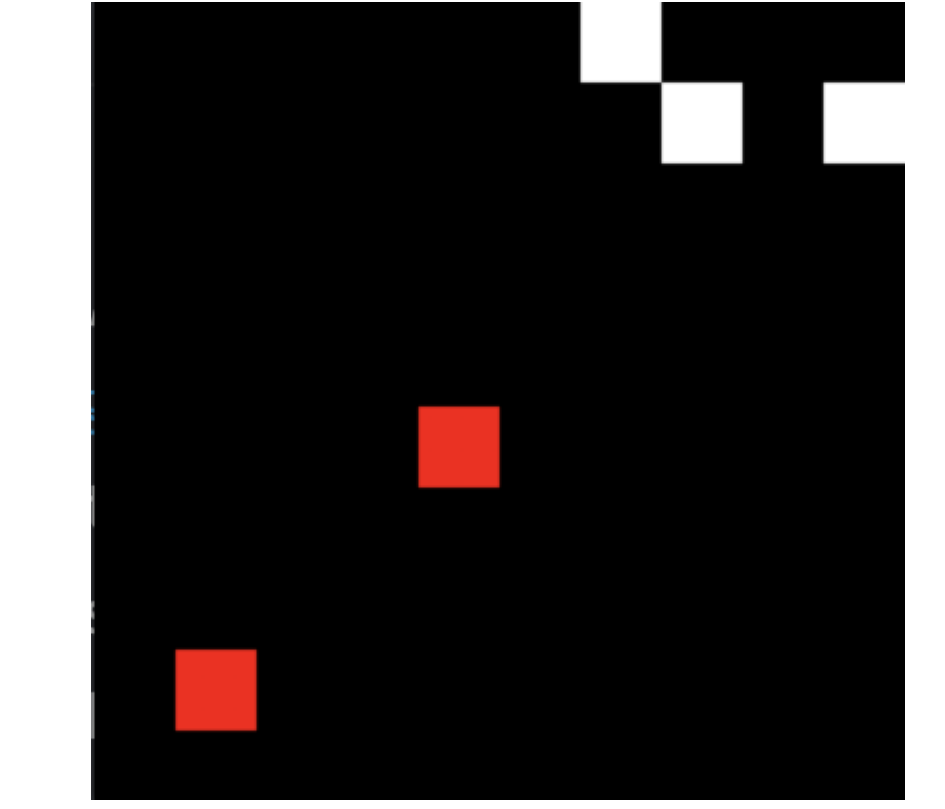}
    }\hfil 
    \subfloat[FortAttack]{%
      \includegraphics[height=0.25\textwidth,trim={0cm 0cm 0cm 0cm}, clip]{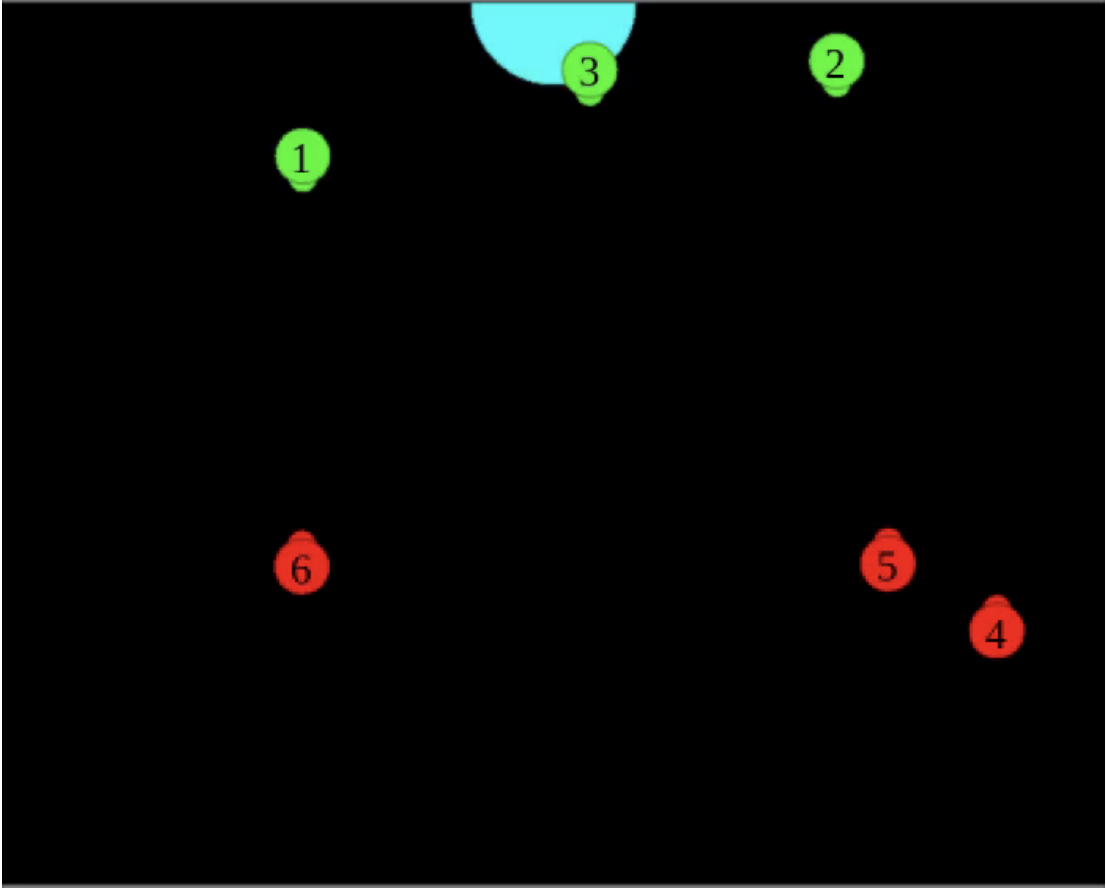}
    }
    \caption{\textnormal{Environment state visualisations.} A visualisation of the state information received by the learner under (a) Level-based Foraging, (b) Wolfpack, and (c) FortAttack, which are the three environments used in our experiments.}
    \label{Fig3:open_ad_hoc_results}
\end{figure}

\textit{Level-Based Foraging (LBF).} LBF is an environment where the learner must retrieve objects that are positioned across an $8\times 8$ grid world. The learner, its teammates, and all objects are each assigned a number as their respective level. All agents are then equipped with actions that enable movement along the four cardinal directions and the retrieval of objects positioned in neighbouring grids. An object is retrieved only if the levels of neighbouring agents which chose the retrieve action has a sum that is not less than the object's own level. After the learner collects an object, the object's level is given to the learner as a reward. 

\textit{Wolfpack.} In Wolfpack, a learner must collaborate with its teammates to hunt a moving prey inside a $10\times10$ grid world. All agents, including the prey, have actions that enable movement across the four cardinal directions. A prey is captured if at least two hunters position themselves adjacent to the prey's current position on the grid. Given a set of hunters $H$ positioned next to a captured prey, the learner is given a reward of $2|H|$ if it is a member of $H$. Conversely, the learner is given a penalty of $-0.5$ if it is next to a prey without any teammates positioned adjacently to the said prey. 

\textit{FortAttack.} FortAttack is an environment where the learner is part of a defending team that must defend a fort 
from advancing attackers. The state space in this environment is continuous and consists of an arena of size $1.8\times2$ in which agents can move around. 
Apart from having actions that enable movement across the four cardinal directions, every agent is equipped with discrete actions that allow them to rotate and shoot opposing team members that venture inside their shooting cones. The episode ends if an attacker reaches the fort, the learner is shot by an attacker, or the attacker fails to reach the fort after a number of 100 timesteps which for each case the learner is given a reward of $-1$, $-0.3$, and $+1$ respectively. Additionally, the learner is given $+0.3$ for successfully shooting an attacker. 

\subsubsection{Environment Openness}
\label{Sec:EnvironmentOpennessSpecification}
In the environments defined in Section~\ref{Sec:Environments}, we define an upper limit to the number of agents in the environment. This upper limit differs during the training and evaluation stages, which allows us to measure the out-of-distribution generalisation capabilities of the proposed method when dealing with open processes that have never been experienced before. In LBF and Wolfpack, the number of agents is limited to three agents during training and five agents during evaluation. On the other hand, there are at most six agents during training and $10$ agents during evaluation for FortAttack.

Environment openness is induced differently for the three environments used in our experiments. In LBF and Wolfpack, a teammate only exists in the environment for a certain number of timesteps. If a teammate has existed for longer than its allocated lifetime, it is immediately removed from the environment. A removed teammate is allocated a waiting period, which is the duration before it is pushed into a reentry queue. Given a non-empty reentry queue, agents in the queue re-enter the environment if the number of agents does not exceed the aforementioned upper limit. It is important to note that the reentry queue is randomised, thus inducing an aleatory team composition during learning. For Wolfpack, teammates' lifetime is sampled uniformly between $25$ and $35$ timesteps while
the waiting period is sampled uniformly between $15$ and $25$ timesteps. By contrast, in LBF teammates' lifetime is sampled uniformly between $15$ to $25$ timesteps while the
waiting period is sampled uniformly between 10 and 20 timesteps.

Unlike LBF and Wolfpack, the changing number of agents in FortAttack is a direct consequence of existing agents' actions. An agent is only removed from the environment if it is shot by a member of the opposing team. After being shot, a shot agent's distance with the shooter determines the waiting time before it can re-enter the environment. An agent is out for 80 timesteps when its distance to the shooter is the closest possible distance between agents. For other distance values, we use a linear interpolation such that shot agents will have less waiting time the larger their distances are with the shooter. Finally, at the beginning of the interaction, the number of agents are initialised according to the previously mentioned maximum number of agents, which are divided equally between the attacking and defending team.

\subsubsection{Teammate Types}
\label{Sec:TeammateTypesFullObs}
We create different teammate types to interact with the learner during our open ad hoc teamwork experiments. For all environments, the policies followed by each teammate type are designed by implementing different behavioural heuristics or using MARL-based methods. Policies from different teammate types are designed to require different policies for effective collaboration\footnote{Further details about teammate types utilised for training are available in \citet{rahman2021towards}.}. 
Finally, during interaction, we randomly choose a type from the set of implemented types and assign it to teammates whenever they re-enter the environment.

\subsubsection{Evaluated Algorithms}
\label{Sec:FullObservedAlgorithms}

The algorithms that we evaluate in the open ad hoc teamwork experiments can be divided into three categories. Algorithms in the first category implement variations of our proposed \ac{GPL} method. The second category is a set of single-agent value-based RL algorithms that act as ablations of GPL. The third category is a set of MARL-based learners. Note that some single-agent and MARL baselines are unable to deal with the changing input size, since they use neural network architectures that receive a fixed-length input. Therefore, we impose a limit on the maximum number of agents allowed in the environment, which allows us to produce fixed-length input vectors for these methods by using placeholder values for features associated to non-existent agents. An overview of the presented algorithms and baselines can be seen in Table \ref{tab:openadhoc_baselines}.

\begin{table}[t]
    \small
    \center
    \caption{\textnormal{Open ad hoc teamwork}: Comparison between algorithms based on value network architecture alongside the usage of agent \& joint action value modelling.}
    \label{tab:openadhoc_baselines}
    \begin{tabular}{|c|c|c|c|}
        \hline
        Models & GNN  & Agent Model & Joint Action-Value \\
        \hline
        QL  & & & \\ \hline
        QL-AM & & \checkmark &  \\ \hline                 
        GNN & \checkmark & &  \\ \hline
        GNN-AM & \checkmark & \checkmark & \\ \hline
        GPL-Q & \checkmark & \checkmark & \checkmark \\ \hline
        GPL-SPI & \checkmark & \checkmark & \checkmark \\ \hline
    \end{tabular}
\end{table}

\textit{Graph-based Policy Learning.}\footnote{These methods appeared originally in \cite{rahman2021towards}.} We define and evaluate two algorithms based on the GPL method defined in Section~\ref{sec:GPLGeneralOverview}. The first algorithm called GPL-Q has its joint action value model trained with Q-learning~\citep{watkins1992q}. The second algorithm called GPL-SPI trains the joint action value model with Soft-Policy Iteration~\citep{haarnoja2018soft} instead. Aside from this subtle difference in the joint action value model training method, both algorithms use the methods described in Section~\ref{sec:GPLActionSelection} and~\ref{sec:GPLLearningObjective} for action selection and learning. 

\textit{Single-agent RL baselines.} In alignment with GPL-Q, the single-agent RL baselines are trained using Q-Learning~\citep{watkins1992q}. These baseline algorithms differ from GPL-Q in terms of the method and model architectures used for action value estimation. At the same time, the single-agent RL baselines also vary in terms of their usage of agent and joint action-value models. While the main characteristics of these baselines are summarised in Table~\ref{tab:openadhoc_baselines}, details of these baselines alongside the insights obtained by comparing them against GPL-Q and GPL-SPI are provided below: 
\begin{itemize}
    \item \textit{QL.} QL estimates the learner's action value by directly passing the representations produced by the type inference model into a multilayer perceptron. Comparisons against QL uncover the effects of not using any GNNs in the learner's model architecture. This comparison also provides insights into direct action value estimation as opposed to using the action value estimation method introduced in Section~\ref{sec:GPLGeneralOverview}. 
    \item \textit{GNN.} The GNN baseline is similar to QL except in its usage of a GNN  that uses multi-head attention~\citep{jiang2019graph} for computing the learner's action value. Comparing QL's performance against GNN enables us to identify the gains resulting from using GNNs for action value estimation. At the same time, a comparison between GNN and GPL-Q's performance enables us to discover the gains from computing the action value following the method described in Section~\ref{sec:GPLGeneralOverview}.
    \item \textit{QL-AM.} Unlike QL, QL-AM has an additional agent model that predicts teammates' actions given the type vectors from the type inference model. For each teammate, their predicted action probabilities derived from the agent model are appended to their type vectors. The collection of concatenated vectors for every teammate are given as input to the multilayer perceptron to compute the action value of the learner. Comparing QL-AM and QL's performance helps us understand the gains achieved by using an agent model. At the same time, comparing QL-AM and GPL-Q's performance provides insights on the advantages of using predicted action probabilities through Equation~\eqref{SingleActionValue} as opposed to using it as input for direct action value estimation.
    \item \textit{GNN-AM.} GNN-AM is QL-AM with GNN's multi-head attention-based action value estimation model. The performance comparison between GPL-Q and GNN-AM helps discover the gains resulting from using Equation~\eqref{SingleActionValue} as opposed to directly utilising predicted action probabilities as input for action value estimation.
\end{itemize}

\textit{MARL baselines.} We compare the performance of the aforementioned algorithms and MARL-based baselines to demonstrate the deficiencies of MARL methods when solving open ad hoc teamwork. While MARL methods' utilisation of joint training and their assumption of knowing teammates' actions prevents it from being a solution for ad hoc teamwork, we can still evaluate the performance of an agent produced by MARL training in open ad hoc teamwork. Our two MARL-based baselines train a group of agents using the MADDPG~\citep{lowe2017multi} and DGN~\citep{jiang2019graph} respectively. We evaluate these methods in open ad hoc teamwork by sampling an agent resulting from the MARL-based training process and letting it interact with previously unseen teammate types. We choose MADDPG and DGN as our baseline MARL methods since they are both designed for MARL in closed and open environments respectively.

\subsubsection{Training \& Evaluation Setup}
Following the previously mentioned details of the environment openness and teammate types, we train every algorithm in Section~\ref{Sec:FullObservedAlgorithms} for open ad hoc teamwork. For LBF and Wolfpack, the algorithms are trained for 6.4 million timesteps. On the other hand, these algorithms are trained for 16 million timesteps for FortAttack. 

At checkpoints which occur every 160000 timesteps the learner's policy is frozen and evaluated in the training and evaluation task, which only differs in terms of their underlying open process as described in Section~\ref{Sec:EnvironmentOpennessSpecification}. This process is repeated across the 8 trained models for each evaluated algorithm, each trained model is initialised with a different random seed. In Section~\ref{Sec:ResultsFullyObservable}, we report the average performance in the training task alongside its 95\% confidence bounds across 8 runs. The performance reported for any algorithm in the evaluation task is based on the optimal checkpoint, which is the checkpoint with the highest average returns across 8 runs.

\subsection{Fully Observable Open Ad Hoc Teamwork Results}
\label{Sec:ResultsFullyObservable}

\begin{figure}[t]
\centering
 \subfloat[Level-based Foraging]{%
      \includegraphics[width=0.47\textwidth]{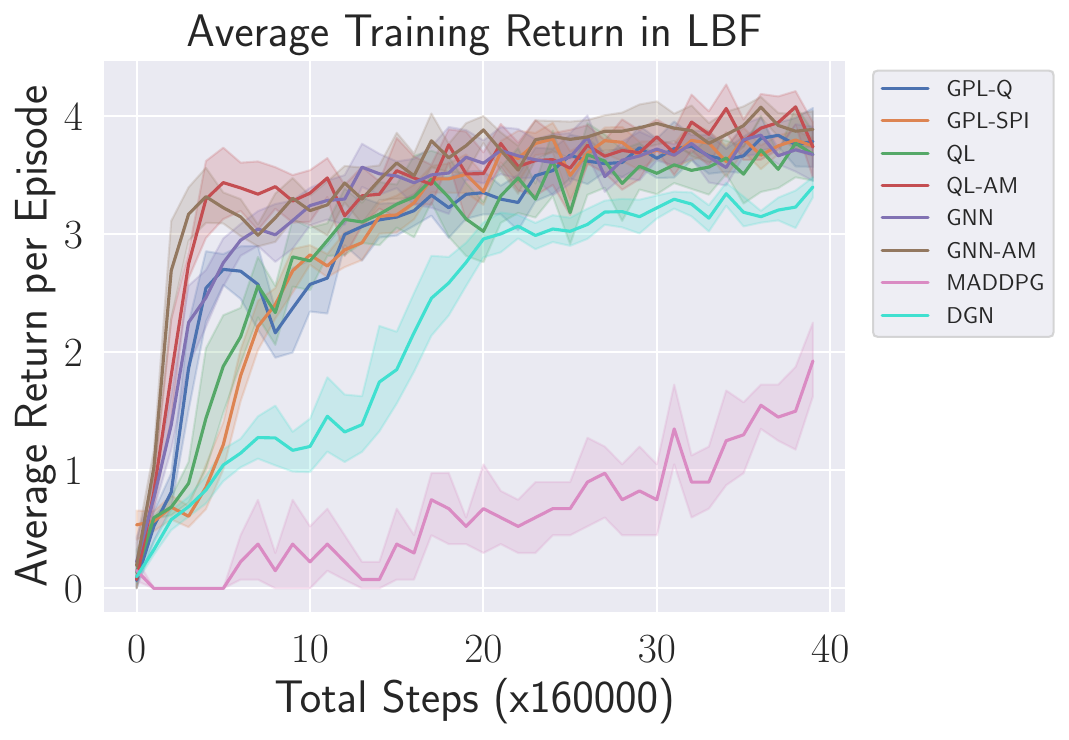}
    }  \hfil
    \subfloat[Wolfpack]{%
      \includegraphics[width=0.47\textwidth]{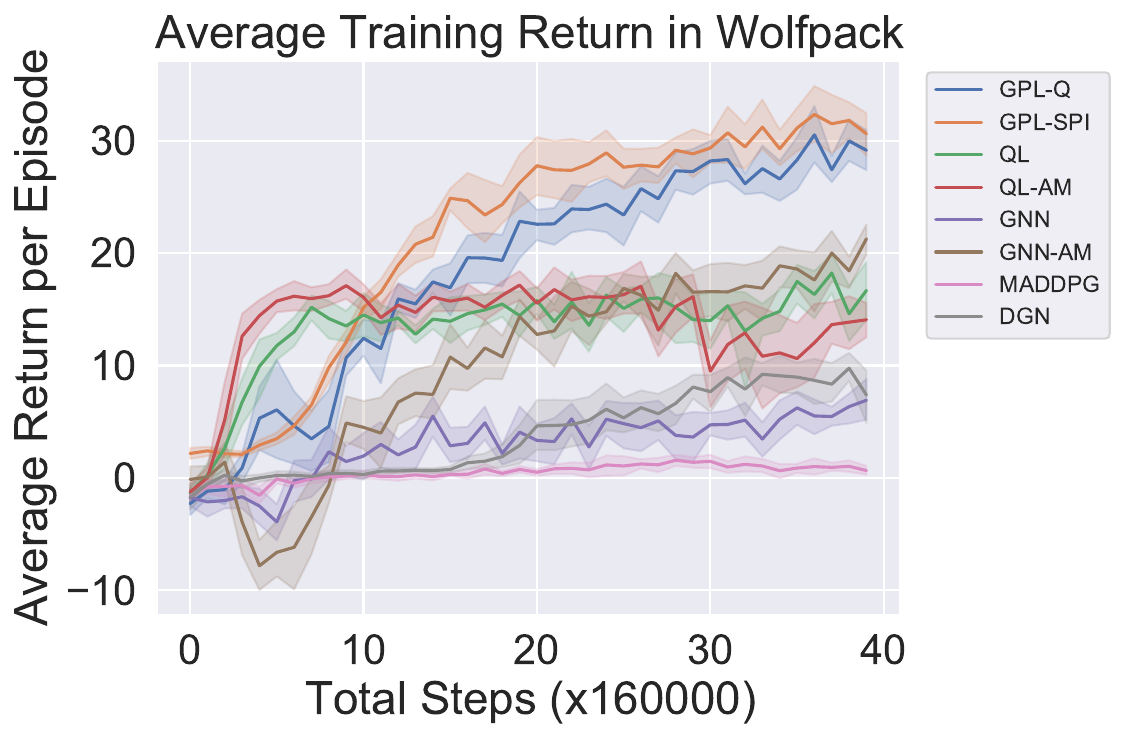}
    }\hfill 
    \subfloat[Fortattack]{%
      \includegraphics[width=0.47\textwidth]{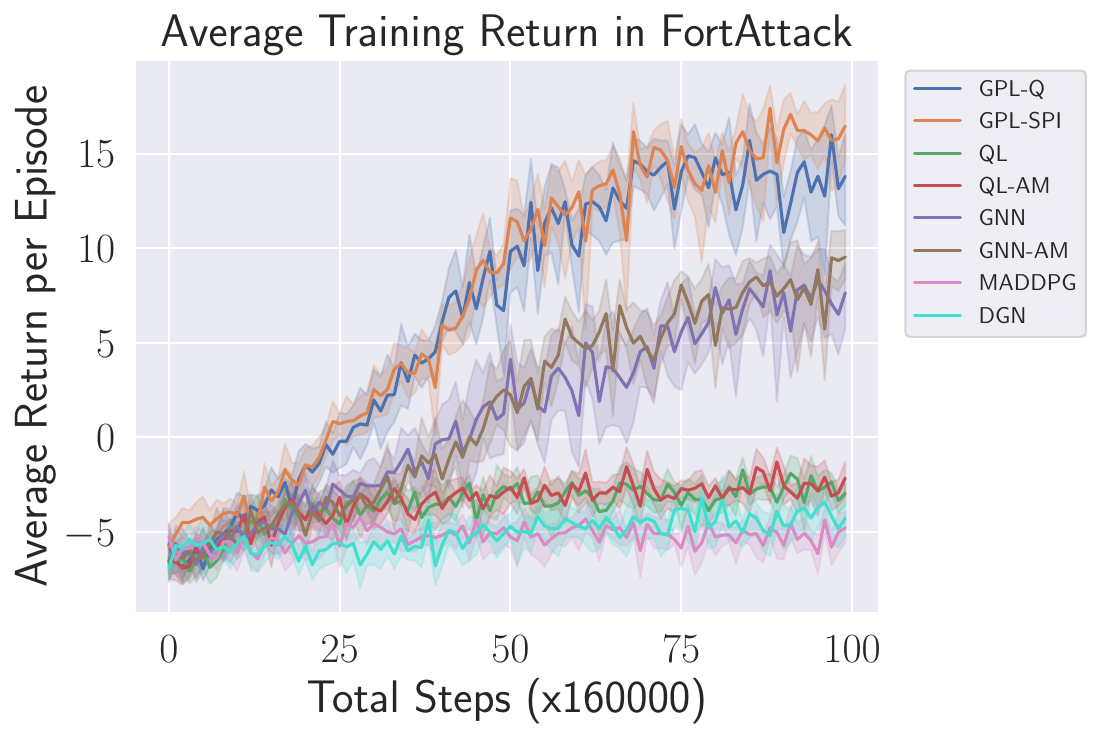}
    }
    \caption{\textnormal{Open ad hoc teamwork results (training).} 
 Obtained returns for all evaluated environments during training. We show the average and $95\%$ confidence bounds utilising 8 seeds.}
    \label{fig:open_ad_hoc_results}
\end{figure}

For every environment described in Section~\ref{Sec:Environments}, Figure~\ref{fig:open_ad_hoc_results} shows the training performance of the evaluated algorithms under the open process encountered during training. The result demonstrates that MARL-based methods, such as MADDPG and DGN, consistently achieve worse performance in all environments when compared to other evaluated algorithms. This is because policies obtained from MARL training are only optimal when interacting with other jointly trained agents. When dealing with previously unseen teammates as in ad hoc teamwork, MARL policies fail to generalise which leads to poor performance.

Figure~\ref{fig:open_ad_hoc_results} also shows the performance gains resulting from two integral designs underlying GPL. By comparing the returns from QL, QL-AM, and the other evaluated algorithms, we first discover that GNN-based architectures deliver improved performance by being more suitable for action value estimation under environment openness. Second, we find the importance of combining joint action value and agent models for estimating the learner's action value based on Equation~\eqref{SingleActionValue}. GPL-Q and GPL-SPI, which are the only evaluated algorithms utilising Equation~\eqref{SingleActionValue} for action value estimation, have significantly higher returns compared to the other algorithms. By comparing the performances between GPL-based algorithms alongside single-agent RL baselines without agent models (e.g. QL and GNN) and with agent models (e.g. QL-AM and GNN-AM), we can also conclude that agent models will not improve returns unless they are combined with joint action value estimates for action value estimation. Further analysis on the importance of GPL's action value estimation method for training will be discussed in Section~\ref{Sec:GPLJointActValAnalysis}.

\begin{table*}[t]
    \centering
    \small	
    \caption{\textnormal{Open ad hoc teamwork results (testing):} We show the average and 95\% confidence bounds during testing utilising 8 seeds. The data was gathered by averaging the returns at the checkpoint which achieved the highest average performance during training. We highlight in bold the algorithm with the highest performance.}
    \label{eval-table}
    \begin{tabular}{|l|c|c|c|}
    \hline 
    Algorithm &  LBF  & Wolf. &  Fort.  \\ \hline
    GPL-Q  & \textbf{2.32$\pm$0.22} & \textbf{36.36$\pm$1.71} & \textbf{14.20$\pm$2.42} \\
    GPL-SPI& \textbf{2.40$\pm$0.16} & \textbf{37.61$\pm$1.69} & \textbf{16.82$\pm$1.92} \\
    QL     &  1.41$\pm$0.14 & 20.57$\pm$1.95 &  -3.51$\pm$0.60 \\
    QL-AM  & 1.22$\pm$0.29  & 14.24$\pm$2.65 & -3.51$\pm$1.51 \\
    GNN    & 2.07$\pm$0.13  &  8.88$\pm$1.57 & 7.01$\pm$1.63 \\
    GNN-AM & 1.80$\pm$0.11 & 30.87$\pm$0.95 &  8.12$\pm$0.74 \\
    DGN    & 0.64 $\pm$ 0.9 & 2.18 $\pm$ 0.66 &  -5.98 $\pm$ 0.82 \\
    MADDPG &    0.91 $\pm$ 0.10 & 19.20 $\pm$ 2.22 &   -4.83 $\pm$ 1.24 \\
    \hline
    \end{tabular}
\end{table*}

The algorithms' performance in the evaluation task depicted in Table~\ref{eval-table} highlights the importance of GNN-based action value estimation to improve generalisation across open processes. Even in LBF, where all but MARL-based baselines deliver similar returns during training, GPL-based methods and GNN-based single-agent RL baselines achieve significantly better generalisation performance compared to QL and QL-AM. However, GPL-based methods still significantly outperform GNN and GNN-AM in terms of generalisation performance.

The reason GPL outperforms other baselines in terms of generalisation performance lies within its action value estimation method. Single-agent RL baselines provide significantly worse generalisation performance because their type and action value network do not learn good representations for estimating the learner's action value in novel open processes. By contrast, GPL estimates the learner's action value following Equation~\eqref{SingleActionValue}.
GPL only suffers in environments where the underlying joint action value and teammates' action distribution does not factorise according to Equation~\eqref{JointActionValueComputation} and Equation~\eqref{GNNequations} as the number of teammates changes.

\subsection{Joint Action Value Analysis in GPL}
\label{Sec:GPLJointActValAnalysis}
In this section, we provide a detailed analysis of the joint action value, which gives insights regarding the higher returns obtained by  GPL-Q and GPL-SPI with respect to the other baselines in the training tasks. For the analysis presented in this section, we focus solely on the more complex FortAttack environment.

\begin{figure*}[t]
\centering
\subfloat[Shooting accuracy in FortAttack]{
  \label{Fig:JointActionValueAnalysisA}
  \includegraphics[height=0.35\textwidth]{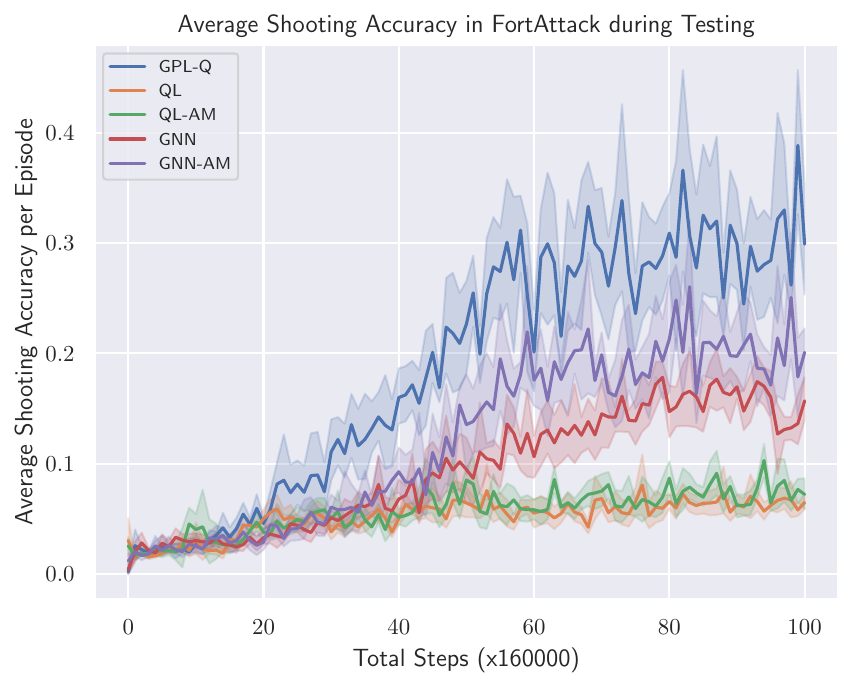}
  } 
\caption{\textnormal{Shooting-related metrics for FortAttack:} (a)
We measure the ratio between the number of times a learner successfully shoots an opponent in relation to the number of times it chooses the shoot action, as defined by Equation~\eqref{Eq:ShootingAccuracy}. This metric is reported for GPL-Q and the single-agent RL baselines, for each training checkpoint in FortAttack.}
\label{Fig:JointActionValueAnalysisShoot}
\end{figure*}

\begin{figure*}[t]
\centering
\subfloat{
  \label{Fig:JointActionValueAnalysisDiagram1}
  \includegraphics[height=0.35\textwidth]{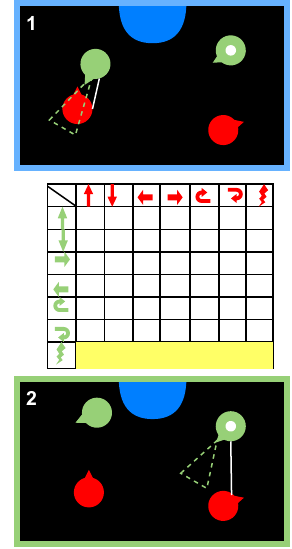}
  \addtocounter{subfigure}{-1}}
  \subfloat[Shooting metrics derived from GPL's pairwise utilites.]{
  \label{Fig:JointActionValueAnalysisBsubfigure}
  \includegraphics[height=0.38\textwidth]{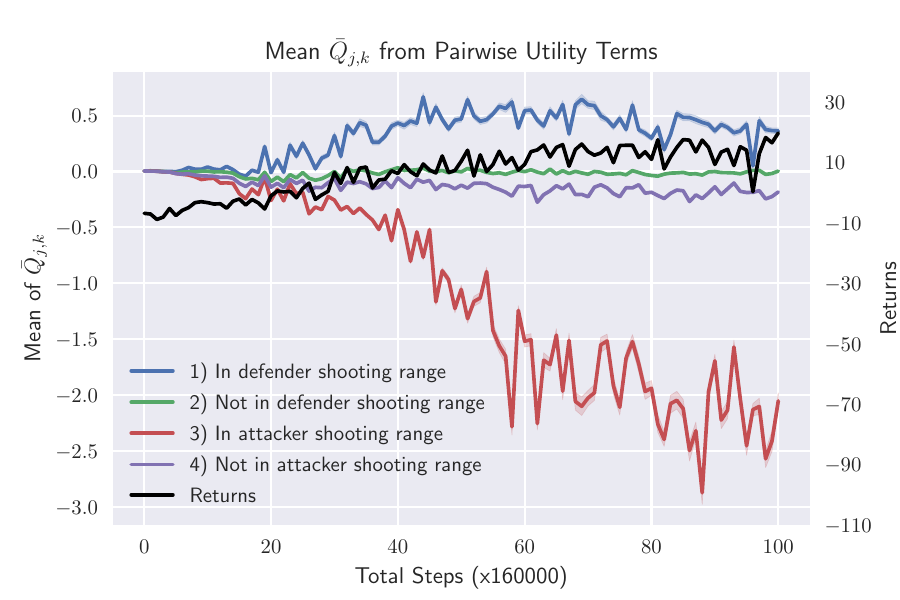}
  } 
  \subfloat{
  \label{Fig:JointActionValueAnalysisDiagram2}
  \includegraphics[height=0.35\textwidth]{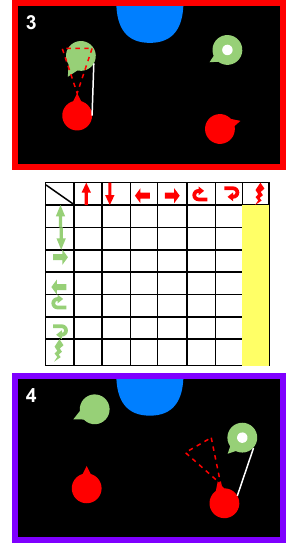}
  } 
\caption{\textnormal{Pairwise utility analysis.} (a)
We measure $\bar{Q}_{j,k}$, which we define in Equation~\eqref{utilmetrics} as GPL-Q's estimate of the contribution towards the learner's returns resulting from agent $j$ shooting an opponent agent, $k$, under four different scenarios defined for this analysis. Lines 1 and 2 represent $\bar{Q}_{j,k}$ for Scenario 1 and Scenario 2, detailed in Equation~\eqref{Eq:SituationQ1} and~\eqref{Eq:SituationQ2}. Lines 3 and 4 represent the value of $\bar{Q}_{j,k}$ in Situation 3 and 4, detailed in Equation~\eqref{Eq:SituationQ3} and~\eqref{Eq:SituationQ4}. Surrounding the main plot, we illustrate the four FortAttack interaction scenarios defined for our analysis and visualize an example interaction under each scenario (represented by the white line in black boxes). Each black box is numbered after the scenario that illustrates. Inside the example visualisation of each scenario, the fort is represented by the blue half circle, attackers by red circles, defenders by green circles, the learner is marked with a white dot, and shooting ranges are indicated with dashed view cones. The square matrices near each black box represent the pairwise utility matrix between attackers and defenders connected by the white line, where the yellow marked fields in each square matrix refer to the matrix entries that are averaged over to compute $\bar{Q}_{j,k}$ for the scenarios depicted above and below it.}
\label{Fig:JointActionValueAnalysisB}
\end{figure*}

We start by analysing the shooting accuracy of the learner which we compute as: 
\begin{equation}
    \label{Eq:ShootingAccuracy}
    \text{SA(Algorithm)} = \dfrac{\sum_{(s,a^{i},s') \in D_{\text{Algorithm}}}\mathbf{1}_{\{\text{True}\}}(\text{OpponentIsHitByLearner}(s'))}{\sum_{(s,a^{i},s') \in D_{\text{Algorithm}}}\mathbf{1}_{\{\text{shoot}\}}(a^{i})},
\end{equation}
In the above expression, $\mathbf{1}_{A}(x)$ denotes the indicator function defined as follows:
\begin{equation}
\label{Eq:Indicator}
    \mathbf{1}_{A}(x)= 
\begin{cases}
    1, & x\in A\\
    0,              & \text{otherwise}
\end{cases}
\end{equation}
while $D_{\text{Algorithm}}$ is defined as a collection of the learner's state, executed actions, and next states resulting from executing the policy produced by the evaluated algorithm. This reported metric is then computed for each checkpoint of the policies in the training process, with the $D_{\text{Algorithm}}$ containing 480000 sample experiences for each algorithm. 

Figure~\ref{Fig:JointActionValueAnalysisShoot} presents the obtained shooting accuracy results. 
Based on Figure~\ref{Fig:JointActionValueAnalysisShoot}, we see that a learner produced via GPL learns to increase its shooting accuracy at a faster rate compared to the baselines. Since shooting is an integral skill for defending the fort, GPL-based learners eventually outperform other learners following its better shooting accuracy. We now analyse various shooting-related metrics and their correlation with the GPL-based learner's returns to highlight why it outperforms other baselines.

Among the many shooting-related metrics that we evaluated, $\bar{Q}_{j,k}(s)$ is the metric with the highest correlation with a GPL learner's returns. Given a set of agents $j$ and $k$ alongside the trained pairwise utility estimator, $Q^{j,k}_{\delta}(a_{t}^{j},a_{t}^{k}|s_{t})$, $\bar{Q}_{j,k}$ is defined as:
\begin{equation}
    \bar{Q}_{j,k}(s) = \dfrac{\sum_{a^{k}\in{A_{k}}}Q^{j,k}_{\delta}(a^{j}=\textrm{shoot}, a^{k}|s)}{|A^{k}|}.
    \label{utilmetrics}
\end{equation}
$\bar{Q}_{j,k}(s)$ is intuitively the estimated pairwise contribution from $j$ towards the learner if $j$ chooses to shoot, which is then averaged across the possible actions of $k$. In FortAttack, note that the learner is always part of the defending team.

We analyse $\bar{Q}_{j,k}(s)$ by first collecting a data set $D$ containing 480000 states, which are obtained by running the learner's frozen policy at every training checkpoint. $D$ is then used to analyse the average of $\bar{Q}_{j,k}$ under four different scenarios. Assuming $N^{\text{att}}(s)$ and $N^{\text{def}}(s)$ denotes the set of existing agents from the attacking and defending team at state $s$, the reported metrics under the different scenarios are defined below:
\begin{itemize}
    \item \textbf{Scenario 1.} We measure the average $\bar{Q}_{j,k}(s)$ when $k$ is an attacking agent who is inside a defender $j$'s shooting range. Formally, this is defined as:
    \begin{equation}
        \label{Eq:SituationQ1}
        \bar{Q}_{j,k}^{S_{1}} = \dfrac{\sum_{s\in{D}}\sum_{j\in N^{\text{def}}(s)}\sum_{k\in N^{\text{att}}(s)}\bar{Q}_{j,k}(s)}{\sum_{s\in{D}}\sum_{j\in N^{\text{def}}(s)}\sum_{k\in N^{\text{att}}(s)}\mathbf{1}_{\{\text{True}\}}(\text{InShootingRange}(j,k))}
    \end{equation}
    \item \textbf{Scenario 2.} This scenario is the opposite of Scenario 1 where $\bar{Q}_{j,k}(s)$ is averaged for instances when an attacker agent $k$ is not in a defender $j$'s shooting range. This is formally defined as:
    \begin{equation}
    \label{Eq:SituationQ2}
        \bar{Q}_{j,k}^{S_{2}} = \dfrac{\sum_{s\in{D}}\sum_{j\in N^{\text{def}}(s)}\sum_{k\in N^{\text{att}}(s)}\bar{Q}_{j,k}(s)}{\sum_{s\in{D}}\sum_{j\in N^{\text{def}}(s)}\sum_{k\in N^{\text{att}}(s)}\mathbf{1}_{\{\text{False}\}}(\text{InShootingRange}(j,k))}
    \end{equation}
    \item \textbf{Scenario 3.} The average of $\bar{Q}_{j,k}(s)$ is computed assuming that $k$ is a defender within an attacker $j$'s shooting range. The evaluated metric in this scenario is defined as:
    \begin{equation}
    \label{Eq:SituationQ3}
        \bar{Q}_{j,k}^{S_{3}} = \dfrac{\sum_{s\in{D}}\sum_{j\in N^{\text{att}}(s)}\sum_{k\in N^{\text{def}}(s)}\bar{Q}_{j,k}(s)}{\sum_{s\in{D}}\sum_{j\in N^{\text{att}}(s)}\sum_{k\in N^{\text{def}}(s)}\mathbf{1}_{\{\text{True}\}}(\text{InShootingRange}(j,k))}
    \end{equation}
    \item \textbf{Scenario 4.} This scenario is similar to Scenario 3 except $\bar{Q}_{j,k}(s)$ is averaged for instances when defender $k$ is not in attacker $j$'s shooting range. The evaluated metric for this scenario is defined below:
     \begin{equation}
     \label{Eq:SituationQ4}
        \bar{Q}_{j,k}^{S_{4}} = \dfrac{\sum_{s\in{D}}\sum_{j\in N^{\text{att}}(s)}\sum_{k\in N^{\text{def}}(s)}\bar{Q}_{j,k}(s)}{\sum_{s\in{D}}\sum_{j\in N^{\text{att}}(s)}\sum_{k\in N^{\text{def}}(s)}\mathbf{1}_{\{\text{False}\}}(\text{InShootingRange}(j,k))}
    \end{equation}
\end{itemize}

We now outline important observations regarding the relationship between $\bar{Q}_{j,k}^{S_{1}}$ alongside the learner's returns and shooting accuracy.
By comparing $\bar{Q}_{j,k}^{S_{1}}$ and the returns of the learner across 100 training checkpoints, we discover that $\bar{Q}_{j,k}^{S_{1}}$ and the learner's returns have a strong positive Pearson correlation coefficient of 0.85. This strong correlation can be seen by comparing the lines associated to $\bar{Q}_{j,k}^{S_{1}}$ and to the learner's returns in Figure~\ref{Fig:JointActionValueAnalysisB}. Comparing Figure~\ref{Fig:JointActionValueAnalysisA} and Figure~\ref{Fig:JointActionValueAnalysisB} also shows that a GPL learner starts to become significantly better than baselines in terms of shooting accuracy after $\bar{Q}_{j,k}^{S_{1}}$ experiences an uptick in its values, which happens around the $20^{\text{th}}$ checkpoint. These observations highlight the importance of GPL's pairwise utility estimator ($MLP_{\delta}$) and more generally its joint action value estimator to achieve high returns in open ad hoc teamwork.

Rather than simply being correlated with the learner's returns, we highlight GPL's joint action value model as the main cause behind GPL's significantly higher returns. The initial increase in value for $\bar{Q}_{j,k}^{S_{1}}$ indicates that $MLP_{\delta}$ starts to see any defender shooting down an attacking team member as advantageous for the learner. Since $MLP_{\delta}$ is shared between the different agents as mentioned in Section~\ref{sec:JointActionValModelling} and the learner itself is a defender, $MLP_{\delta}$ also increases the value of the learner shooting down attacking team members. 
This is an important point as it shows that the learner is able to derive knowledge directly from its teammates.
As learning progresses, we see $MLP_{\delta}$ further increasing the estimated value of $\bar{Q}_{j,k}^{S_{1}}$. This further contrasts the difference between the estimates of $\bar{Q}_{j,k}^{S_{1}}$ and $\bar{Q}_{j,k}^{S_{2}}$, which drives the learner's policy to more frequently get attackers inside the learner's shooting range. 
These results show that GPL's joint action value model and its parameter sharing configuration improves the shooting accuracy and the returns of the learner, and it does so by observing teammate behaviour.

Aside from learning to more accurately shoot attackers, GPL's joint action value model is also responsible for enabling the learner to avoid being shot by attackers. Despite learning this rather late compared to shooting down attackers, Figure~\ref{Fig:JointActionValueAnalysisB} shows the line associated to $\bar{Q}_{j,k}^{S_{3}}$ decreasing as learning progresses. As the value of $\bar{Q}_{j,k}^{S_{3}}$ keeps decreasing relative to the value of $\bar{Q}_{j,k}^{S_{4}}$, the learner's policy learns to avoid getting inside any attacker's shooting range.

We show in the next section that learners resulting from baseline algorithms cannot learn to shoot or evade attackers by observing teammate defenders. In the absence of a joint action value estimation model, a learner can only learn to shoot by experiencing firsthand shooting down attackers. For an initially untrained learner, successfully shooting trained attackers is difficult since getting close to attackers and discovering the right orientation alone is difficult to randomly achieve during exploration. Even if a learner manages to get closer to an attacker, their inexperience will more likely result in the learner being shot down by the attackers instead.

\subsection{Action Value Analysis in Single-Agent RL Baselines}

\begin{figure*}[t]
\centering
\subfloat[State values for QL.]{
  \label{Fig:BaselineStateValuesA}
  \includegraphics[width=0.43\textwidth,trim={2cm 0cm 2cm 0cm}, clip]{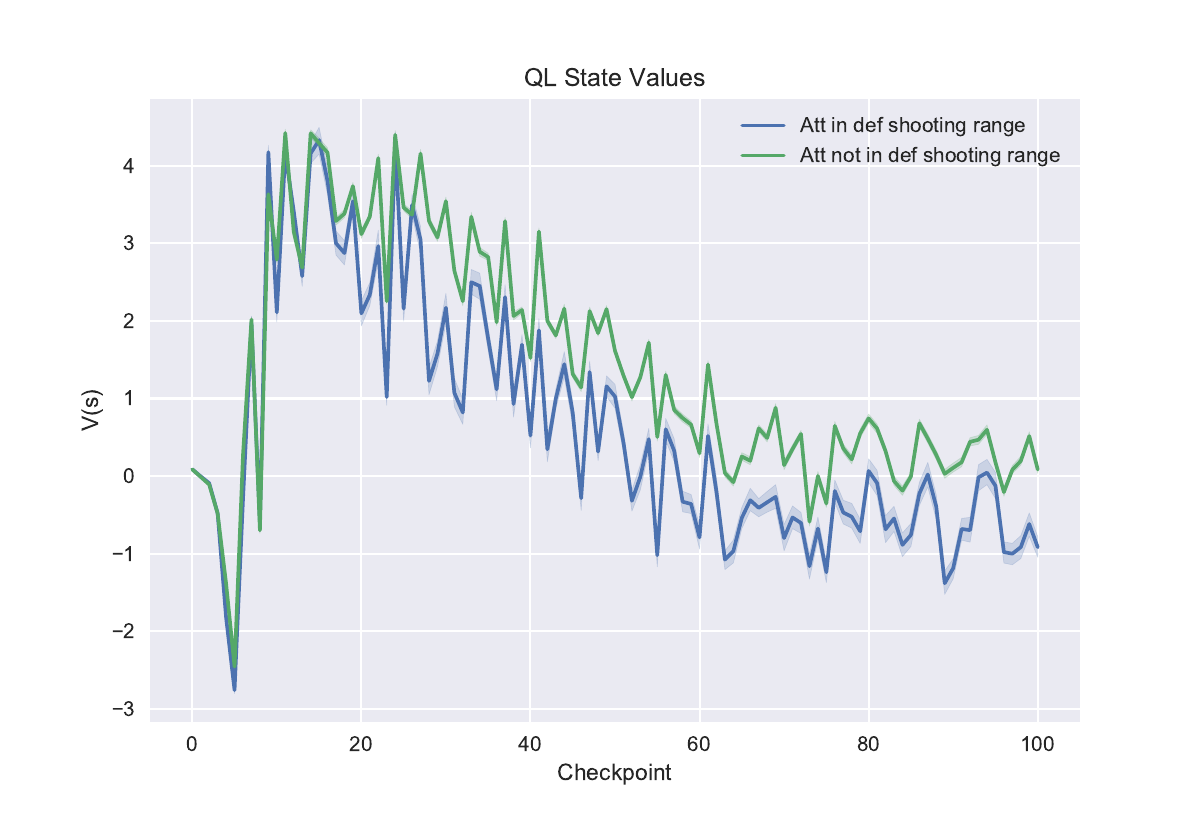}
  }  \hfil
\subfloat[State values for QL-AM.]{
  \label{Fig:BaselineStateValuesB}
  \includegraphics[width=0.43\textwidth,trim={2cm 0cm 2cm 0cm}, clip]{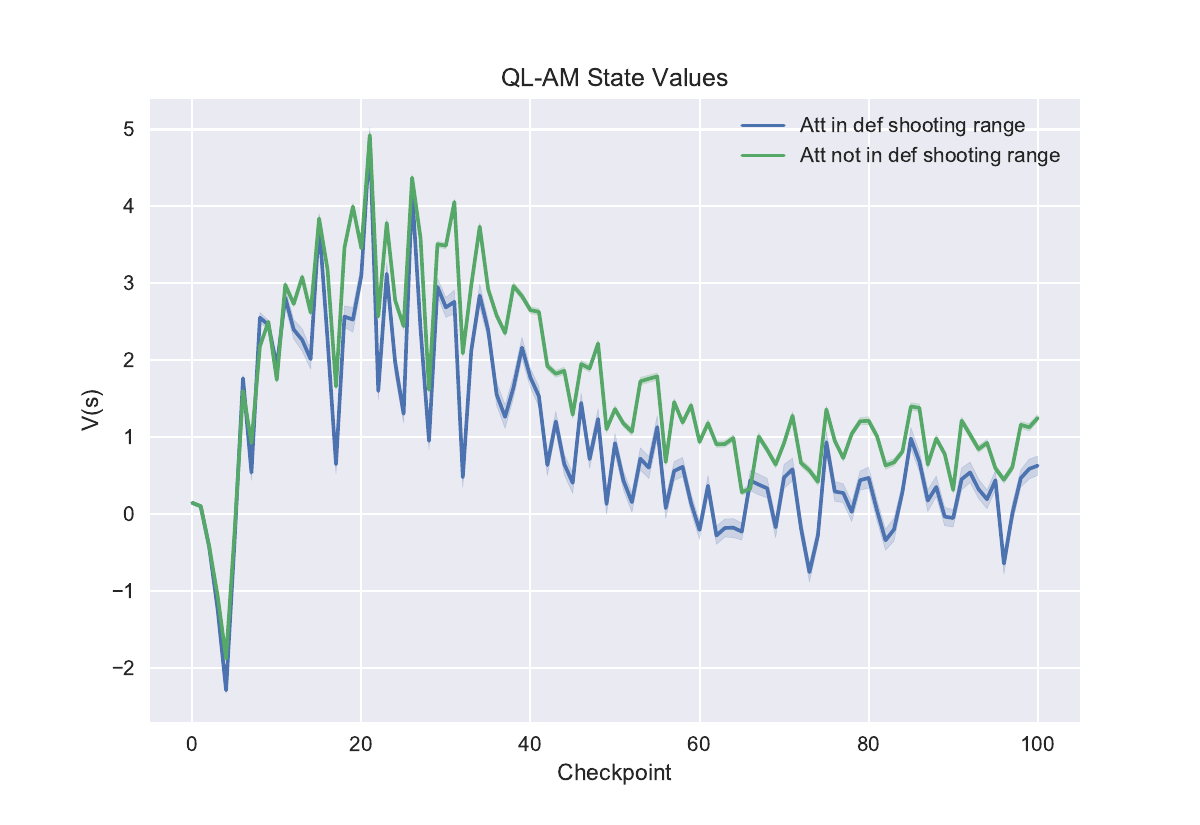}
}  \hfil
\\
\subfloat[State values for GNN.]{
  \label{Fig:BaselineStateValuesC}
  \includegraphics[width=0.43\textwidth,trim={2cm 0cm 2cm 0cm}, clip]{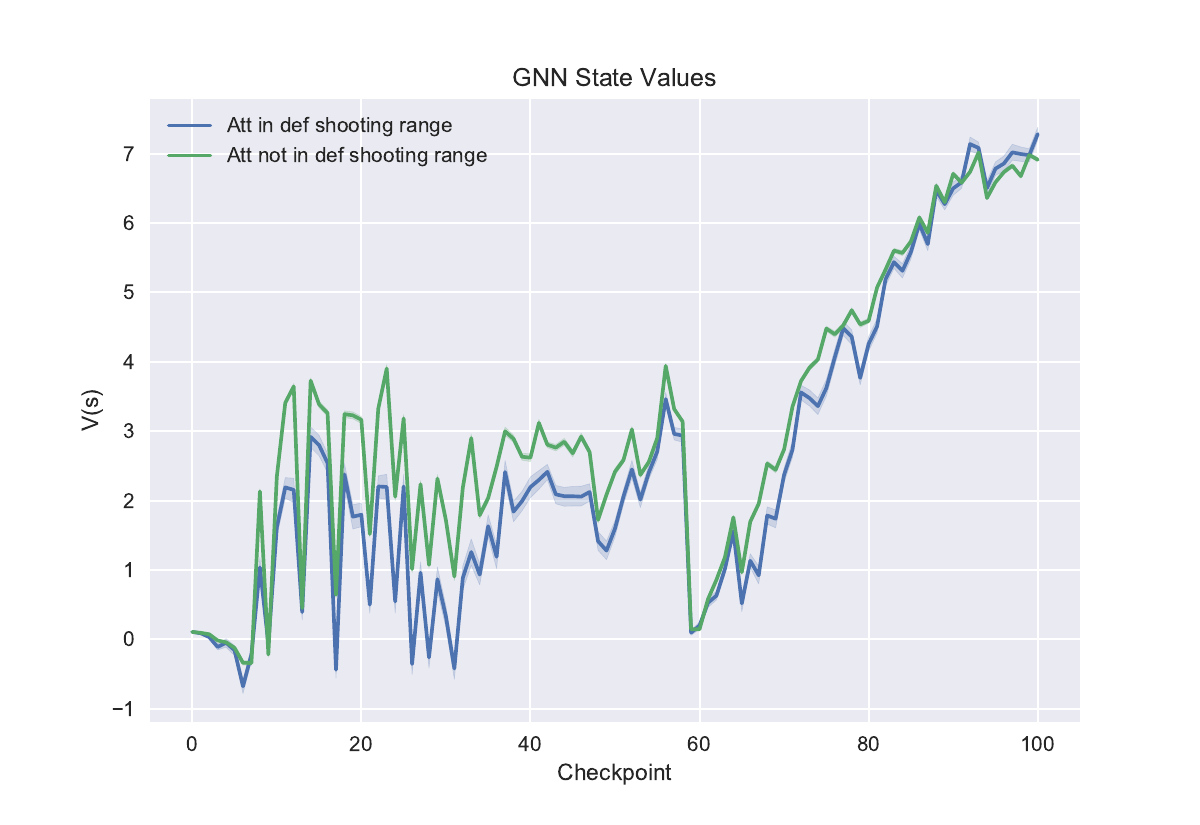}
  }  \hfil
\subfloat[State values for GNN-AM.]{
  \label{Fig:BaselineStateValuesD}
  \includegraphics[width=0.43\textwidth,trim={2cm 0cm 2cm 0cm}, clip]{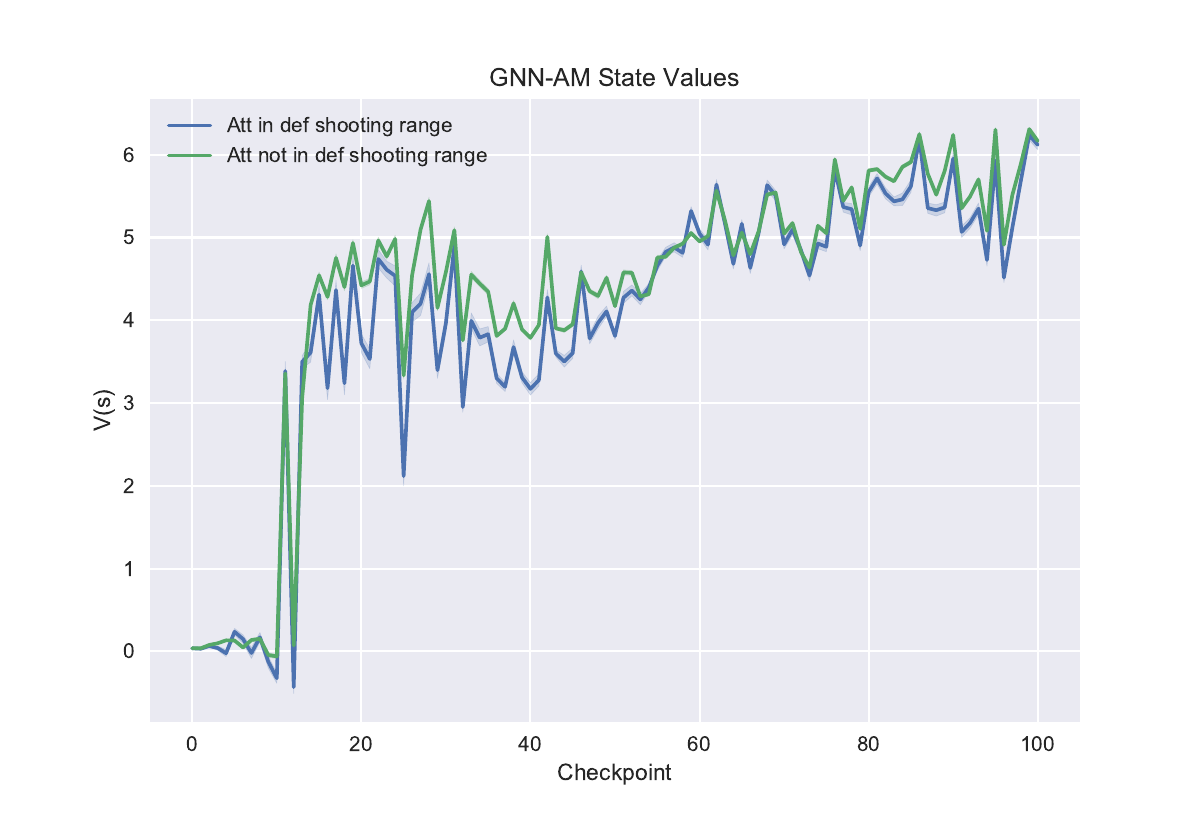}
}

\caption{\textnormal{State values for all single-agent RL baselines. }This visualisation compares the state values in Scenario 1 and 2 for (a) QL, (b) QL-AM, (c) GNN, and (d) GNN-AM. The blue line in each plot shows the average and 95\% confidence bounds of $V(s)$ under Scenario 1. By contrast, the green line shows the average and 95\% confidence bounds of $V(s)$ under Scenario 2. This figure demonstrates that neither single-agent RL baselines manage to learn the effects of other agents' actions on the learner.}
\label{Fig:JointActionValueAnalysisSARL}
\end{figure*}

Following the absence of a joint action value model, this section demonstrates that the single-agent RL baseline algorithms are incapable of learning the effects of teammates' actions to subsequently improve the learner's performance. Our analysis follows Section~\ref{Sec:GPLJointActValAnalysis} by being limited to FortAttack. As in the analysis with GPL, we collect a data set of 480000 states at every training checkpoint by running the frozen learner's policy. We subsequently report measures related to the action value estimates produced by each baseline algorithm.

While it is not possible to directly compute $\bar{Q}_{j,k}$ for baseline algorithms due to the absence of CG-based joint action value models, we can use a Monte Carlo estimate to compute state values under specific scenarios. Assuming a set of states that fulfil a specific criterion $S$, the state value under that specific criteria is estimated as:
\begin{equation}
    V(S) = \dfrac{1}{|S|}\sum_{s\in{S}}V(s),
\end{equation}
with $V(s) = \text{max}_{a}Q(s,a)$ being the action value of the optimal action at $s$ according to the model produced by the evaluated baseline. We now outline the two evaluation scenarios of interest, defined as the following:
\begin{itemize}
    \item \textbf{Scenario 1.} The first scenario evaluates $V(S_{1})$ for a collection of states where an attacker is within the shooting range of any defender. 
    \item \textbf{Scenario 2.} The second scenario computes $V(S_{2})$ for states where no attacker is within the shooting range of any defender.
\end{itemize}
These two scenarios correspond to Scenario 1 and 2 in Section~\ref{Sec:GPLJointActValAnalysis} respectively.
We limit our analysis in this section to these two scenarios, since Section~\ref{Sec:GPLJointActValAnalysis} specifically attributed GPL's learning performance towards the joint action value model's ability to evaluate the pairwise utility in these two scenarios.
Intuitively, $V(S_{1})$ and $V(S_{2})$ can be viewed as an approximation of $\bar{Q}_{j,k}^{S_{1}}$ and $\bar{Q}_{j,k}^{S_{2}}$, defined in Section~\ref{Sec:GPLJointActValAnalysis}. By evaluating the difference between $V(S_{1})$ and $V(S_{2})$, we can determine whether the single-agent RL baselines learn to recognize the value of any defender being in a position to shoot down opposing attackers.

 The value of $V(S_{1})$ and $V(S_{2})$  across the different baselines are reported in Figure~\ref{Fig:JointActionValueAnalysisSARL}. The results in Figure~\ref{Fig:JointActionValueAnalysisSARL} demonstrate that the single-agent RL baselines fail to recognize the advantages of having attackers in the shooting range of any defender. QL and QL-AM instead assign lower average values to states where any defender can shoot down attackers. This negative view of states in Scenario 1 may explain why QL and QL-AM learners have very low shooting accuracy and perform poorly during training. On the other hand, GNN and GNN-AM's average estimate of $V(S_{1})$ and $V(S_{2})$ also do not highlight the inherent positive difference between the values in Situation 1 and 2, which indicates the failure of both baselines to learn the effects of teammates' actions to the learner. This inability to understand the effects of others' actions prevents the baselines from learning important knowledge required to perform well in FortAttack.

%% file: sections/methods_2.tex
\section{Open Ad Hoc Teamwork in Partially Observable Environments}
\label{sec:POGPL}

In the previous sections, we discussed the necessary main components to solve the open ad hoc teamwork problem in the fully observable setting. 
We now relax the previous assumptions about full state observability, and discuss ways to solve the open ad hoc teamwork problem under partial observability. 
We start by providing an overview of the problem in Section~\ref{sec:POGPLOverview}. Then we discuss three different models for inferring the unobserved state variables alongside their usage in computing the learner's optimal action in Section~\ref{sec:VAEBasedStateInference}, Section~\ref{sec:AutoencoderBasedStateInference}, and Section~\ref{sec:ParticleBasedStateInference}. We then discuss the learning objectives to train these belief inference models for open ad hoc teamwork in Section~\ref{sec:PseudocodeAndLearningObj}.

\subsection{General Overview}
\label{sec:POGPLOverview}
Under partial observability, a learner cannot observe certain information about unobserved teammates, such as their existence $e_{t}$ and state features $s_{t}$. Additionally, as in the fully observable case, the agent has no information regarding teammates' types $\theta_{t}$ and their previous actions $a^{-i}_{t-1}$. The unobserved information is important for decision-making and must be inferred to solve partially observable open ad hoc teamwork problems. In this section, we define components to infer the value of these latent variables solely based on the learner's perceived observations and executed actions $H_{t} = \{o_{\leq{t}}, a^{i}_{<t}\}$. We then use the inferred values of these latent variables to estimate the learner's optimal policy,  defined in Definition~\ref{def:LearningObjPOOSBG}.

Given $H_{t}$, there are potentially multiple values of inferred latent variables that are plausible given $H_{t}$. It is therefore useful to maintain a probabilistic belief over the plausible latent variable values given $H_{t}$. As in the case with POMDPs, we call our probabilistic belief over the latent variables given $H_{t}$ the \textit{belief state}.
In PO-OSBGs, at each timestep, the previous belief state estimate can be updated following the learner's most recent observation, $o_{t}$, and executed action, $a^{i}_{t-1}$. Using the Bayes rule, the updated belief state can be found with the following expression:
    \begin{flalign}
    \label{Eq:BayesUpdate}
p(a_{t-1}^{-i}, e_{t}, s_{t},\theta_{t}|H_{t}) \propto \,  &p(o_{t}|e_{t}, s_{t}) &&\text{(Observation likelihood)} \nonumber \\ &p(e_{t},s_{t},\theta_{t}|a^{-i}_{t-1},a^{i}_{t-1},e_{t-1},s_{t-1},\theta_{t-1}) &&\text{(State likelihood)}\nonumber\\&p(a^{-i}_{t-1}|e_{t-1},s_{t-1},\theta_{t-1}) &&\text{(Joint action likelihood)}\\
& p(a_{t-2}, e_{t-1}, s_{t-1},\theta_{t-1}|H_{t-1}). &&\text{(Previous belief state)} \nonumber
\end{flalign}    
Equation~\eqref{Eq:BayesUpdate} intuitively corresponds to the interaction process between a learner and its teammates, which we elaborated on in Section~\ref{sec:OSBG} and~\ref{sec:POOSBG}. An exact evaluation of $p(a_t^{-i}, e_{t}, s_{t},\theta_{t}|H_{t})$ requires integrating the right-hand side expression in Equation~\eqref{Eq:BayesUpdate} over the latent variables, which may not have a closed form expression. In such cases, approximate belief updates can be used to estimate Equation~\eqref{Eq:BayesUpdate}. 

\begin{figure}[p]
\centering
    \subfloat[Particle-based representation]{%
      \includegraphics[height=5cm, trim={0.5cm 8cm 5.5cm 0.25cm},clip]{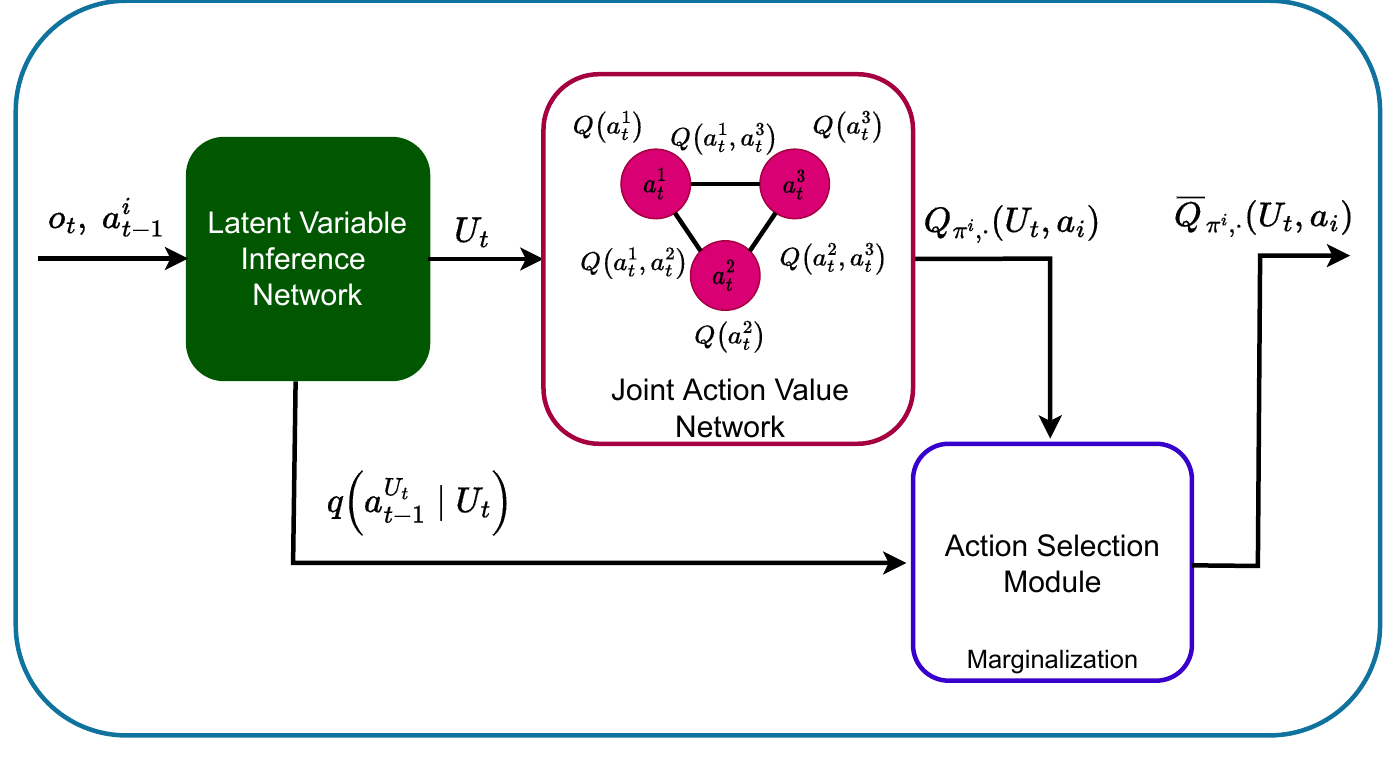} \label{fig:PO_Diagrams_particle}
      } \hfil
           \subfloat[Autoencoder representation]{%
      \includegraphics[height=5cm, trim={1.25cm 6cm 7.5cm 1cm},clip]{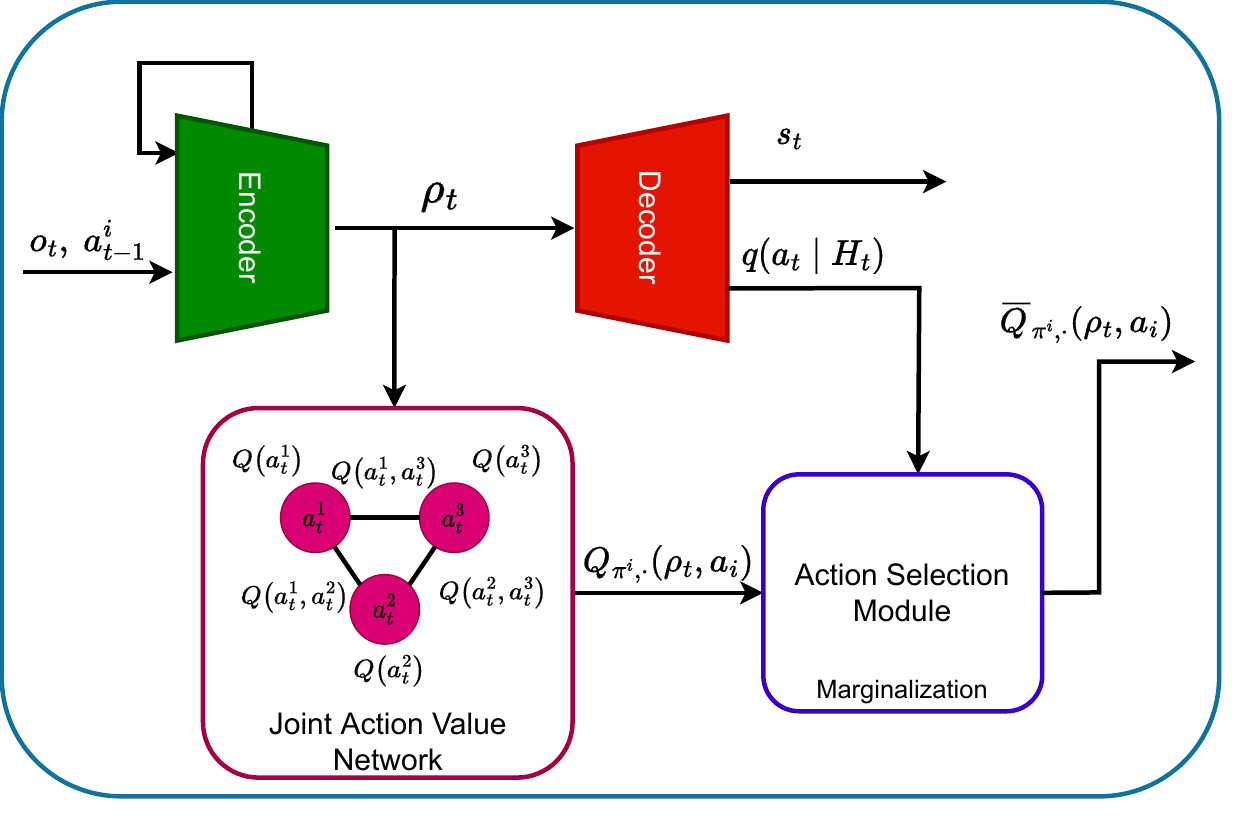}\label{fig:PO_Diagrams_AE}
    }  \hfil
       \subfloat[Variational Autoencoder representation]{%
      \includegraphics[height=5cm, trim={1.cm 6cm 7.75cm 1cm},clip]{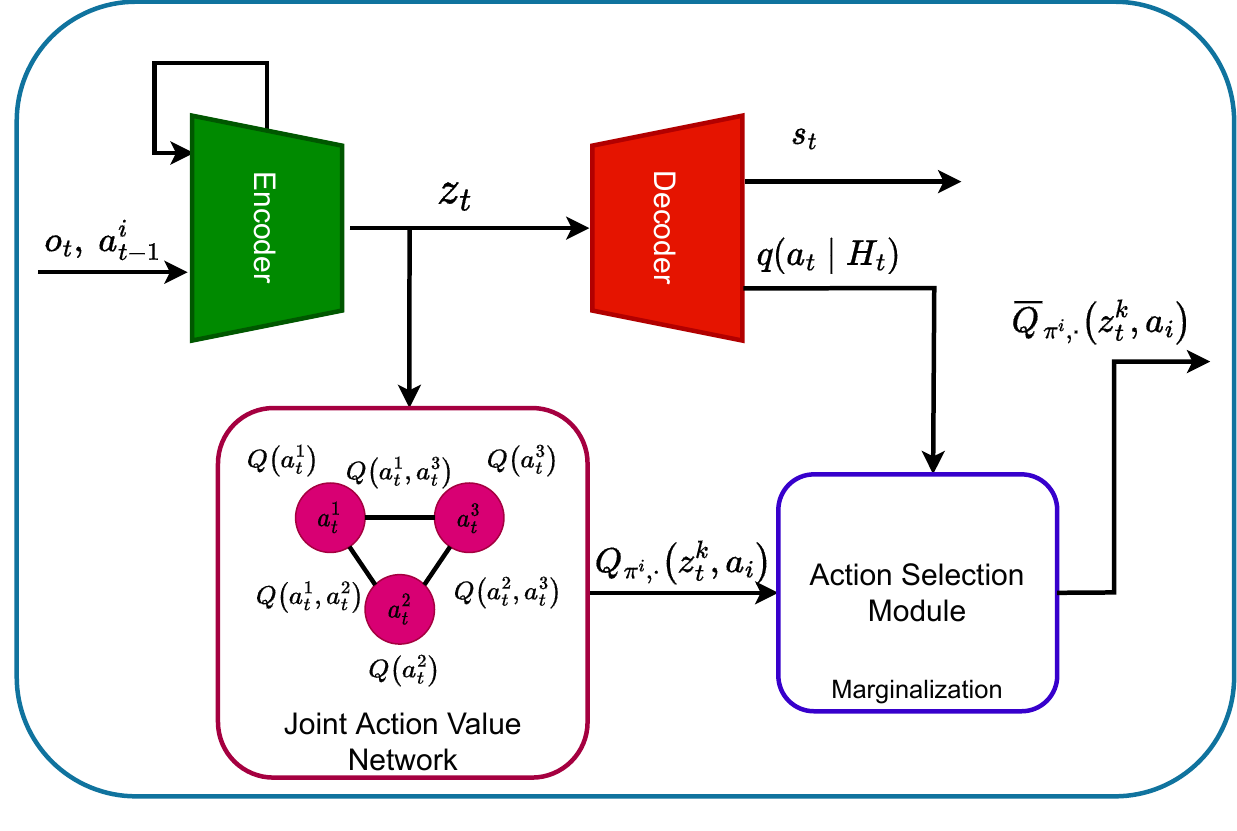}\label{fig:PO_Diagrams_VAE}
    } 
    \caption{\textnormal{Overview of partially observable methods.} (a) Particle-based methods take observation $o_t$ and past actions $a_{t-1}^i$ and produces a set of particles $U_t$ which provides a belief over the latent variables. The particles are then taken as input by a joint action value network, which estimates $Q_{\pi^{i,*}}(U_t, a^i)$. The action selection module then combines the output of the joint action value network and the estimated action coming from the action inference module to obtain $\bar{Q}_{\pi^{i,*}}(U_t, a^i)$.
    (b) For Autoencoder architectures, the observation and actions are encoded into a fixed length vector $\rho_t$ (Section~\ref{sec:AutoencoderBasedStateInference}). This representation is then sent to a joint action value network to obtain $Q_{\pi^{i,*}}(\rho_t, a^i)$. This value together with the teammates' actions, as estimated by the decoder network ($q(a_t|\rho_t)$), are used in the action selection module to estimate $\bar{Q}_{\pi^{i,*}}(\rho_t, a^i)$ via marginalisation.
    (c) In Variational Autoencoders-based belief, we encode the observation and past action to generate a representation $z_t$,  which consists of a distribution over representation (Section~\ref{sec:VAEBasedStateInference}). This representation is then sent to a joint action value network to obtain $\bar{Q}_{\pi^{i,*}}(H_t, a^i)$ by marginalisation.}
    \label{fig:PO_Diagrams}
\end{figure}

In the following sections, we identify three major ways of approximating beliefs: (i) maintaining a fixed length vector which contains information about each teammate, (ii) maintaining a particle-based approach which estimates the belief state as a set of particles, and (iii) maintaining a distribution over representation vectors that contain information about all latent variables. Figure~\ref{fig:PO_Diagrams} presents an overview of the three methods presented in this section. 
These methods allow the learner to infer the latent information required for solving the ad hoc teamwork problem under partial observability. The remaining step left is for the learner to integrate the inferred latent information during action selection. 

Given a representation that encodes a value of the inferred latent variables ($e_{t}, s_{t}, \theta_{t}$, $ a^{-i}_{t-1}$), we can estimate the optimal action-value function under the latent variables' value, $\bar{Q}_{\pi^{i,*}}(e_{t}, s_{t},\theta_{t},a^{i})$. This can be done by combining such a representation with a joint action value module, an agent model module, and an action selection module, obtaining a structure similar to that of GPL (Section~\ref{sec:GPLGeneralOverview}). 

For methods that produce a single representation $\rho$ to infer the latent variables, such as in the autoencoder method shown in Figure~\ref{fig:PO_Diagrams_AE}, the resulting representation can be directly used in combination with the other modules. In this case, the representation vector $\rho$ could be seen as the input vector $B_t$ in GPL (Figure~\ref{Fig:GPL_overview}). Or in the case of an RNN-based autoencoder, we can view the encoder as a type inference module, and thus the vector $\rho$ can be treated as a type vector. The optimal action can then be computed as in GPL.

In contrast, methods that maintain a probabilistic belief over the latent variables must compute the optimal action-value function as the expected value of $\bar{Q}_{\pi^{i,*}}(e_{t}, s_{t},\theta_{t},a^{i})$ under the belief state following this expression:
\begin{equation}
    \label{Eq:FullToPart}
    \bar{Q}_{\pi^{i,*}}(H_{t}, a^{i}) = \int_{a^{-i}_{t-1}, e_{t}, s_{t},\theta_{t}} \bar{Q}_{\pi^{i,*}}(e_{t}, s_{t},\theta_{t},a^{i}) p(a^{-i}_{t-1},e_{t}, s_{t},\theta_{t}|H_{t}) \, da^{-i}_{t-1} \, de_{t} \, ds_{t} \, d\theta_{t}.
\end{equation}
Equation~\eqref{Eq:FullToPart} intuitively expresses $\bar{Q}_{\pi^{i,*}}(H_{t}, a^{i})$ as the expected value of the state-action value estimate given the belief over the teammate's existence $e_t$, the state $s_t$ and teammates’ types $\theta_{t}$, all resulting from the perceived observations and actions $H_t$.

In this work, we develop two methods that use a probabilistic belief over the latent variable for decision-making. The first is the particle-based representation shown in Figure~\ref{fig:PO_Diagrams_particle}. This method produces a set of particles $U_t$, based on the observations and past actions, that provides a belief state estimate. The particle representation can be used by the joint action value network to estimate a particle-based joint-action value $Q_{\pi^{i,*}}(U_t, a^i)$.
The last method presented in this section utilises a variational autoencoder to maintain a distribution over representation vectors $z_t$ to estimate the latent variables. An overview of the method is given in Figure~\ref{fig:PO_Diagrams_particle}. The structure is quite similar to the autoencoder method shown in Figure~\ref{fig:PO_Diagrams_AE}, with the difference that instead of a representation vector $\rho$, the variational autoencoder outputs a distribution over representation
vectors, $z_t$.

In the following sections, the definition of each belief approximation model is accompanied by a method that approximates Equation~\eqref{Eq:BayesUpdate} to update the belief over latent variables.
For clarity, we present each of these methods in different subsections. Section~\ref{sec:AutoencoderBasedStateInference} presents the fixed length vector representation produced by using autoencoders. Section~\ref{sec:ParticleBasedStateInference} presents particle-based methods. Finally, Section~\ref{sec:VAEBasedStateInference} presents methods based on variational autoencoders. 
Note that for the description of each belief inference method, we also detail the way to incorporate their inferred latent variable information for decision-making under a PO-OSBG.

\subsection{Representation-based State Inference}
\label{sec:AutoencoderBasedStateInference}

One initial approach to represent information on teammates' latent variables is to use a fixed-length vector, or embedding, $\rho_t$ for each teammate. This fixed length representation provides a straightforward solution when compared to the other methods used in this section. However, similar approaches have been utilised successfully in the literature before~\citep{Papoudakis2021AgentMU}. In order to learn this representation, we utilise an autoencoder architecture. The encoder takes as input the previous actions of the agent $a_{t-1}^i$ and current observations $o_t$, and provides as output an embedding  $\rho_t$, as can be seen in Figure~\ref{fig:PO_Diagrams_AE}. This encoding is then passed to a decoder, which, given the embedding, provides an estimation of all agents' existence $e_{t}$ and state $s_t$, alongside with the teammates' actions $a_t^{-i}$. Both the decoder and encoder are parameterised by recurrent neural networks. Further details regarding the architecture of models for the autoencoder-based belief inference method are provided in Appendix~\ref{Sec:AEInputPreprocessingAndArchitecture}.

\subsubsection{Action selection}
\label{sec:IntRepresentationBasedInferenceModels}

Given the fixed-representation $\rho_t$ that encodes the inferred values, we can then utilise a similar architecture as that of GPL in order to estimate the optimal action value. We utilise a joint action value model, followed by an action decision module, to compute the action value estimation. This process can be observed in Figure~\ref{fig:PO_Diagrams_AE}. It is important to note that the decoder network can be used as a substitute for GPL's agent model, since the decoder also predicts teammates' actions given its input representation. Therefore, combining the decoder with the joint action value model results in a similar structure to that of GPL. Once the value of $\bar{Q}_{\pi^{i,*}}(\rho_t, a^i)$ is obtained, the learner chooses the actions that greedily maximise $\bar{Q}_{\pi^{i,*}}(\rho_t, a^i)$ at each timestep.

\subsection{Particle-based Belief}
\label{sec:ParticleBasedStateInference}

We provide here a different method to estimate the belief state, in this case by means of a collection of sampled particles from $p(a^{-i}_{t-1}, e_{t}, s_{t},\theta_{t}|H_{t})$. We first provide an overview of how the belief can be represented utilising a graph-based approach in Section~\ref{Sec:ParticleBeliefRepresentation}. Section~\ref{Sec:ParticleBeliefUpdate} outlines a method to update the belief representation using neural network models which receive the learner's most recent observation and action as input.
Then, in Section~\ref{sec:ActionSelectionParticles} we outline a method to select optimal actions based on the particle representation. Further details on how the state is preprocessed and the architectures of the models are provided in Appendix~\ref{Sec:ParticleBeliefModelArchitecture}.

\subsubsection{Belief Representation}
\label{Sec:ParticleBeliefRepresentation}

The particle-based belief representation~\citep{IglICML2018} estimates belief over latent variables at time $T$ as a collection of particles denoted by $U_{t}$. There are two motivations for representing belief as a collection of particles. First, it provides the flexibility to estimate belief states that do not belong to any particular family of distributions. Second, it enables a tractable optimal policy computation.

Previous works have utilised particle representations for solving single agent reinforcement learning problems under partial observability~\citep{IglICML2018,SWB2021}. However, these works have not been extended to open ad hoc teamwork problems where it is necessary to maintain a belief not only on the state of the system but on the existence of other agents, their types, and their actions. Furthermore, the belief representation needs to be able to account for environmental openness. 

We extend the particle-based approach for solving partial observability in open ad hoc teamwork by defining a particle $u_{k}\in{U_{t}}$ as a 5-tuple  $<a^{u_{k}}_{t-1}, e^{u_{k}}_{t}, s^{u_{k}}_{t}, \theta^{u_{k}}_{t}, w^{u_{k}}_{t}>$. We assume knowledge over the set of all agents ($N$) that can exist in a PO-OSBG  and encode a possible value of their latent information in each particle. For a particle $u_{k}$, each of its components are defined as follows:
\begin{itemize}
    \item $a^{u_{k}}_{t-1} \in \boldsymbol{A_{N}}$ is the joint action of agents in $N$ at the previous timestep.
    \item $s^{u_{k}}_{t} \in \mathbb{R}^{|N|\times{m}}$ is a collection of vectors of length $m$, which represents the inferred state feature of each agent in $N$.
    \item $e^{u_{k}}_{t} \in \{0,1\}^{|N|}$ are indicator variables indicating the existence of each agent in $N$. 
    \item $\theta^{u_{k}}_{t} \in \mathbb{R}^{|N|\times{n}}$ are vectors that denote the inferred types of each agent in $N$. 
    \item $w^{u_{k}}_{t} \in \mathbb{R}$ is the log likelihood of $<a^{u_{k}}_{t-1}, e^{u_{k}}_{t}, s^{u_{k}}_{t}, \theta^{u_{k}}_{t}>$ given $H_{t}$.
\end{itemize}
Note that our particle representation represents agents' states and types as vectors of length $m$ and $n$ since we assume no knowledge regarding the underlying state and type space of the PO-OSBG. Furthermore, the state, type, and joint actions associated to agents that are deemed non-existent are set to default values. Given $U_{t}$, the belief state at $t$ is then estimated as:
\begin{equation}
    \label{Eq:ParticleLikelihood}
     p(a^{-i}_{t-1}, e_{t}, s_{t}, \theta_{t}|H_{t}) = \sum_{u_{k}\in{U_{t}}}\left(\dfrac{\mathbf{1}_{\{<a^{u_{k}}_{t-1}, e^{u_{k}}_{t}, s^{u_{k}}_{t}, \theta^{u_{k}}_{t}>\}}({<a^{-i}_{t-1}, e_{t}, s_{t}, \theta_{t}>})\text{exp}(w^{u_{k}}_{t})}{\sum_{u_{j}\in{U_{t}}} \text{exp}(w^{u_{j}}_{t})}\right),
\end{equation}
with $\mathbf{1}_{A}(x)$ denoting the indicator function defined in Equation~\eqref{Eq:Indicator}.

\begin{figure*}[t]
    \centering
       \includegraphics[width=0.95\textwidth, trim={1cm 24cm 61cm 10cm},clip]{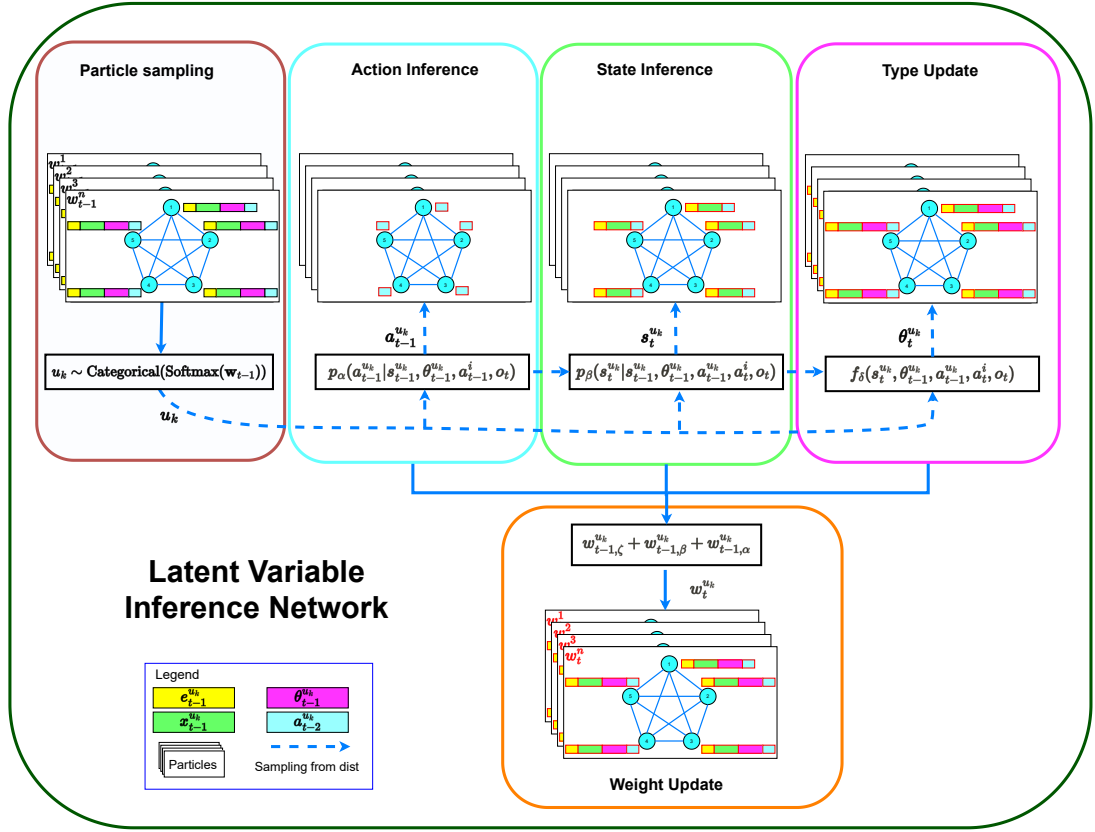}
     \caption{\textnormal{Overview of Particle-based Belief Update.} Given the learner's most recent observation ($o_{t}$) and executed action ($a^{i}_{t-1}$), at timestep $t$ our proposed approach approximates the distribution over all agents' existence ($e_{t}$), feature representations ($s_{t}$), types ($\theta_{t}$), and past joint actions ($a_{t-1}$) as a collection of graph-based particles produced by the \textit{latent variable inference network}. The learner's belief is updated at each timestep by recomputing the contents of each particle from $t-1$ through a sequential execution of the following steps: (i) sampling previous particles based on their log weights ($w_{t-1}$) (ii) a prediction step, which consists of action inference, state inference and type update (iii) a particle weight update step. The sampling operations and deterministic updates %
     produce updated contents ($a_{t-1},(e_{t},x_{t}),\theta_{t},w_{t}$) for the sampled particles. %
     }
     \label{Fig:ParticleBeliefUpdateArchitecture}
\end{figure*}

\subsubsection{Belief Update}
\label{Sec:ParticleBeliefUpdate}

The particle-based belief representation is updated by applying the AESMC technique~\citep{le2018auto}, which is an approximate inference technique to update particle-based latent variable estimates in stochastic processes, such as PO-OSBGs. This update utilises a collection of distributions, which perform stochastic updates to the latent variable estimates from each particle based on $o_{t}$ and $a^{i}_{t-1}$. The log likelihood of each particle is recomputed based on the updated latent variable estimates' likelihood according to an estimate of the right-hand side of Equation~\eqref{Eq:BayesUpdate}. An illustration of the AESMC-based update is provided in Figure~\ref{Fig:ParticleBeliefUpdateArchitecture}.

The models used for updating the particle-based belief estimate are grouped together in the \textit{latent variable inference network}, which approximates the update in Equation~\eqref{Eq:BayesUpdate} following three steps: i) particle sampling, ii) prediction step, and iii) particle log likelihood update. 

\textit{Particle Sampling.} Given $\mathbf{w}_{t-1} = \{w^{u_{k}}_{t-1}|u_{k}\in{}U_{t-1}\}$, the first step is to sample particles from $U_{t-1}$ with replacement based on their log likelihood:
\begin{equation}
\label{Eq:ParticleSampling}
 u_{k} \sim \text{Categorical}(\text{Softmax}(\mathbf{w}_{t-1})).
\end{equation}
We denote the collection of $K$ sampled particles as $\bar{U}_{t-1}$. For each $u_{k}\in{\bar{U}_{t-1}}$, the contents of $u_{k}$ are updated in the subsequent steps.

\textit{Prediction Step.} 
The prediction step updates the estimated values of the state, action, existence and type of each agent in every particle $u_{k}$ in ${\bar{U}_{t-1}}$ at time $t$. This process is sequential and starts with the action, as seen in Figure~\ref{Fig:ParticleBeliefUpdateArchitecture}. The action update is followed by a process that updates the state representations and existence of agents. Finally, the type representation of each agent is updated. For each component of the particle, we utilise proposal distributions that enable us to incorporate important information on the updated value of each particle component. 
Specifically, we incorporate the learner's observation $o_t$ and most recent action $a^i_{t-1}$ when updating the particle representation.

To update the joint action component of each particle, given $o_{t}$ and $a^{i}_{t-1}$, we introduce a \textit{proposal action distribution}, $p_{\alpha^{p}}(a^{u_{k}}_{t-1}|e^{u_{k}}_{t-1}, s^{u_{k}}_{t-1},\theta^{u_{k}}_{t-1},a^{i}_{t-1},o_{t})$. For each particle $u_{k}\in{\bar{U}_{t-1}}$, we draw a sample from the proposal distribution such that,
\begin{equation}
   a^{u_{k}}_{t-1} \sim  p_{\alpha^{p}}(a^{u_{k}}_{t-1}|e^{u_{k}}_{t-1},s^{u_{k}}_{t-1},\theta^{u_{k}}_{t-1},a^{i}_{t-1},o_{t}),
\end{equation}
and use $a^{u_{k}}_{t-1}$ as the updated joint action of each particle $u_{k}$. In $a^{u_{k}}_{t-1}$, note that actions of teammates deemed to not have existed in the previous timestep by $u_{k}$ are set to a default value of no action. Furthermore, the learner's known previous action is set to its observed value $a^{i}_{t-1}$.%

After updating the joint actions, $e_{t-1}^{u_{k}}$ and $s_{t-1}^{u_{k}}$ are updated according to $o_{t}$, $a^{i}_{t-1}$, and the newly updated value of $a_{t-1}$ for all $u_{k}\in\bar{U}_{t-1}$. This update is based on sampling from the updated teammate existence and state representation from the \textit{proposal state distribution} such that:
\begin{equation}
    e^{u_{k}}_{t}, \hat{s}^{u_{k}}_{t} \sim p_{\beta}(e^{u_{k}}_{t}, s^{u_{k}}_{t}|e^{u_{k}}_{t-1}, s^{u_{k}}_{t-1},a^{u_{k}}_{t-1},a^{i}_{t-1},o_{t}).
\end{equation}
Like the proposal action distribution for joint action inference, we sample from the proposal distribution to account for $o_{t}$ and $a^{i}_{t-1}$ in updating $e^{u_k}_{t-1}$ and $s^{u_k}_{t-1}$. It is also important to note that the state is updated based on the predicted existence following $s^{u_{k}}_{t} = e^{u_{k}}_{t}\cdot\hat{s}^{u_{k}}_{t}$.

The next step is to update the inferred teammate types $\theta^{u_{k}}_{t}$ for each particle in $\bar{U}_{t-1}$. Teammate types are updated based on the sampled $a^{u_{k}}_{t-1}$, $e^{u_{k}}_{t-1}$ and $s^{u_{k}}_{t-1}$. While teammates deemed non-existent ($e^{u_{k},j}_{t-1} = 0$) are assigned a type vector of $\mathbf{0}$, existing agents' types undergo a deterministic update using the \textit{type update network} parameterised by $\delta$ following this expression:
\begin{equation}
    \theta^{u_{k}}_{t} = f_{\delta}(s^{u_{k}}_{t},\theta^{u_{k}}_{t-1},a^{u_{k}}_{t-1}).
\end{equation}

\textit{Particle Weight Update.} The final step in the particle-based belief update is to update the log likelihood of particles in $\bar{U}_{t-1}$. Note that particles' log likelihood cannot be updated based on the aforementioned proposal distributions alone. Specifically, the approximated belief update in Equation~\eqref{Eq:BayesUpdate} is defined over target distributions that are not conditioned on the learner's most recent observation. To compensate for the way particle values are not updated through the estimated target distributions, we apply importance sampling correction when updating the weights of each particle.

After sampling $a^{u_{k}}_{t-1}$, the likelihood of this sampled joint action is incorporated when updating the log likelihood of the new set of particles in $\bar{U}_{t-1}$. The likelihood of $a^{u_{k}}_{t-1}$ is evaluated based on the \textit{target action distribution}, $p_{\alpha^{t}}(a^{u_{k}}_{t-1}|e^{u_{k}}_{t-1},s^{u_{k}}_{t-1},\theta^{u_{k}}_{t-1})$, which is used to update the belief in Equation~\eqref{Eq:BayesUpdate}. Since we sample from a different distribution to incorporate $o_{t}$ and $a^{i}_{t-1}$ to update $a_{t-2}$, additional corrections are done to the likelihood computation, which results in the following joint action likelihood expression:
\begin{equation}\label{eq:w_action_update}
w_{t-1,\alpha}^{u_{k}} = \text{log}\left( \dfrac{p_{\alpha^{t}}(a^{u_{k}}_{t-1}|e^{u_{k}}_{t-1},s^{u_{k}}_{t-1},\theta^{u_{k}}_{t-1})}{p_{\alpha^{p}}(a^{u_{k}}_{t-1}|e^{u_{k}}_{t-1},s^{u_{k}}_{t-1},\theta^{u_{k}}_{t-1},a^{i}_{t-1},o_{t})}\right).
\end{equation}

The sampled $e^{u_k}_{t}$ and $s^{u_k}_{t}$ is also utilised for updating the log likelihood of each particle in $\bar{U}_{t-1}$. We specifically compute the likelihood of $e^{u_k}_{t}$ and $s^{u_k}_{t}$ according to the target state distribution,  $q_{\beta}(s^{u_{k}}_{t}, e^{u_{k}}_{t}|e^{u_{k}}_{t-1},s^{u_{k}}_{t-1},a^{u_{k}}_{t-1})$, which is used for the Bayesian belief update in Equation~\eqref{Eq:BayesUpdate}. To account for sampling $e^{u_k}_{t}$ and $s^{u_k}_{t}$ from the proposal distribution, the likelihood of $s^{u_{k}}_{t}$ under both distribution is then evaluated following this expression:
\begin{equation}\label{eq:w_state_update}
w_{t-1,\beta}^{u_{k}} = \text{log}\left( \dfrac{q_{\beta}(e^{u_{k}}_{t}, s^{u_{k}}_{t}|e^{u_{k}}_{t-1}, s^{u_{k}}_{t-1},a^{u_{k}}_{t-1})}{p_{\beta}(e^{u_{k}}_{t}, s^{u_{k}}_{t}|e^{u_{k}}_{t-1}, s^{u_{k}}_{t-1},a^{u_{k}}_{t-1},a^{i}_{t-1},o_{t})}\right).
\end{equation}

An additional term is taken into consideration for updating the particle weight, which is based on the observations $o_t$, and is defined as: 
\begin{equation}
    w_{t-1,\zeta}^{u_{k}} = \text{log}(q_{\zeta}(o_{t}|s^{u_{k}}_{t},a^{u_{k}}_{t-1})), 
\end{equation}
with $q_{\zeta}(o_{t}|s^{u_{k}}_{t},a^{u_{k}}_{t-1})$ being the \textit{observation likelihood distribution}, which evaluates the likelihood of a learner's observation given $s^{u_{k}}_{t}$ and $a^{u_{k}}_{t-1}$ resulting from the state and joint action inference step during the update.

Finally, and following Equation~\eqref{Eq:BayesUpdate}, a sampled particle's log likelihood is updated following:
\begin{equation}
    \label{Eq:ParticleWeightUpdate}
    w^{u_{k}}_{t} = w_{t-1,\zeta}^{u_{k}} + w_{t-1,\beta}^{u_{k}} + w_{t-1,\alpha}^{u_{k}}.
\end{equation}
\noindent The updated content of each particle $u_{k}\in\bar{U}_{t-1}$ is then used as an estimate of the current belief state, $U_{t}=\{(a^{u_{k}}_{t-1},s^{u_{k}}_{t},\theta^{u_{k}}_{t},w^{u_{k}}_{t} )|{u_{k}\in\bar{U}_{t}}\}$. 
Note that Equation~\eqref{Eq:ParticleWeightUpdate} estimates Equation~\eqref{Eq:BayesUpdate} while accounting for the usage of samples generated from target distributions that incorporate $o_{t}$ and $a^{i}_{t-1}$ to update the particles. Finally, $\mathbf{w}_{t-1}$ is not considered in Equation~\eqref{Eq:ParticleWeightUpdate} since the particle sampling step implicitly accounts for the particles' weights from the previous timestep.

\subsubsection{Action selection}
\label{sec:ActionSelectionParticles}

\begin{figure*}[t!]
    \centering
    \includegraphics[width=\textwidth, trim={8cm 79cm 87cm 4cm},clip]{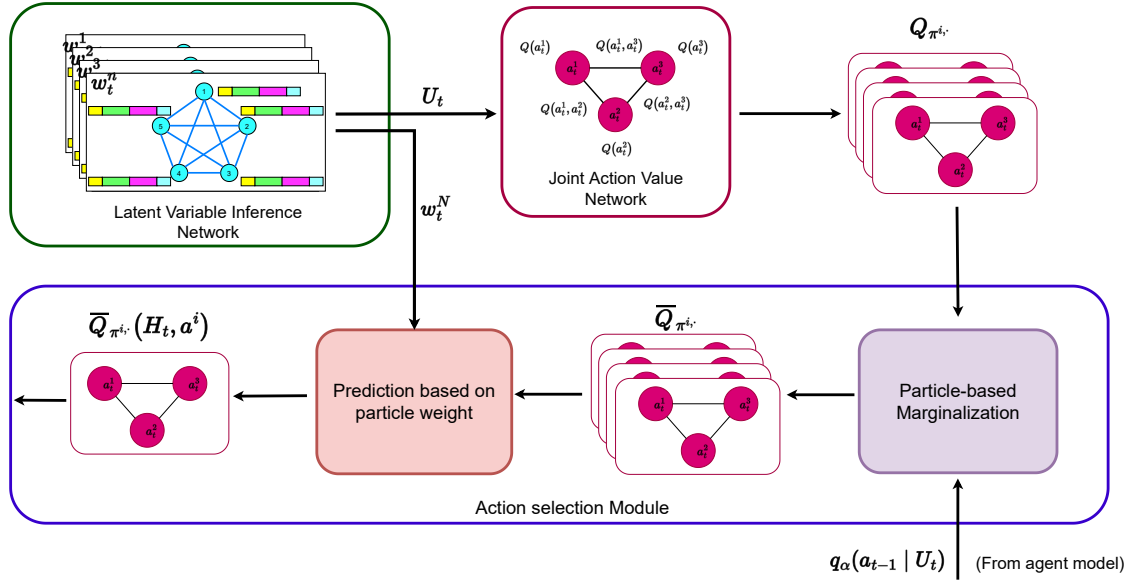}
     \caption{\textnormal{Overview of Particle-based Action Selection.} Given the updated set of particles ($U_t$) at time $t$, the Joint Action Value Network utilises this representation to provide a particle based approximation of $Q_{\pi^{i,*}}(e_{t}^{u_{k}}, s_{t}^{u_{k}},\theta_{t}^{u_{k}},a^{i})$. The Action Selection Module carries a two-step process. First, it marginalises over $Q_{\pi^{i,*}}(e_{t}^{u_{k}}, s_{t}^{u_{k}},\theta_{t}^{u_{k}},a^{i})$, with the teammate action probability $q_{\alpha}(a_{t-1} | U_t)$ coming from the action inference module in the Latent Variable Inference Network.  
     Second, the resulting particle-based state value $\bar{Q}_{\pi^{i,*}}(e_{t}^{u_{k}}, s_{t}^{u_{k}},\theta_{t}^{u_{k}},a^{i})$ is then collapsed to a single representation based on the particle weight $w_t$ to obtain $\bar{Q}_{\pi^{i,*}}(H_t,a^i)$, following Equation~\eqref{Eq:BeliefSumActVal}.}
     \label{Fig:ParticleActionSelection}
\end{figure*}

We circumvent a challenge in evaluating the learner's optimal action-value function by representing our belief estimates as a collection of particles. As mentioned in Section~\ref{sec:POGPLOverview}, $p(a_{t-1}, e_{t}, s_{t},\theta_{t}|H_{t})$ can be combined with the different modules of GPL to compute the learner's optimal action-value function under partial observability. A problem arises for the exact evaluation of Equation~\eqref{Eq:ParticleLikelihood} when $\bar{Q}_{\pi^{i,*}}(e_{t}, s_{t},\theta_{t},a^{i})$ is implemented as a neural network since the integral generally does not have a closed form expression. 

By using a particle-based belief representation, we avoid integrating over all possible values of the latent variables. This process is detailed in Figure~\ref{Fig:ParticleActionSelection}. Substituting Equation~\eqref{Eq:ParticleLikelihood} into $p(a_{t-1}, e_{t}, s_{t},\theta_{t}|H_{t})$ in Equation~\eqref{Eq:FullToPart} results in the following expression:
\begin{equation}
    \label{Eq:BeliefSumActVal}
    \bar{Q}_{\pi^{i,*}}(H_{t}, a^{i}) = \sum_{u_{k}\in{U_{t}}}\left( \dfrac{\text{exp}(w^{u_{k}}_{t})}{\sum_{u_{j}\in{U_{t}}} \text{exp}(w^{u_{j}}_{t})}\right) \bar{Q}_{\pi^{i,*}}(e_{t}^{u_{k}}, s_{t}^{u_{k}},\theta_{t}^{u_{k}},a^{i}),
\end{equation}
which is only a summation over functions defined over the contents of the particles. In the above expression, we estimate $\bar{Q}_{\pi^{i,*}}(e_{t}^{u_{k}}, \allowbreak s_{t}^{u_{k}}, \allowbreak \theta_{t}^{u_{k}},a^{i})$ by marginalising over $Q_{\pi^{i,*}}(e_{t}^{u_{k}}, s_{t}^{u_{k}}, \allowbreak \theta_{t}^{u_{k}},a^{i})$ as output by the joint action value model as seen in Figure~\ref{Fig:ParticleActionSelection}.
The learner then greedily chooses actions that maximise $\bar{Q}_{\pi^{i,*}}(H_{t}, a^{i})$ at any timestep.

\subsection{Variational Autoencoder-based Belief}
\label{sec:VAEBasedStateInference}

A problem occurs under particle-based approaches as more particles are required to estimate a distribution when the dimension of inferred latent variables increases or if distributions required for updating the particle contents have high variance~\citep{murphy2001rao}. Ensuring an accurate representation of the belief posterior with a large number of particles is computationally expensive to maintain and update. In this section, we provide an alternative method which does not maintain a collection of particles to represent belief.

\subsubsection{Belief Representation \& Update}
The alternative approach is to instead represent belief as a distribution over representation vectors, $z_{t}\in\mathbb{R}^{|N|\times{m}}$. The belief over $z_{t}$, $p(z_{t}|H_{t})$, is then evaluated given the learner's interaction experience $H_{t}$. We prevent having to maintain a large collection of particles by ensuring this distribution is a parametric distribution with low variance, which parameters are estimated by a trained model that receives $H_{t}$ as input. The model is trained to ensure that higher likelihood is associated to sampling representations $z_{t}$ that are more informative of the interaction experience $H_{t}$. Sampled values of $z_{t}$ then provide relevant information for action value computation.

We achieve our goal of training a model for estimating $p(z_{t}|H_{t})$ using variational autoencoders (VAEs)~\citep{kingma2013auto}. To ensure $z_{t}$ is informative of $H_{t}$, VAEs assume the existence of an underlying generative model, $p(H_{t}|z_{t})$, that determines the way $H_{t}$ is generated from $z_{t}$. Given a prior distribution on $z_{t}$, the true posterior over $z_{t}$, $p(z_{t}|H_{t})$, may then be evaluated via the Bayes theorem:
\begin{equation}
    \label{Eq:BayesVAE}
    p(z_{t}|H_{t}) = \dfrac{p(H_{t}|z_{t}) p(z_{t})}{\int_{z_{t}}p(H_{t}|z_{t}) p(z_{t}) dz_{t}}.
\end{equation}

The exact evaluation of Equation~\eqref{Eq:BayesVAE} is generally intractable, since the integral operation does not have a closed form expression. Instead, VAEs estimate the posterior with a variational parametric distribution, $q(z_{t}|H_{t}) = \mathcal{N}(z_{t};\mu,\Sigma)$. The variational parametric distribution is optimised to minimise the Kullback-Liebler divergence between the two distributions, $D_{KL}(q(z_{t}|H_{t})||p(z_{t}|H_{t}))$. 

Both $p(H_{t}|z_{t})$ and $q(z_{t}|H_{t})$ are represented by VAEs as parametric distributions which parameters are estimated by neural network models called the decoder and encoder respectively. At each timestep, updates to the learner's belief over the latent variables are done by computing the distribution parameters of $q(z_{t}|H_{t})$ based on $H_{t-1}$ and the learner's most recent observation and action. Details of the network architectures that we use to represent the encoder and decoder are provided in Appendix~\ref{Sec:VAEInputPreprocessingAndArchitecture}. The objective functions for training the VAE's encoder and decoder are then provided in Section~\ref{Sec:VAEBasedBeliefObj}.

\subsubsection{Action Selection}

Given the variational parametric distribution, the action value under partial observability is computed as:
\begin{equation}
    \label{Eq:FullToPartVAE}
    \bar{Q}_{\pi^{i,*}}(H_{t}, a^{i}) = \int_{z_{t}} \bar{Q}_{\pi^{i,*}}(z_{t},a^{i}) q(z_{t}|H_{t}) dz_{t}.
\end{equation}
$\bar{Q}_{\pi^{i,*}}(z_{t},a^{i})$ denotes the action value estimate based on Equation~\eqref{SingleActionValue} given $z_{t}$ as input. However, exact evaluation of Equation~\eqref{Eq:FullToPartVAE} is not possible since the integral generally does not have a closed form expression when $\bar{Q}_{\pi^{i,*}}(z_{t},a^{i})$ is represented as a neural network.  

To approximate Equation~\eqref{Eq:FullToPartVAE}, we instead adopt a Monte Carlo approach. We sample $n$ samples from $q(z_{t}|H_{t})$,
\begin{equation}
    \label{Eq:VAEBeliefSampling}
    z^{1}_{t}, z^{2}_{t}, ..., z^{n}_{t} \overset{\mathrm{iid}}{\sim} q(z_{t}|H_{t}),
\end{equation}
and estimate $\bar{Q}_{\pi^{i,*}}(H_{t}, a^{i})$ based on the following Equation:
\begin{equation}
    \label{Eq:IntegralApprox}
    \bar{Q}_{\pi^{i,*}}(H_{t}, a^{i}) = \dfrac{\sum_{k=1}^{n} \bar{Q}_{\pi^{i,*}}(z^{k}_{t},a^{i}) q(z^{k}_{t}|H_{t})}{\sum_{l=1}^{n}q(z^{l}_{t}|H_{t})}.
\end{equation}

\subsection{Learning Objective}
\label{sec:PseudocodeAndLearningObj}

The aforementioned latent variable inference models are trained alongside GPL to infer important latent information for decision-making and use it for action selection. 
During execution, the learner has only access to its own observations and past actions. However,  during training, we assume that the learner also has access to the environment state and the observed teammates' joint actions to train its modules.
Having knowledge of the full state of the system during training is a common assumption in partially observable environments~\citep{gu2021online,Papoudakis2021AgentMU}. Therefore, given a set of interaction experiences, $D = \{\{(s^{n}_{t}, o^{n}_{t}, a^{V,n}_{t}, r^{n}_{t}, o^{n}_{t+1})\}_{t=1}^{T_{n}}\}_{n=1}^{|D|}$, we train the models on the following loss function:  
\begin{equation}
    \label{Eq:PartOverallLoss}
    L_{P_{inf}, P_{st}, P_{ag}, P_{val}}(D) = L^{INF}_{P_{inf}}(D) +
    L^{SR}_{P_{inf}\cup{}P_{st}}(D) +
    L^{NLL}_{P_{ag}}(D) + L^{RL}_{P_{val}}(D).
\end{equation}
In the above equation, $P_{inf}, P_{st}, P_{ag}$ and $P_{val}$ denote the collection of model parameters for latent variable inference, state reconstruction, GPL's agent model and joint action value models respectively.

While its computation may differ across the latent variable inference model being used, each of the terms on the right-hand side of Equation~\eqref{Eq:PartOverallLoss} fulfil an important role in the optimisation process. $L^{INF}_{P_{inf}}(D)$ serves as the loss function that is optimised by the latent variable inference models to produce representations for decision-making. $L^{SR}_{P_{inf}\cup{}P_{st}}(D)$ is the state reconstruction loss, which aligns with previous works that use privileged state information to train the belief inference model to produce representations that are more informative of the state~\citep{Papoudakis2021AgentMU}. On the other hand, $L^{NLL}_{P_{ag}}(D)$ and $L^{RL}_{P_{val}}(D)$ are the negative log likelihood and value-based RL losses which we introduce in Section~\ref{sec:GPLLearningObjective} to train GPL for solving open ad hoc teamwork. We provide details regarding the computation of $L^{INF}_{P_{inf}}(D)$ and $L^{SR}_{P_{inf}\cup{}P_{st}}(D)$ across the previously defined belief inference models in the following sections. In Table~\ref{tab:partialadhoc_loss} we provide a summary of the loss functions utilised for each method. While details of the remaining loss terms that are based on Equation~\eqref{ActionModelLoss} and Equation~\eqref{ValueLoss} are provided in Appendix~\ref{sec:AgentJointActionLoss}.

\subsubsection{Particle-based Belief Models}
\label{Sec:ParticleBasedBeliefObj}
\textit{Belief Inference Loss Function.} In the model introduced in Section~\ref{sec:ParticleBasedStateInference}, the negative ELBO loss is defined as a function of the belief model parameters, $P_{inf}=(\alpha,\beta,\delta,\zeta)$. Following AESMC~\citep{le2018auto}, $P_{inf}$ is trained to minimise the negative ELBO defined as:
\begin{equation}
    \label{Eq:ELBO}
    L^{ELBO}_{P_{inf}}(D) = -\sum_{H_{n}\in{D}}\text{log}\left(\sum_{u_{k}\in{U_{n}}} \text{exp}(w^{u_{k}}_{T_{n}})\right),
\end{equation}
assuming that $U_{n}$ is the collection of particles resulting from applying the belief inference procedure in Section~\ref{Sec:ParticleBeliefUpdate} to $H_{n}$,
\begin{equation}
    \label{Eq:ParticleBeliefUpdate}
    U_{n} = \text{BeliefUpdate}_{P_{inf}}(H_{n}).
\end{equation}

\begin{table}[t]
    \small
    \center
    \caption{\textnormal{Loss functions}: A description of the loss functions used for training each method.}
    \label{tab:partialadhoc_loss}
        \begin{tabular}{|c|c|c|c|c|}
        \hline
         \quad  Models \quad \quad  & \thead{Belief Inference \\ ($P_{inf}$)} & \thead{State Reconstruction \\ ($P_{st}$) } & \thead{Agent Model \\ ($P_{ag}$)} & \thead{Joint Action Value Model \\ ($P_{val}$)}   \\ \hline
         PF-GPL & Eq.~\eqref{Eq:ELBO} & Eq.~\eqref{Eq:Particlestatereconsloss}  & Eq.~\eqref{ActionModelLoss} & Eq.~\eqref{ValueLoss} \\ \hline
        VAE-GPL  & Eq.~\eqref{Eq:VAEBeliefLossFunction} & Eq.~\eqref{Eq:VAEStateLossFunction}  & Eq.~\eqref{ActionModelLoss} & Eq.~\eqref{ValueLoss}  \\ \hline
        AE-GPL   &   Eq.~\eqref{Eq:AEBeliefLossFunction}  &  Eq.~\eqref{Eq:AEStateLossFunction}  & Eq.~\eqref{ActionModelLoss} & Eq.~\eqref{ValueLoss} \\ \hline
        GPL-Q  & -- & -- & Eq.~\eqref{ActionModelLoss} & Eq.~\eqref{ValueLoss}\\
        \hline
    \end{tabular}
    \label{tab:PO_methods}
\end{table}

\textit{State Reconstruction Loss Function.} The state reconstruction loss is computed based on the set of particles, $U_{n}$, computed in Equation~\eqref{Eq:ParticleBeliefUpdate}. Given $U_{n}$ and a \textit{state reconstruction distribution} parameterised by $P_{st}=\{\theta\}$, the state reconstruction loss function is defined as:
\begin{equation}
 \label{Eq:Particlestatereconsloss}
    L^{SR}_{P_{inf}\cup{}P_{st}}(D) = -\sum_{H_{n}\in{D}}\left(\sum_{u_{k}\in{U_{n}}}\text{log}(q_{\theta}(s_{T_{n}}^{n}|s^{u_{k}}_{T_{n}},a^{u_{k}}_{T_{n}-1}))\right).
\end{equation}
In the above equation, we maximise the likelihood of the state information given the state representation and the teammate predicted action information contained in each particle.

\subsubsection{Variational Autoencoder-based Belief Models}
\label{Sec:VAEBasedBeliefObj}
\textit{Belief Inference Loss Function.} The ELBO loss function that we define to train the variational autoencoder-based belief model is defined below:
\begin{equation}
\begin{split}
\label{Eq:VAEBeliefLossFunction}
    L^{ELBO}_{P_{inf}}(D) = & -\sum_{H_{n}\sim{D}} \mathbb{E}_{z_{T_{n}}\sim{q_{P_{inf}}(z_{T_{n}}|H_{n})}}\big[\text{log}(p_{P_{inf}}(B_{obs}(o^n_{T_{n}})|z_{T_{n}})) \\ &+ \text{log}((p_{P_{inf}}(a^{V}_{T_{n}}|z_{T_{n}})))\big]\\ &- D_{KL}(q_{P_{inf}}(z_{T_{n}}|H_{n})||p(z_{t})).
\end{split}
\end{equation}
The distributions involved in the computation of this loss function are defined following the network architectures described in Section~\ref{Sec:VAEInputPreprocessingAndArchitecture}, which are parameterised by $P_{inf}=\{\alpha,\beta,\gamma\}$. To enable backpropagation through the sampling operation on $q(z_{T_{n}}|H_{n})$, we use reparameterisation tricks that are commonly used in optimising variational autoencoders~\citep{kingma2013auto}.

\textit{State Reconstruction Loss Function.} Like in the particle-based inference method, we define another model parameterised by $P_{st}=\{\zeta\}$ to parameterise the \textit{state reconstruction distribution}, $p_{P_{st}}(B_{obs}(s^n_{T_{n}})|z_{T_{n}})$. Given representations sampled from the encoder, $z_{T_{n}}$, the state reconstruction loss function is defined as:
\begin{align}
\label{Eq:VAEStateLossFunction}
   L^{SR}_{P_{inf}\cup{}P_{st}}(D) = -\sum_{H_{n}\sim{D}} \mathbb{E}_{z_{T_{n}}\sim{q_{P_{inf}}(z_{T_{n}}|H_{n})}}\big[\text{log}(p_{P_{st}}(B_{obs}(s^n_{T_{n}})|z_{T_{n}}))].
\end{align}

\subsubsection{Representation-based Models}
\label{Sec:AEBasedBeliefObj}

\textit{Belief Inference Loss Function.} The encoder and decoder are trained to minimise the following reconstruction loss function:
\begin{align}
\label{Eq:AEBeliefLossFunction}
    L^{RECONS}_{P_{inf}}(D) &= -\sum_{H_{n}\in{D}} \Big(||B_{pred}^{P_{inf}}(\rho_{T_{n}}) - B_{obs}(o^n_{T_{n}})||^{2} + \text{log}(p_{P_{inf}}(a^{V}_{T_{n}}|\rho_{T_{n}}))\Big),
\end{align}
assuming that $P_{inf} = \{\alpha, \beta\}$ are the parameters of the encoder and decoder model introduced in Section~\ref{Sec:AEInputPreprocessingAndArchitecture}. The first term in Equation~\eqref{Eq:AEBeliefLossFunction} ensures the encoder produces representations, $\rho_{T_{n}}$, containing observed teammate information. The second term enforces $\rho_{t}$ to be predictive of teammates' actions. As in the ELBO loss for variational autoencoders, the above loss function enables the encoder to produce representations that are informative of teammates' behaviour during interaction. 

\textit{State Reconstruction Loss Function.} Similar to the optimisation of our VAE-based model, we define a state reconstruction model parameterised by $P_{st}=\{\zeta\}$. This model is used to reconstruct the state from the representation produced by the encoder. Both the autoencoder and the state reconstruction model are then trained to minimise the following loss function:
\begin{align}
\label{Eq:AEStateLossFunction}
L^{SR}_{P_{inf}\cup{}P_{st}}(D) &= -\sum_{H_{n}\in{D}} ||B_{pred}^{P_{inf}\cup{}P_{st}}(\rho_{T_{n}}) - B_{obs}(s^n_{T_{n}})||^{2}.    
\end{align}

%% file: sections/results_2.tex
\section{Partially Observable Open Ad Hoc Teamwork Experiments}
\label{sec:results_2}

In this section, we describe different experiments performed with the methods introduced in the previous Section. We evaluate the methods in several open ad hoc teamwork tasks under partial observability. We first start by describing the environments and algorithms used in our evaluation (Section~\ref{Sec:PartObsExpSetup}). We then present a performance of our algorithms followed by a reconstruction analysis that seeks to evaluate the performance of the different belief methods.

\subsection{Experimental Setup}
\label{Sec:PartObsExpSetup}

Following, we describe the environments (Section~\ref{Sec:PartObsEnvs}) and the algorithms (Section~\ref{Sec:Baselines}) used for our evaluation. Our experimental setup in this section with respect to environment openness and teammates types follows that of Section~\ref{sec:exp_setup}.

\subsubsection{Environments}
\label{Sec:PartObsEnvs}
We utilised two of the previously described environments, for which we induce partial observability by means of different observation functions. Finally, we also incorporate a new environment for the partial observable case only, namely \ac{PCN}.   

\textit{Level-Based Foraging.} In LBF we induce partial observability by only allowing the learner to see objects and teammates within a certain region surrounding the learner's current grid. For this particular test setup, we utilised a grid world of size $12\times 12$ and only allow the learner to observe entities within a $5\times5$ grid centred in the learner.

\textit{Wolfpack.} In Wolfpack, partial observability is induced by restricting the learner to only observe agents and prey whose Manhattan distance is less than a certain value relative to itself. We set the grid world as a $10\times10$ square and limit the learner's observation to entities within a Manhattan distance of 3 from itself.

\textit{Penalized Cooperative Navigation (PCN).} Similar to the cooperative navigation environment~\citep{tacchetti_relational_2018}, multiple players must navigate through a $12\times12$ grid world to simultaneously cover two destination grids to get a reward of $1$. However, the learner is given a $-0.2$ penalty if it arrives at a destination without other teammates covering the other. We make reasoning through partial observability a necessity by frequently positioning the destination grids far away from each other. To avoid penalties, the player must then reason whether teammates are about to arrive at a destination outside its observation. After a pair of agents arrive at the destinations, we randomly choose a new pair of destination grids. Similar to LBF, the learner can only see the destination grids or teammates if they are inside a $5\times5$ region surrounding the learner. 

\subsubsection{Algorithms}
\label{Sec:Baselines}

We present here the different algorithms developed based on the belief representation methods described in Section~\ref{sec:POGPL}. Each of these algorithms uses GNNs to produce representations that characterise the latent environment state, which is inputted to the joint action value and agent model for optimal action-value function estimation. Table \ref{tab:PO_methods_evaluated} provides a summary of the different algorithms by describing their main components.

\begin{table}[t]
    \small
    \center
    \caption{\textnormal{Evaluated Belief Inference Algorithms}: Belief inference algorithms evaluated in this work are based on the usage of separate representations for different latent variables, the addition of state reconstruction loss for training, and the approximate belief inference method being used.}
    \label{tab:partialadhoc_baselines}
    \begin{tabular}{|c|c|c|c|c|c|c|}
        \hline
        Models & Separate variable &  State & \multicolumn{4}{c|}{State Inference method} \\ \cline{4-7} 
        & representation &  reconstruction & RNN  & AE-based & Particle-based & VAE   \\ \hline
        GPL-Q  &  &  & \checkmark & & & \\ \hline 
        AE-GPL   &  & \checkmark & & \checkmark & & \\ \hline
        PF-GPL & \checkmark & \checkmark & & & \checkmark  & \\ \hline
        VAE-GPL  &  & \checkmark & & & & \checkmark   \\ %
        \hline
    \end{tabular}
    \label{tab:PO_methods_evaluated}
\end{table}

\textit{Representation-based State Inference (AE-GPL)} We utilise an autoencoder architecture, as presented in Section~\ref{sec:AutoencoderBasedStateInference}, to create the AE-GPL algorithm. AE-GPL learns an embedding $z_t$ that contains information about the teammates' policies and the environment dynamics. This embedding $z_t$ can then be used for decision-making by the learner. We named this algorithm Autoencoder GPL (AE-GPL).

\textit{Particle-based Belief (PF-GPL)} 
We introduce the particle filter graph-based belief and policy learning (PF-GPL) algorithm, which utilises the particle-based representation as presented in Section~\ref{sec:ParticleBasedStateInference}. PF-GPL allocates separate representations to model the different latent variables required for decision-making. We formulate different ablations to identify the differences that result from utilising different numbers of particles in the belief inference of \ac{SMC}-based methods. We, therefore, run PF-GPL ablations with ten (PF-GPL-10), five (PF-GPL-5) and one particle (PF-GPL-1) to see the effects of using less and even a single sampled vector to represent the agent's belief. 

\textit{Variational Autoencoder-based Belief (VAE-GPL)}
Variational Autoencoder GPL (VAE-GPL) is an algorithm based on the method presented in Section~\ref{sec:VAEBasedStateInference}. VAE-GPL utilises a variational autoencoder to maintain a distribution of latent variables $z_t$ that encodes the belief about the current state of the system. Note that we also train VAE-GPL to reconstruct the state information, which we assume to be known during training.

\textit{Graph-based Policy Learning (GPL-Q)} 
\label{Sec:GPLBaselineDescription}
While it assumes full observability of the state, GPL-Q can still be used under partial observability. We apply GPL-Q in our experiments by treating the learner's observations as input states from the environment. Following the effectiveness of RNNs for learning policies for POMDPs~\citep{hausknecht2015deep}, GPL-Q's RNN-based type inference network should still facilitate the learning of reasonable policies from $o_{\leq{t}}$ and $a^{i}_{<t}$. Unlike other evaluated algorithms, GPL-Q is not trained to reconstruct the state information during training.

\textit{Single-agent RL baselines.} In addition to the previously described methods, we also evaluated two single-agent RL baselines. Unlike our proposed methods, these single-agent RL baselines do not perform any agent modelling or joint action computation. Comparing these methods can then shed light on the improvements our methods bring.  
\begin{itemize}
    \item Proximal Policy Optimization (PPO): We utilise PPO as a single agent baseline \citep{schulman2017proximal}. The original PPO method is intended for fully observable environments so it does include any method to deal with partially observable environments. Comparing against such a baseline can provide information regarding the value of models that infer the unobserved state variables. 
    \item Deep Variational Reinforcement Learning (DVRL): The DVLR baseline is also a single-agent RL method \citep{IglICML2018}, but unlike PPO, it has a method to estimate the unknown state variables based on the agent's observations. DVRL utilises sequential Monte Carlo, with 10 particles. More details of the hyperparameters are given in the Appendix~\ref{annex:PO_results}. 
\end{itemize}

\begin{figure}[t]
\centering
 \subfloat[Level-based Foraging]{%
      \includegraphics[width=0.475\textwidth]{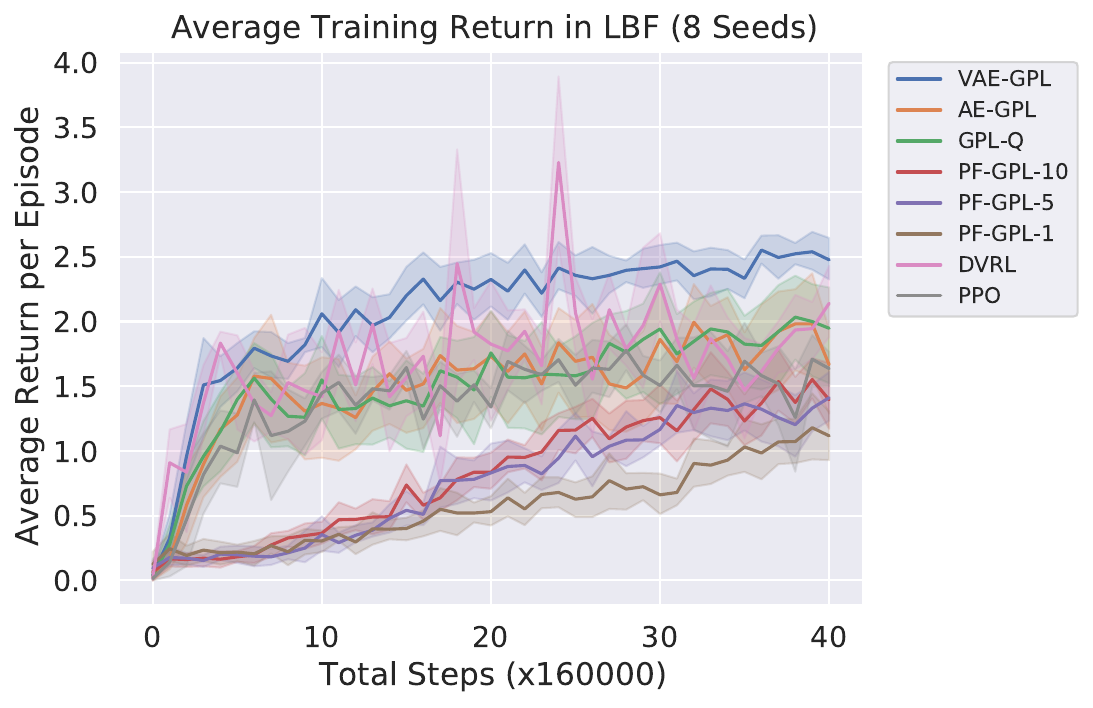}
    }  \hfil
    \subfloat[Wolfpack ]{%
      \includegraphics[width=0.475\textwidth]{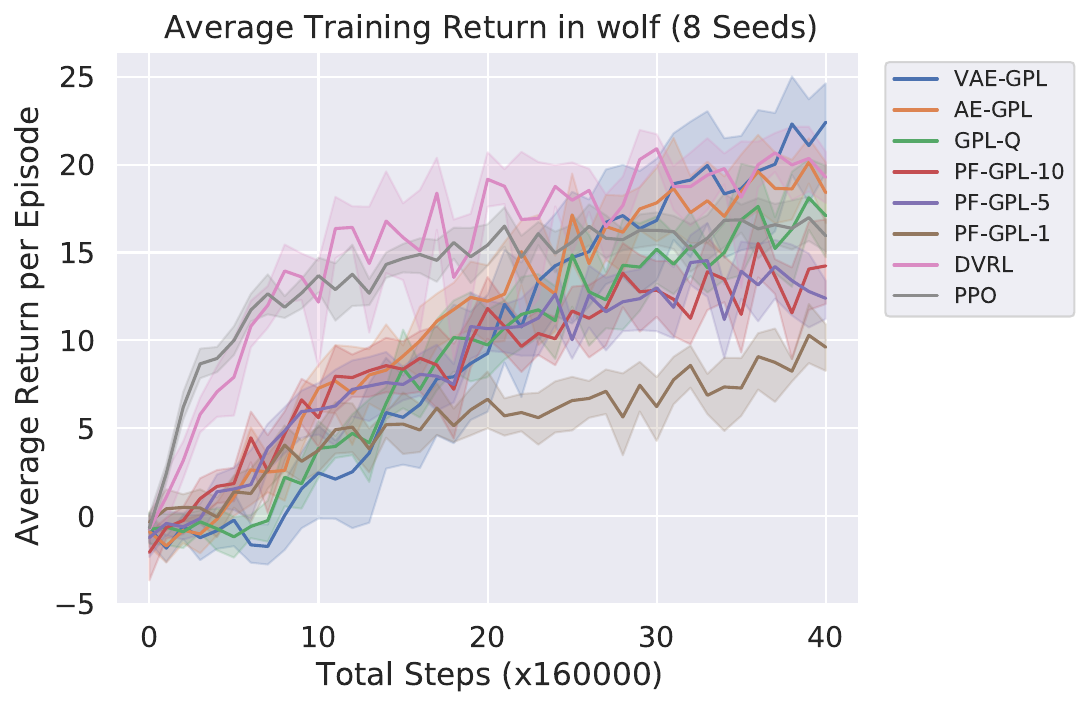}
    } \hfill
    \subfloat[Cooperative Navigation]{%
    \label{fig:PO_open_ad_hoc_results_coop}
      \includegraphics[width=0.47\textwidth]{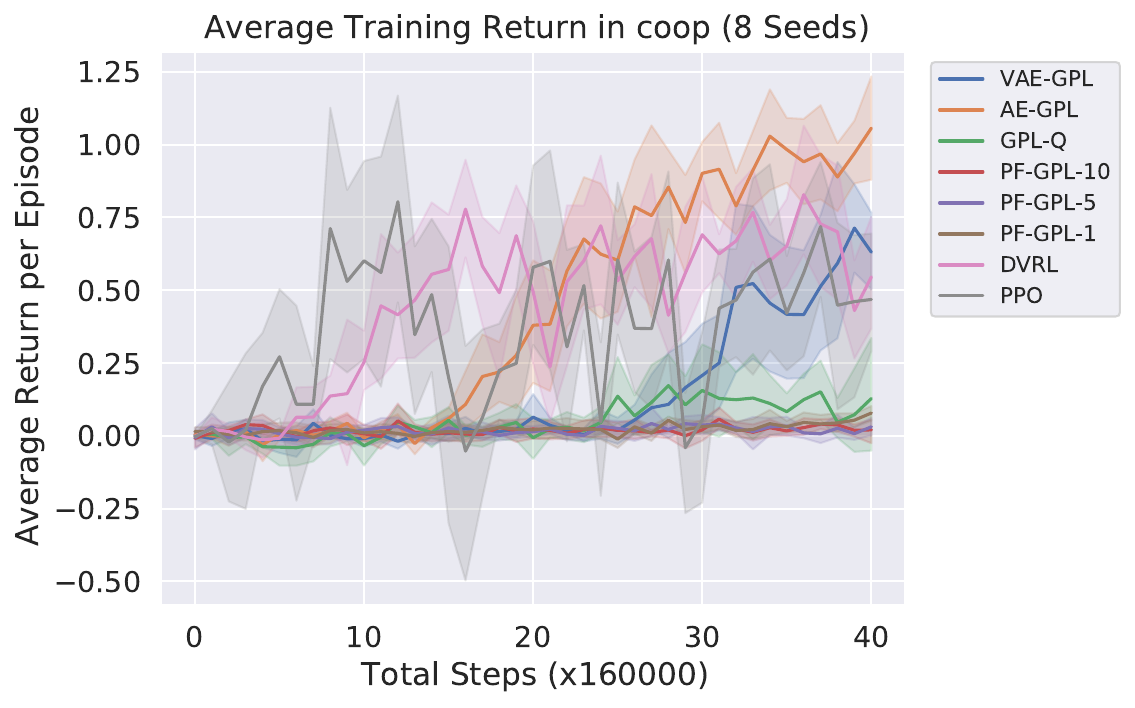}
    }
    \caption{\textnormal{Partially observable open ad hoc teamwork results (training).}  Obtained returns for all evaluated environments during training. We show the average and $95\%$ confidence bounds utilising 8 seeds. }
    \label{fig:PO_open_ad_hoc_results}
\end{figure}

\subsection{Partially Observable Open Ad Hoc Teamwork Results}

The returns obtained by the proposed methods in the partially observable open ad hoc teamwork experiments are provided in Figure~\ref{fig:PO_open_ad_hoc_results}. These are the training results, without generalising to unseen agents. 
In all three environments, we see that the autoencoder and variational autoencoder-based methods learn to achieve significantly higher returns than other methods, closely followed by DVRL. This is particularly true in the cooperative navigation environment, in which PF-GPL-based methods achieve a return close to zero. Nonetheless, PF-GPL-based methods improve their returns in LBF and Wolfpack as the number of particles used during inference increases. This aligns with previous results from other particle-based methods~\citep{albrecht2016causality}, which demonstrates the need for using a larger number of particles to increase belief representation accuracy. PPO, while having no mechanism to estimate belief states, has a performance that is comparable to the other methods, even surpassing the PF-GPL baselines. However, further analysis shows that PPO lacks generalisation capabilities as its performance degrades when collaborating with unseen teammates (in Section~\ref{sec:PO_generalization_results}). 

The suboptimal performance of PF-GPL in Figure~\ref{fig:PO_open_ad_hoc_results} suggests that the proposed graph-based particle belief representation is not able to generate useful representations for decision-making. This contrasts with DVRL, which also utilises a particle belief representation, but is able to achieve higher returns in all environments. We believe that this is due to the major number of network models used to estimate teammates' information in PF-GPL. These additional terms increase the complexity of the network of PF-GPL, which potentially requires more environment interactions and a higher number of particles to achieve comparable performance. %

We can see that VAE-GPL is the best-performing method in LBF. While GPL-Q, AE-GPL and DVRL achieved comparable returns. PPO performance, while lower than VAE-GPL, still surpasses PF-GPL methods. This tendency is maintained in Wolfpack as can be seen in Figure~\ref{fig:PO_open_ad_hoc_results}. It is important to note that in both environments, GPL-Q and PPO achieve comparable performance despite not having models that are specifically designed for belief inference. %
In the case of GPL-Q, the RNN-based type inference model still enables the discovery of important information for decision-making based on the sequence of observations experienced by the learner. We can view the changing number of teammates resulting from the learner's partial observability as another open process, which the learner can still solve as long as the sequence of observations contains useful information for action selection. 
While in the case of PPO, these findings are in line with other works that show that PPO is able to achieve similar returns when compared to methods specifically tailored to partially observable problems~\citep{morad2023popgym}. %

\begin{table*}[t]
    \centering
    \small
    \caption{\textnormal{Partially observable open ad hoc teamwork results (testing):} We show the average and 95\% confidence bounds during testing utilising 8 seeds. The data was gathered by averaging the returns at the checkpoint which achieved the highest average performance during training. We highlight in bold the algorithm with the highest average returns.}
    \label{eval-table-partial-obs}
    \begin{tabular}{|l|c|c|c|}
    \hline 
    Algorithm & LBF & Wolf. & Coop. \\ \hline
    VAE-GPL & 0.99$\pm$0.18  & \textbf{27.36$\pm$2.9} & 0.64$\pm$0.17 \\
    AE-GPL  & 0.88$\pm$0.23 &  25.62$\pm$1.06 & \textbf{0.96$\pm$0.11} \\
    GPL-Q  & \textbf{1.03$\pm$0.15} & 23.18$\pm$1.3 & 0.10$\pm$0.10 \\
    PF-GPL-10  & 0.75$\pm$0.09  & 19.67$\pm$2.0 &   0.02$\pm$0.02 \\
    PF-GPL-5  & 0.73$\pm$0.12 & 19.06$\pm$1.5 & 0.03$\pm$0.04 \\
    PF-GPL-1   & 0.57$\pm$0.11 & 14.73$\pm$1.2 &  0.05$\pm$0.04 \\
    DVRL & \textbf{1.12$\pm$0.61} & 20.26$\pm$1.1 & 0.59$\pm$0.12 \\
    PPO  & 0.95$\pm$0.36 &  20.06$\pm$1.2 & 0.42$\pm$0.20 \\ \hline
    \end{tabular}
\end{table*}

In contrast to their results in LBF and Wolfpack, GPL-Q and PPO perform poorly in cooperative navigation. This is because the observations perceived by the learner contains the least useful information compared to other environments. It is important to note that the most important information in cooperative navigation is whether another teammate is positioned nearby another destination grid, which is usually unobserved. As such, methods with an additional state reconstruction loss will certainly produce better representations for decision-making compared to GPL-Q and PPO.

AE-GPL, VAE-GPL and DVRL are the only methods that consistently achieve high returns in all three environments. In the Cooperative Navigation environment where reasoning capabilities on unobserved teammates are most important, AE-GPL, VAE-GPL and DVRL still achieve high returns. AE-GPL and VAE-GPL's significantly higher returns than the other methods suggest the importance of using recurrent neural networks for latent variable inference and observation reconstruction to create useful representations for decision-making. Other methods that are not equipped with observation reconstruction, such as GPL-Q, cannot consistently achieve high returns.
In the next Section, we perform a generalisation analysis where we show that GPL-based methods are able to outperform the other baselines due to the use of agent modelling techniques.

\begin{figure}[t]
\centering
 \subfloat[Level-based Foraging]{%
      \includegraphics[width=0.47\textwidth]{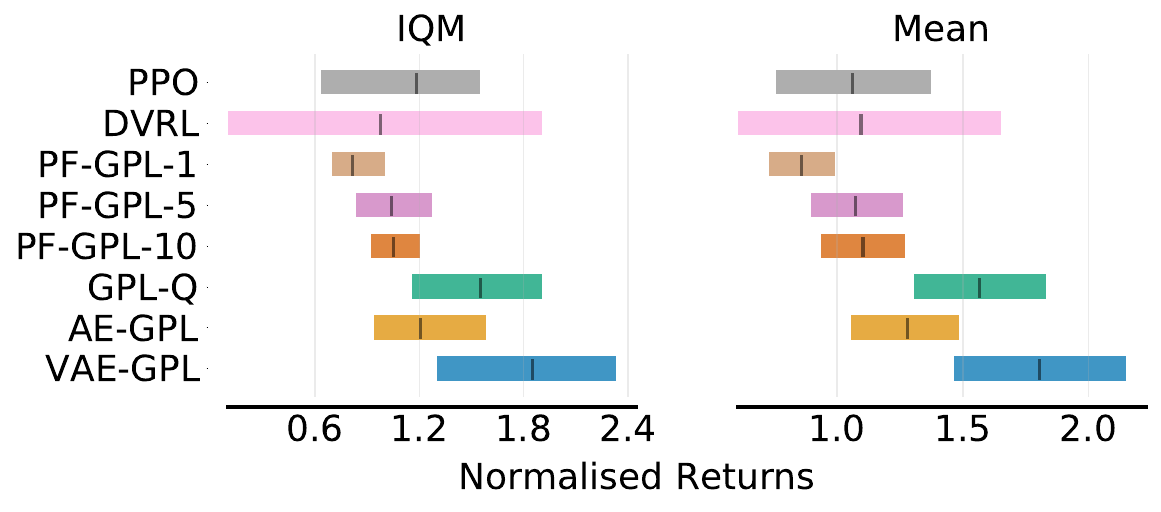}
    }  \hfil
    \subfloat[Wolfpack]{%
      \includegraphics[width=0.47\textwidth]{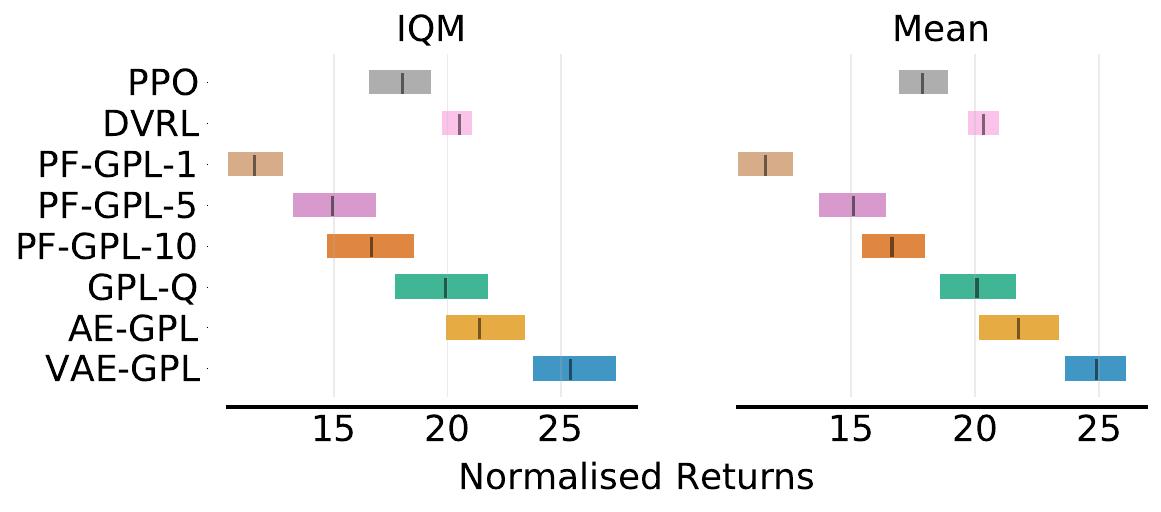}
    } \hfill
    \subfloat[Cooperative Navigation]{%
      \includegraphics[width=0.47\textwidth]{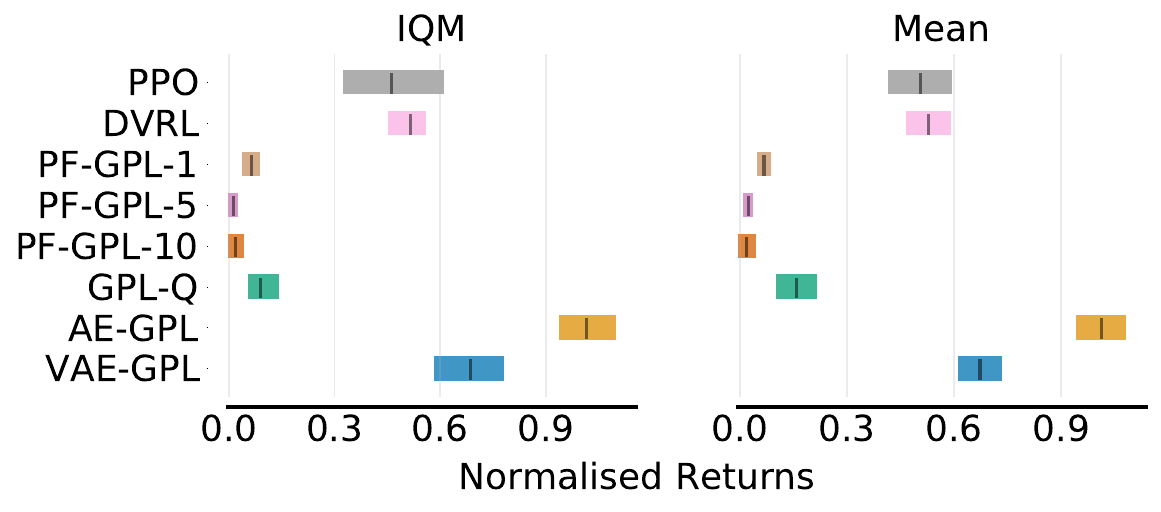}
    }
    \caption{\textnormal{Aggregated performance when collaborating with a different number of teammates.} We aggregated the performance of the different algorithms when collaborating with three and five teammates using the last training checkpoint.}        
    \label{fig:IQM}
\end{figure}

\begin{table*}[ht]
    \centering
    \small	
    \caption{\textnormal{Unseen teammates evaluation (testing):} We show the average and 95\% confidence bounds during testing for the partially observable open ad hoc teamwork utilising 8 seeds. The data was gathered by averaging the returns at the checkpoint which achieved the highest average performance during training. We highlight in bold the algorithm with the highest average returns.}
    \label{unseen-teammates-table-partial-obs}
    \begin{tabular}{|l|c|c|c|}
    \hline 
    Algorithm & LBF & Wolf. & Coop. \\ \hline
    VAE-GPL & \textbf{0.80$\pm$0.08} & 23.32$\pm$3.48 & 0.17$\pm$0.15 \\
    AE-GPL  &  0.76$\pm$0.05 & \textbf{23.78$\pm$1.69} & \textbf{0.24$\pm$0.10} \\
    GPL-Q  &   0.77$\pm$0.10 &  21.28$\pm$1.58 & 0.07$\pm$0.13 \\
    PF-GPL-10  & 0.61$\pm$0.10  &  18.12$\pm$2.38 &  0.02$\pm$0.03 \\
    PF-GPL-5  & 0.63$\pm$0.06 & 15.85$\pm$0.68 &  0.004$\pm$0.02 \\
    PF-GPL-1   &  0.61$\pm$0.10 & 12.88$\pm$1.36 &  0.002$\pm$0.03 \\
    DVRL & 0.07$\pm$0.05 & 19.29$\pm$0.96 & -0.11$\pm$0.09 \\
    PPO  & 0.03$\pm$0.03 &   18.43$\pm$0.89 &   -0.24$\pm$0.12 \\ \hline
    \end{tabular}
\end{table*}

\subsection{Generalisation results}
\label{sec:PO_generalization_results}

Similarly to our generalisation experiments under the fully observable setting, we present the generalisation capabilities of the agents to different numbers of teammates in Table~\ref{eval-table-partial-obs}. Unsurprisingly, methods that achieve low returns during training, such as PF-GPL, will also achieve subpar performance when generalising to different open processes. 
However, although VAE-GPL, and AE-GPL are still the top performers, it seems that none of the methods outperforms the other. 
To achieve a more clear picture, we evaluated the performance of the algorithms for different numbers of teammates and aggregated the results utilising the IQM metric \citep{agarwal2021deep}. We show the results in Figure~\ref{fig:IQM}.
These results  indicate that VAE-GPL and AE-GPL are able to achieve higher returns when collaborating with a different number of teammates.

Finally, we evaluated the generalisation capabilities of the methods to unseen teammates. To achieve this, we included new teammates that were not seen during training and evaluated the performance of each of the methods. More information about the generated teammates can be found in Appendix~\ref{annex:unseen_teammates}. Table~\ref{unseen-teammates-table-partial-obs} summarises the obtained results. This evaluation is a critical point, as collaborating with unseen teammates is one of the main requirements of ad hoc teamwork. It can be seen that methods that have a way of estimating other agent types and their actions obtain higher returns. While DVRL and PPO achieved comparable returns in the previous evaluation, they fail to generalise to teammates that are outside of their training distribution. Similarly, PF-GPL methods are not able to achieve high returns in any of the evaluated environments. On the other hand, VAE-GPL and AE-GPL are still able to achieve high returns in all environments. This highlights the advantage of using recurrent neural networks for latent variable inference and observation. But more importantly the need for type inference and agent modelling for optimal action selection in ad hoc teamwork.

\subsection{Reconstruction Results}

\begin{figure}[t]
\centering
 \subfloat[Level-based Foraging]{%
      \includegraphics[width=0.47\textwidth]{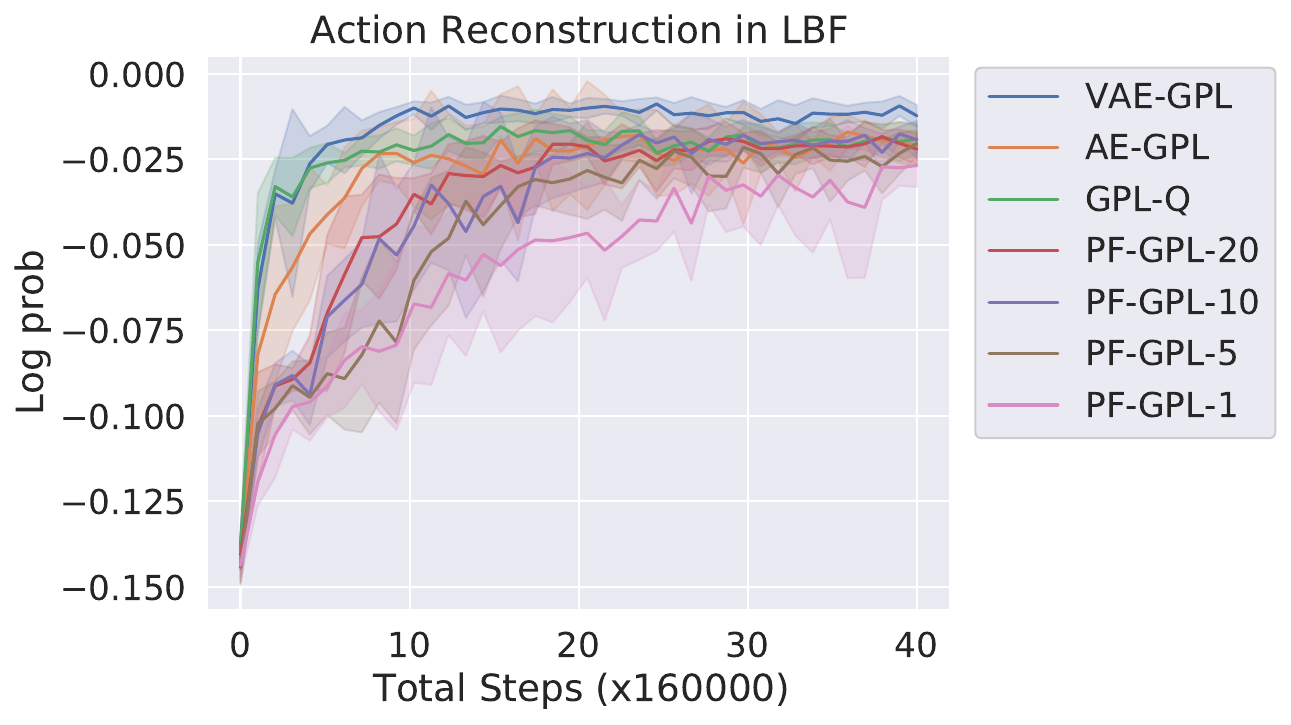}
    }  \hfil
    \subfloat[Wolfpack]{%
      \includegraphics[width=0.47\textwidth]{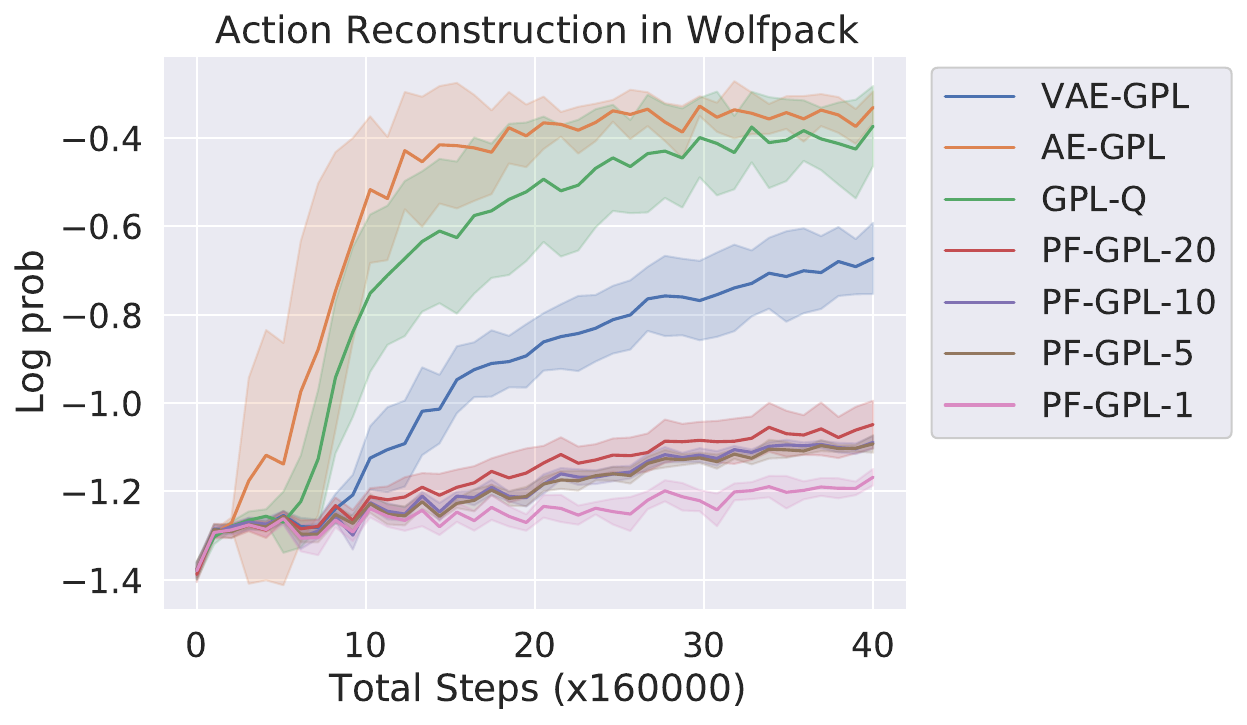}
    } \hfill
    \subfloat[Cooperative Navigation]{%
      \includegraphics[width=0.47\textwidth]{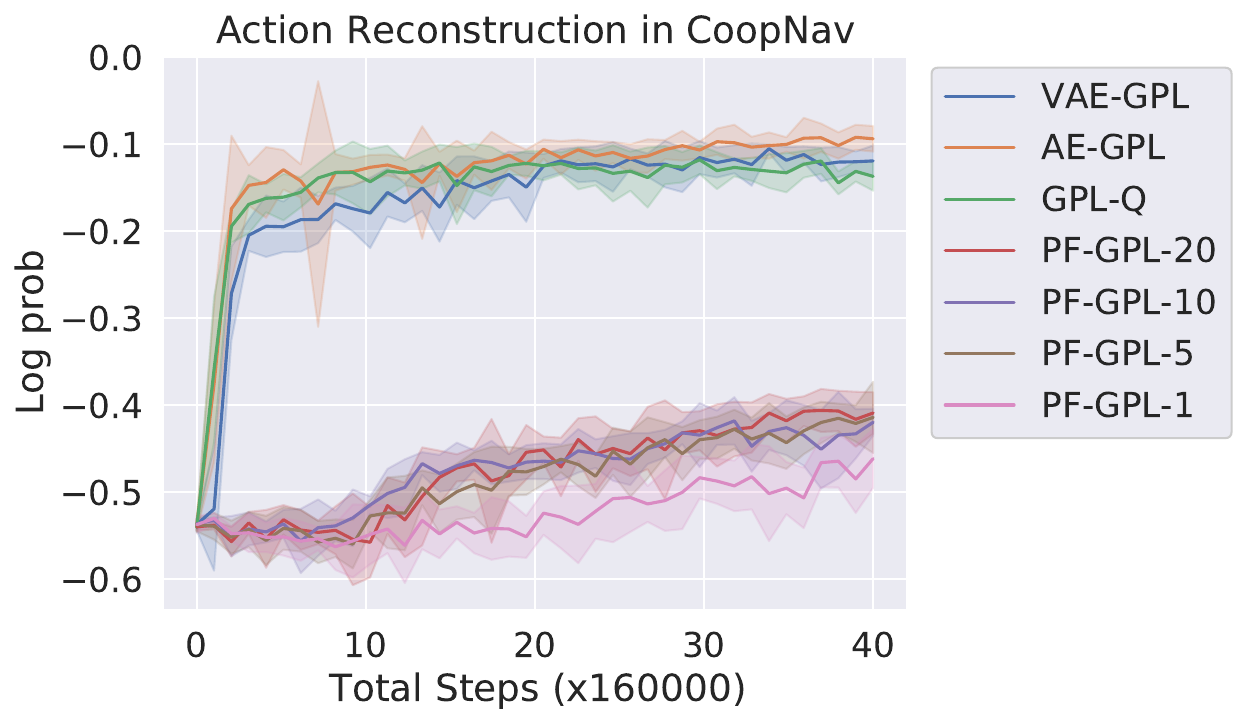}
    } 
    \caption{\textnormal{Action reconstruction accuracy.} We evaluate the log probability between the predicted actions and the true actions taken by the teammates. We ran the inference modules for each of the algorithms over a fixed episode in which actions were predetermined. We evaluated the log probability  n times over the fixed episode for each checkpoint and report the mean and $95\%$ confidence bounds.}     
    \label{fig:action_reconstruction_accuracy}
\end{figure}

\begin{figure}[t]
\centering
 \subfloat[Level-based Foraging]{%
      \includegraphics[width=0.47\textwidth]{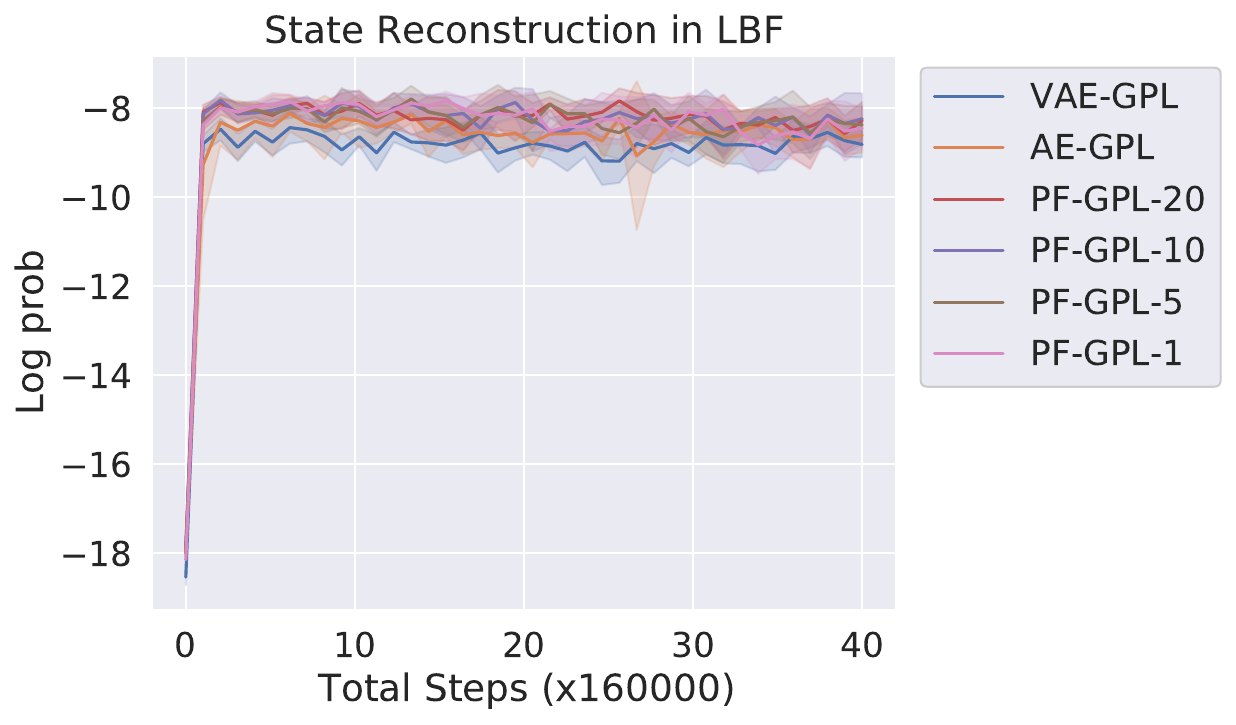}
    }  \hfil
    \subfloat[Wolfpack]{%
      \includegraphics[width=0.47\textwidth]{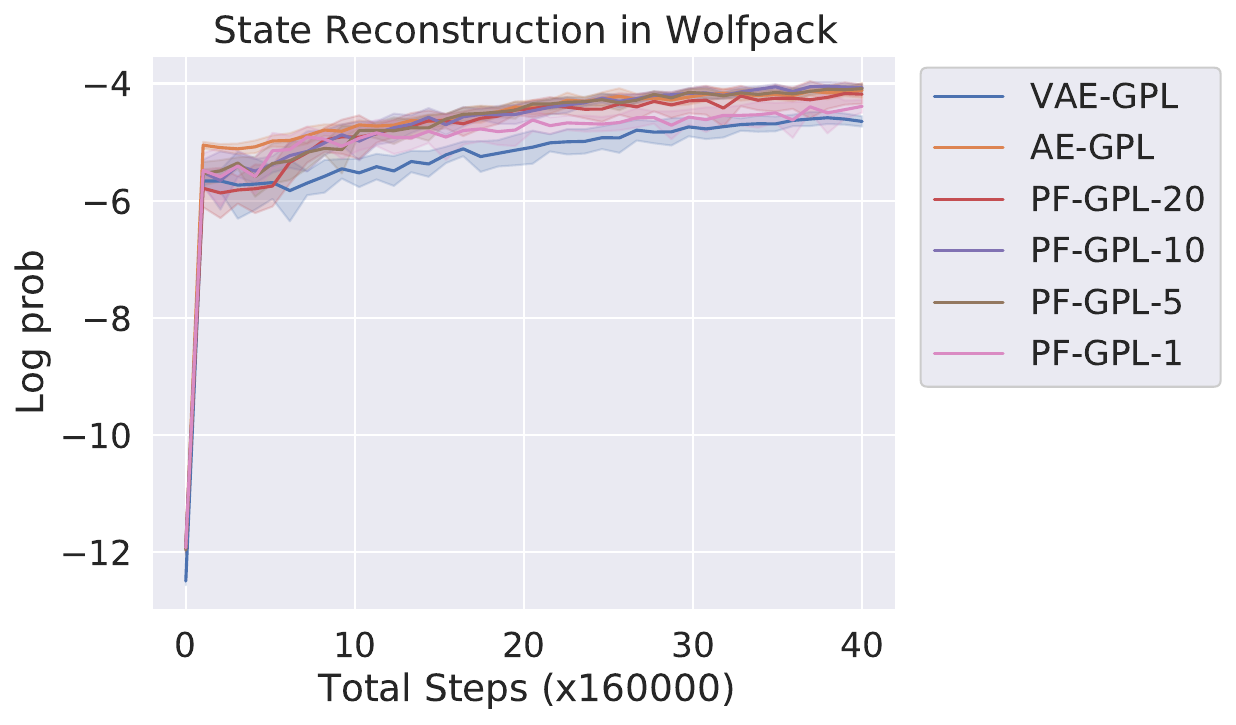}
    } \hfill
    \subfloat[Cooperative Navigation]{%
      \includegraphics[width=0.47\textwidth]{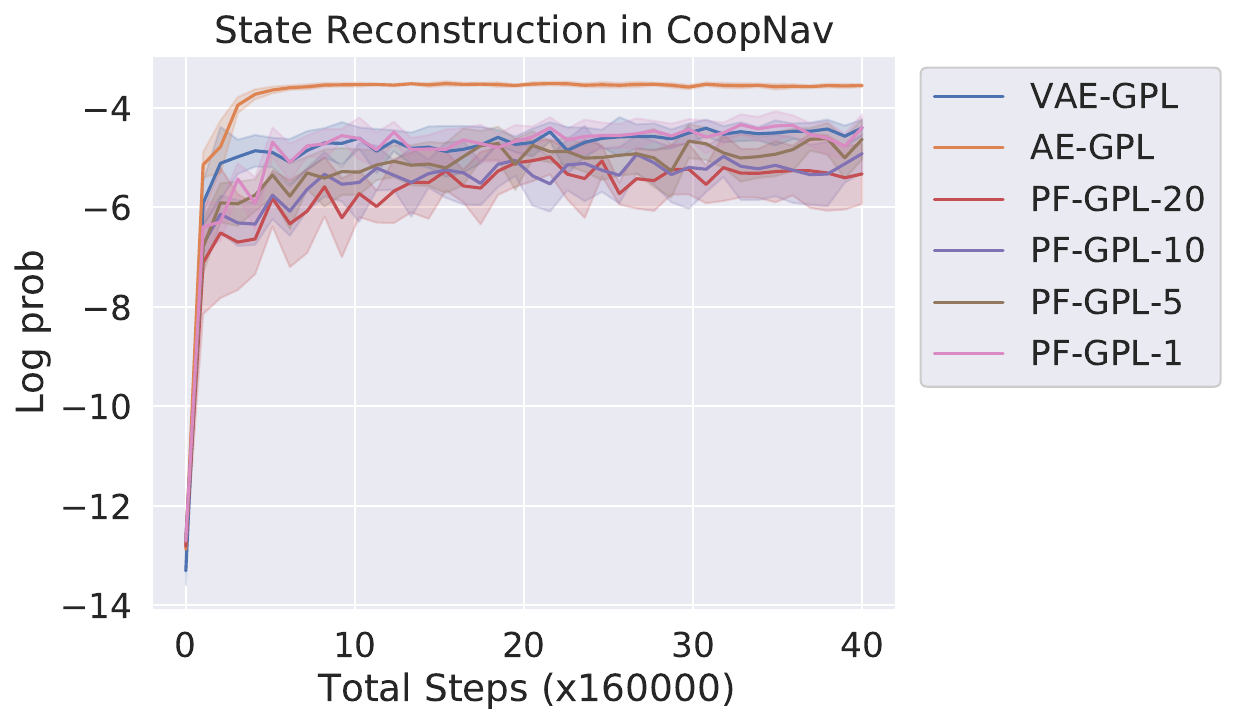}
    }
    \caption{\textnormal{State reconstruction accuracy.} We evaluate the log probability between the predicted state and the true state of the system. We ran the inference modules for each of the algorithms over a fixed episode in which actions were predetermined. We evaluated the log probability n times over the fixed episode for each checkpoint and report the mean and $95\%$ confidence bounds.}     
    \label{fig:state_reconstruction_accuracy}
\end{figure}

\begin{figure}[t]
\centering
 \subfloat[Level-based Foraging]{%
      \includegraphics[width=0.47\textwidth]{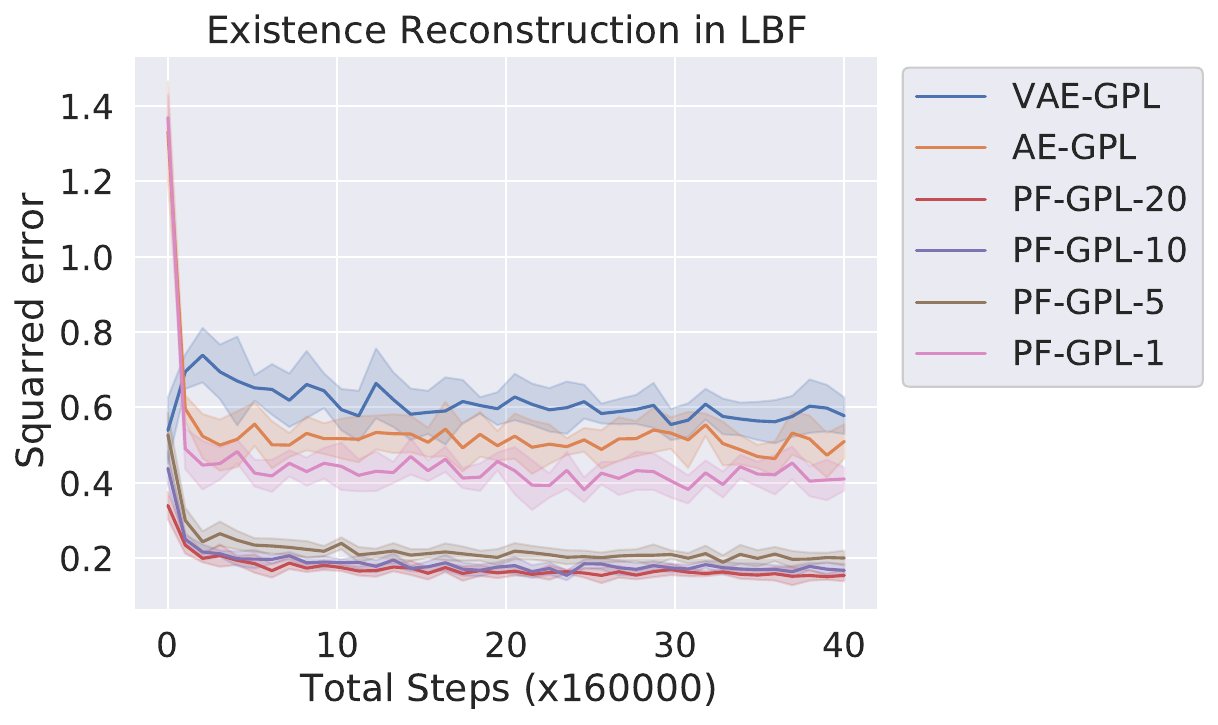}
    }  \hfil
    \subfloat[Wolfpack]{%
      \includegraphics[width=0.47\textwidth]{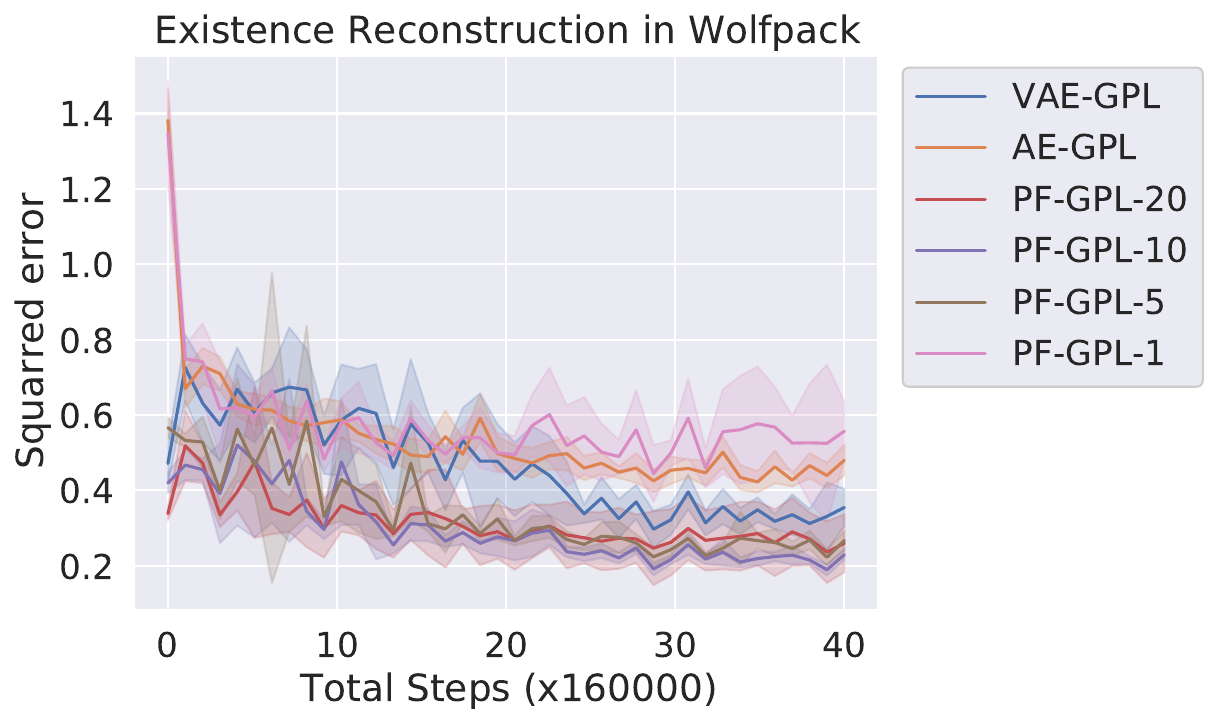}
    } \hfill
    \subfloat[Cooperative Navigation]{%
      \includegraphics[width=0.47\textwidth]{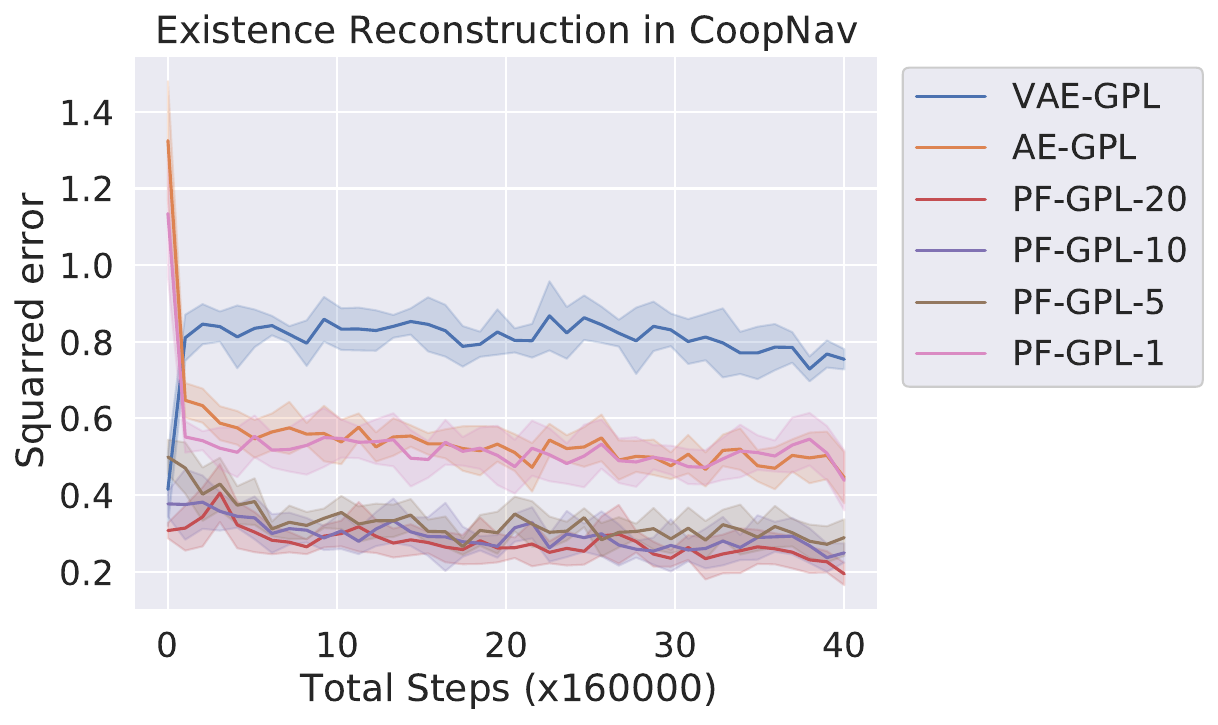}
    } 
    \caption{\textnormal{Existence reconstruction accuracy.} We measure how accurately the belief inference network can predict the existence of other agents in the environment. We calculated the squared error between the true number of existent agents in the environment ($\sum{e_t}$) and the estimation. We ran the inference modules for each of the algorithms over a fixed episode in which actions were predetermined. We evaluated the log probability n times over the fixed episode for each checkpoint and report the mean and $95\%$ confidence bounds.}     
    \label{fig:existence_econstruction_accuracy}
\end{figure}

In this section, we evaluate the reconstruction capabilities of the methods proposed in Section~\ref{Sec:Baselines}. We do this evaluation for two reasons. First, we want to examine whether the methods are capable of representing useful information for decision-making. Second, we also aim to elucidate which learned information is most useful in improving the returns of the learner. This evaluation is done on the environments defined in Section~\ref{Sec:PartObsEnvs}. 

The reconstruction evaluation was done over a single episode. We collect an episode of interaction data $H=\{o_{t}, a_{t}\}_{t=1}^{T}$, by executing the policy resulting from the algorithm with the highest training returns in each respective environment. At every training checkpoint, we utilise the single-episode interaction data to evaluate each method's reconstruction capabilities for different measures such as the environment state, teammates' joint actions, and teammates' existence. 

The resulting reconstruction performance for teammates' joint actions, state reconstruction, and teammates' existence reconstruction are provided in Figure~\ref{fig:action_reconstruction_accuracy}, Figure~\ref{fig:state_reconstruction_accuracy}, and Figure~\ref{fig:existence_econstruction_accuracy} respectively. To evaluate action reconstruction, at each checkpoint we report the average log likelihood of all teammates' joint actions, which includes teammates that are not observed by the learner. We then evaluate the state reconstruction capabilities of the methods by reporting the log probability they assign to the unobserved state of the environment. Assuming existing teammates are denoted by a binary value of one while non-existent teammates are assigned a value of zero, we report the sum of the squared error between the predicted and real teammate existence for all agents. 

Among the evaluated measures, the capability of the methods in terms of teammate action prediction is the best indicator of their achieved returns during training. This is mainly because a method incapable of accurately predicting the teammates' joint actions will lead the learner to produce worse action value estimates. Following its significantly worse action prediction performance compared to other methods, it is unsurprising to see PF-based methods' failure in achieving high returns during training. Meanwhile, GPL-Q, AE-GPL, and VAE-GPL, produce higher returns resulting from having better teammate joint action prediction.

An improved state reconstruction capability of a method also leads towards improved returns during training. While the state reconstruction performance of the methods under this measure are similar to each other in LBF and FortAttack, AE-GPL is significantly better than other methods in cooperative navigation. Improving state prediction capabilities in cooperative navigation is crucial for producing high returns, since estimating whether teammates are close to an unobserved destination grid is the only way for the learner to avoid being penalised. As a result, AE-GPL outperforms other methods in cooperative navigation even if it has similar performances with GPL-Q  and VAE-GPL in terms of action reconstruction.

Finally, the results suggest that reconstructing agent existence is the least important for producing high returns during training. PF-based methods significantly achieve the lowest squared error for this particular measure. Despite its ability to very accurately predict the existence of agents, its inability to accurately predict the state and joint actions of teammates prevents PF-based methods from achieving higher returns.

%% file: sections/conclusion.tex
\section{Conclusions}
\label{sec:conclusions}

In this work, we addressed the challenging problem of open ad hoc teamwork, both under full and partial observability. 
We first addressed the open ad hoc teamwork problem under full observability. To this end, we introduced different algorithms based on our proposed framework \acf{GPL}. GPL consists of three main components: a type inference model, an agent model and a joint-action value network. 
We evaluated GPL in three different environments, in which the agent has access to the full state of the environment, and compared it against a set of single-agent RL and MARL baselines. Our results show that our proposed approach was able to learn an optimal policy, successfully outperforming all baselines. 
Further analysis showed that the agent modelling module in GPL plays a crucial role in the learner's performance.  
Our analysis demonstrated that the joint action value model allows the learner to identify, and emulate, effective behaviour directly from other well-performing teammates.

We then addressed the problem of open ad hoc teamwork under partial observability. We explored how different methodologies could provide a belief estimate of the state. Specifically, we evaluated three different methodologies, i) autoencoders architectures, ii) variational autoencoders, and iii) particle belief methods. 
Similar to the full observability case, we evaluate our proposed algorithms in three different environments in which agents have only partial access to the state of the system, and compared it against state-of-the-art single-agent baselines, PPO and DVRL. The results obtained show that variational autoencoder methods are able to outperform the other baselines in LBF and Wolfpack, while autoencoder-based methods were able to surpass the performance of other belief inference methods for Cooperative Navigation. Our reconstruction analysis shows that methods capable of improved accuracy in predicting the teammate's actions are able to achieve higher returns, while an improved state estimation, such as in autoencoder methods, explains the difference in performance for the Cooperative Navigation environment. While the single agent RL baselines were able to achieve comparable results in the training environment, our generalisation evaluation showed that these methods fail to generalise to agents that are outside of the training distribution. On the other hand, our proposed methods based on autoencoder architectures were able to achieve higher returns.

In future work, we propose to extend GPL and its extensions to problems with continuous actions, as it will allow the learner to tackle a more diverse set of problems~\citep{carlucho2022cooperative}.
Another issue that needs to be investigated is the scalability of the methods to larger groups of agents. This is an open research question in the multi-agent literature, especially in MARL \citep{Gronauer2022,gogineni2023scalability}.
Additionally, in our present work, the learner is only trained and evaluated in settings where teammates have a fixed policy. However, this assumption might not hold in real-world environments as the team might have agents that are also learning or updating their policy over time. This can potentially cause non-stationarity issues, similar to what occurs in MARL \citep{Papoudakis2021AgentMU}. This issue will require further adaptability from our agent, as its type inference method will have to adapt to these changes during learning. An interesting approach in this regard is to identify from interactions with a teammate whether its policy is adequately represented by the types known to the learner~\citep{albrecht2015criticising}.

More work is also needed to efficiently estimate beliefs in the partially observable setting.  In a \acf{PO-OSBG}, not all necessary information may be available in the observation. Therefore, it may be possible to design learners that take specific actions to improve the accuracy of their belief states, or that try to communicate with other agents in the team to gather additional information about the true state of the environment. 
Furthermore, since our experiments indicate that the proposed belief inference models are performing well in inferring different types of latent variables, exploring a combination of our proposed approaches to improve the inference of all important latent variables for decision-making is also a promising research direction.
Finally, one area that is worth exploring in future works is how other learning algorithms, such as PPO, could be leveraged to achieve better ad hoc agents. Our initial results show that PPO is an effective baseline, even when utilising a simple MLP network, which presents interesting opportunities for future research.

%% file: sections/appendix.tex
\appendix

\section{GPL Pseudocode Under Full Observability}
\label{sec:GPLPseudocode}
Before we describe the full GPL pseudocode, we first define important functions that we will use in the pseudocode. First, we denote the observation and hidden vector preprocessing method described in Section~\ref{sec:GPLInputPreprocessing} as the \textbf{PREPROCESS} function. Furthermore, we denote the action-value and joint-action value computation through Equation~\eqref{SingleActionValue} and ~\eqref{JointActionValueComputation} as the \textbf{MARGINALIZE} and \textbf{JOINTACTEVAL} functions respectively. Based on these functions, we define the \textbf{QV} function that preprocesses the input and computes the action-values for given joint-action values and agent networks. The computations in \textbf{QV} is provided in Algorithm~\ref{alg:ActValComputation}.
\begin{algorithm}[!htb]
   \caption{GPL Action Value Computation}
   \label{alg:ActValComputation}
\begin{algorithmic}[1]
   \STATE {\bfseries Input:} state $s$, \\ joint-action value model parameters ($\alpha_{Q}, \beta,\delta$),\\ agent model parameters ($\alpha_{q}, \eta,\zeta$), \\ agent model LSTM hidden vectors  $h_{t-1,q}$, \\ joint-action value model LSTM hidden vectors $ h_{t-1,Q}$
   \FUNCTION{\textbf{QV}($s, \alpha_{Q}, \alpha_{q}, \beta, \delta, \eta, \zeta, h_{t-1,Q}, h_{t-1,q}$)}
   \STATE $\mathrm{B}, \theta_{Q}, c_{Q} \leftarrow \mathrm{\textbf{PREPROCESS}}(s, h_{t-1,Q})$
   \STATE $\mathrm{B}, \theta_{q}, c_{q} \leftarrow \mathrm{\textbf{PREPROCESS}}(s, h_{t-1,q})$
   \STATE $\theta'_{Q}, c'_{Q} \leftarrow \mathrm{LSTM}_{\alpha_{Q}}(B, \theta_{Q}, c_{Q})$
   \STATE $\theta'_{q}, c'_{q} \leftarrow \mathrm{LSTM}_{\alpha_{q}}(B, \theta_{q}, c_{q})$
   \STATE $\forall{j}, \bar{n}_{j} \leftarrow (RFM_{\zeta}(\theta'_{q}, c'_{q}))_{j}$
   \STATE $\forall{j}, q_{\eta,\zeta, \alpha_{q}}(.|s_{t}) \leftarrow \mathrm{Softmax}(\mathrm{MLP}_{\eta}(\bar{n}_{j}))$
   \STATE $\forall{j, a^{j}}, Q^{j}_{\beta, \alpha_{Q}}(a^{j}|H_{t}) \leftarrow \textrm{MLP}_{\beta}(\theta_{Q}^{'j}, \theta_{Q}^{'i})(a^{j})$
   \STATE $\forall{j, a^{j}, a^{k}}$,
   $$
   Q^{j,k}_{\delta, \alpha_{Q}}(a^{j},a^{k}|H_{t}) \leftarrow \textrm{MLP}_{\delta}(\theta_{Q}^{'j},\theta_{Q}^{'k}, \theta_{Q}^{'i})(a^{j},a^{k})$$
   \STATE Compute $\bar{Q}(H_{t},a^{i})$ using Equation~\eqref{SingleActionValue}
   \STATE $\bar{Q}(H,.) \leftarrow \mathrm{\textbf{MARGINALIZE}}($
   $$q_{\eta,\zeta, \alpha_{q}}(.|s_{t}), Q_{\beta, \alpha_{Q}}(.|H_{t}), Q_{\delta, \alpha_{Q}}(.,.|H_{t})$$
   )
   \STATE \textbf{return} $\bar{Q}(H,.), (\theta'_{Q}, c'_{Q}), (\theta'_{q}, c'_{q})$
   \ENDFUNCTION
\end{algorithmic}
\end{algorithm}

Aside from these functions, we define \textbf{QJOINT} and \textbf{PTEAM}, which output is required to compute the loss functions, $L_{\beta,\delta}$ and $L_{\eta,\zeta}$, in Equation~\eqref{ValueLoss} and~\eqref{ActionModelLoss}. \textbf{QJOINT} is a function that computes the predicted joint action value for an observed state and joint actions. On the other hand, \textbf{PTEAM} computes the joint teammate action probability at a state. Both \textbf{QJOINT} and \textbf{PTEAM} are further defined in Algorithm~\ref{alg:jointactval} and~\ref{alg:jointactprobcomp}.

\begin{algorithm}[!htb]
   \caption{GPL Joint-Action Value Computation}
   \label{alg:jointactval}
\begin{algorithmic}[1]
   \STATE {\bfseries Input:} state $s$, observed joint action a, \\ joint-action value model parameters ($\alpha_{Q}, \beta,\delta$),\\ joint-action value model LSTM hidden vectors $ h_{t-1,Q}$
   \FUNCTION{\textbf{QJOINT}($s, a, \alpha_{Q}, \beta, \delta, h_{t-1,Q}$)}
   \STATE $\mathrm{B}, \theta_{Q}, c_{Q} \leftarrow \mathrm{\textbf{PREPROCESS}}(s, h_{t-1,Q})$
   \STATE $\theta'_{Q}, c'_{Q} \leftarrow \mathrm{LSTM}_{\alpha_{Q}}(B, \theta_{Q}, c_{Q})$
   
   \STATE $\forall{j, a^{j}}, Q^{j}_{\beta, \alpha_{Q}}(a^{j}|H_{t}) \leftarrow \textrm{MLP}_{\beta}(\theta_{Q}^{'j}, \theta_{Q}^{'i})(a^{j})$
   \STATE $\forall{j, a^{j}, a^{k}}$,
   $$
   Q^{j,k}_{\delta, \alpha_{Q}}(a^{j},a^{k}|H_{t}) \leftarrow \textrm{MLP}_{\delta}(\theta_{Q}^{'j},\theta_{Q}^{'k}, \theta_{Q}^{'i})(a^{j},a^{k})$$
   \STATE Compute $Q(s,a)$ using Equation~\eqref{JointActionValueComputation}\\ $Q(s,a) \leftarrow \mathrm{\textbf{JOINTACTEVAL}}($\\
   \begin{center}
       $a, Q_{\beta,\alpha_{Q}}(.|H_{t}), Q_{\delta,\alpha_{Q}}(.,.|H_{t})$
   \end{center}
   )
   \STATE \textbf{return} $Q(s,a)$
   \ENDFUNCTION
\end{algorithmic}
\end{algorithm}

\begin{algorithm}[!htb]
   \caption{GPL Teammate Action Probability Computation}
   \label{alg:jointactprobcomp}
\begin{algorithmic}[1]
   \STATE {\bfseries Input:} state $s$, observed joint actions $a$,\\ agent model parameters ($\alpha_{q}, \eta,\zeta$), \\ agent model LSTM hidden vectors  $h_{t-1,q}$
   \FUNCTION{\textbf{PTEAM}($s, a, \alpha_{q}, \eta, \zeta, h_{t-1,q}$)}
   \STATE $\mathrm{B}, \theta_{q}, c_{q} \leftarrow \mathrm{\textbf{PREPROCESS}}(s, h_{t-1,q})$
   \STATE $\theta'_{q}, c'_{q} \leftarrow \mathrm{LSTM}_{\alpha_{q}}(B, \theta_{q}, c_{q})$
   \STATE $\forall{j}, \bar{n}_{j} \leftarrow (RFM_{\zeta}(\theta'_{q}, c'_{q}))_{j}$
   \STATE $\forall{j}, q^{j}_{\eta,\zeta, \alpha_{q}}(.|s) \leftarrow \mathrm{Softmax}(\mathrm{MLP}_{\eta}(\bar{n}_{j}))$
   \STATE $q_{\eta,\zeta}(a^{-i}|s, \theta^{-i}) \leftarrow \prod_{j\in{-i}}q^{j}_{\eta,\zeta}(a^{j}|s)$
   \STATE \textbf{return} $q_{\eta,\zeta}(a^{-i}|s, \theta^{-i})$
   \ENDFUNCTION
\end{algorithmic}
\end{algorithm}

Using the functions we previously defined, we finally describe GPL's training algorithm. GPL collects experience from parallel environments through the modified Asynchronous Q-Learning framework~\cite {mnih2016asynchronous} where asynchronous data collection is replaced with a synchronous data collection instead. Despite this, it is relatively straightforward to modify the pseudocode to use an experience replay instead of a synchronous process for data collection. As in the case of existing deep value-based RL approaches, we also use a separate target network whose parameters are periodically copied from the joint action value model to compute the target values required for optimising Equation~\ref{ValueLoss}. We finally optimise the model parameters in the pseudocode to optimise the loss function provided in Section~\ref{sec:GPLLearningObjective} using gradient descent. GPL's training process is finally described in Algorithm~\ref{alg:fullGPL}.

\begin{algorithm*}[p]\footnotesize
   \caption{GPL Training}
   \label{alg:fullGPL}
\begin{algorithmic}[1]
 \STATE \textbf{Input: }Number of training steps $T$, time between updates $t_{update}$, time between target network updates $t_{targ\_update}$.
   \STATE Initialize the joint-action value model parameters, $\alpha_{Q}, \beta, \delta$.
   \STATE Initialize the agent model parameters, $\alpha_{q}, \eta, \zeta$.
   \STATE Create target joint-action value networks.\\
   \begin{center}
       $\alpha'_{Q}, \beta', \delta' \leftarrow \alpha_{Q}, \beta, \delta$ 
   \end{center}
   
   \STATE $\theta_{Q}, c_{Q}, \theta^{targ}_{Q}, c^{targ}_{Q} \leftarrow \mathbf{0,0,0,0}$
   \STATE $\theta_{q}, c_{q} \leftarrow \mathbf{0,0}$
   \STATE $d\alpha_{Q}, d\alpha_{q}, d\beta, d\delta, d\eta, d\zeta \leftarrow \mathbf{0, 0, 0, 0, 0, 0}$
   \STATE Observe $s$ from environment
   \FOR{$t=1$ {\bfseries to} $T$}
   \STATE $h_{Q}, h_{q}, h_{Q}^{targ} \leftarrow (\theta_{Q}, c_{Q}),(\theta_{q}, c_{q}), (\theta^{targ}_{Q}, c^{targ}_{Q})$
   \STATE $\bar{Q}(H,.), h'_{Q}, h'_{q} \leftarrow \mathrm{\textbf{QV}}(s, \alpha_{Q}, \alpha_{q},\beta, \delta, \eta, \zeta, h_{Q}, h_{q})$
   \STATE Sample action according to the learning algorithm being used, \\\begin{center}
   $a^{i}_{t} \sim \begin{cases}
    \text{eps-greedy}(\epsilon, \bar{Q}(H,.)),& \text{if Q-Learning}\\
    p_{\text{SPI}}(\bar{Q}(H,.),\tau)             & \text{if SPI}
    \end{cases}
    $
    \end{center}
    \STATE Execute $a^{i}$ and observe $a, r$ and $s'$.
    \STATE Compute predicted joint-action value for $a_{t}$,\\\begin{center}
        $Q_{\beta,\delta,\alpha_{Q}}(H,a) \leftarrow \mathrm{\mathbf{QJOINT}}(s, a, \alpha_{Q}, \beta, \delta, h_{Q})$
    \end{center}
    \STATE Compute the action-value of the next state using the target network. \\ \begin{center}
        $\bar{Q}'\left(H,a^{i}\right),h^{targ}_{Q},\_ \leftarrow \mathrm{\textbf{QV}}(s', \alpha'_{Q}, \alpha_{q}, \beta', \delta', \eta, \zeta, h^{targ}_{Q}, h'_{q})$
    \end{center}
    
    \STATE Compute target value for updating the joint-action value model with, \\\begin{center}
    $ y\left(r, H'\right) = 
    r + \gamma \text{max}_{a^{i}}\bar{Q}'\left(H',a^{i}\right),$\\\end{center} if Q-Learning is used, or \\ \begin{center}$
     y\left(r, H'\right) = r + \gamma \sum_{a^{i}}p_{\mathrm{SPI}}(a^{i}|H')\bar{Q}'\left(H',a^{i}\right), 
    $\\\end{center} if using SPI.
    \STATE Compute predicted action probabilities of teammates using the agent models,\\ \begin{center}
        $q_{\eta,\zeta, \alpha_{q}}(a^{-i}|s, \theta^{-i}) \leftarrow \mathrm{\mathbf{PTEAM}}(s, a, \alpha_{q}, \eta, \zeta, h_{q})$
    \end{center}
    
    \STATE Using $Q_{\beta,\delta,\alpha_{Q}}(H_{t},a_{t}), y\left(r_{t}, H_{t+1}\right)$, and $q_{\eta,\zeta, \alpha_{q}}(a^{-i}|s, a^{i})$, compute $L_{\zeta,\eta, \alpha_{q}}$ and $L_{\beta,\delta, \alpha_{Q}}$ with Equation~\eqref{ActionModelLoss} and~\eqref{ValueLoss}.
    \STATE Accumulate parameter gradients for updates\\\begin{center}
        $d\alpha_{Q} = d\alpha_{Q} + \nabla_{\alpha_{Q}} L_{\beta,\delta}$, $d\alpha_{q} = d\alpha_{q} + \nabla_{\alpha_{q}} L_{\eta,\zeta}$\\
        $d\beta = d\beta + \nabla_{\beta} L_{\beta,\delta}$, $d\delta = d\delta + \nabla_{\delta} L_{\beta,\delta}$\\
        $d\eta = d\eta + \nabla_{\eta} L_{\eta,\zeta}$, $d\zeta = d\zeta + \nabla_{\zeta} L_{\eta,\zeta}$
    \end{center}
    \IF{$t$ mod $t_{\mathrm{update}}$ = 0}
        \STATE Update $\alpha_{Q}, \alpha_{q}, \beta, \delta, \eta, \zeta$ using gradient descent based on $d\alpha_{Q}, d\alpha_{q}, d\beta, d\delta, d\eta, d\zeta$.
        \STATE $d\alpha_{Q}, d\alpha_{q}, d\beta, d\delta, d\eta, d\zeta \leftarrow \mathbf{0, 0, 0, 0, 0, 0}$
    \ENDIF
    \IF{$t$ mod $t_{\mathrm{targ\_update}}$ = 0}
        \STATE $\alpha'_{Q}, \beta', \delta' \leftarrow \alpha_{Q}, \beta, \delta$ 
    \ENDIF
     \STATE $(\theta_{Q}, c_{Q}),(\theta_{q}, c_{q}), s \leftarrow h'_{Q}, h'_{q}, s' $
   \ENDFOR
\end{algorithmic}
\end{algorithm*}

\section{GPL Pseudocode Under Partial Observability}
\label{app:GPLPseudocodePartObs}

This section focuses on providing pseudocodes that illustrate the way the learner updates its belief representations alongside the usage of belief representations to estimate the learner's optimal action-value function under partial observability. Note the general training and decision-making procedure under partial observability highly resembles their respective counterparts under full observability provided in Algorithm~\ref{alg:fullGPL}. Therefore, we avoid rewriting the entire training and decision-making pseudocode by highlighting the main differences between instances of Algorithm~\ref{alg:fullGPL} under the partial and fully observable scenarios.

Instances of Algorithm~\ref{alg:ActValComputation} in partial and fully observable environments have three main differences. The first two differences are related to how action values are computed following Algorithm~\ref{alg:ActValComputation}. The final difference is then related to the additional loss functions to train the belief inference models under partially observable scenarios.

First, the action value computation in fully and partially observable scenarios differ in how to input representations for the joint action value and agent model are computed. Note that in the third and fourth lines under the function name in Algorithm~\ref{alg:ActValComputation}, under full observability input representations for these models are computed via an LSTM. By contrast, input representations for the joint action value and agent model are computed via the belief inference model that we have introduced in Section~\ref{sec:POGPLOverview}. This process of computing the input representation given our belief inference models is provided in Algorithm~\ref{alg:pred}. Therefore, we replace the LSTM-based representation evaluation with the belief inference models that have been introduced previously for decision-making under partial observability.

\begin{algorithm*}
\caption{Belief Inference}\label{alg:pred}
\begin{algorithmic}[1]
\STATE \textbf{Input:} \text{Observation received by learner} $o_t$,\\ \text{The learner's previous action} $a^{i}_{t-1}$,\\
\text{The belief inference algorithm}, $\text{alg}\in \{\text{PF, AE, VAE}\},$ \\
\text{Representations resulting from previous step},
\[
    \rho_{t-1}= 
\begin{cases}
    \{\left(a^{u_{k}}_{t-2}, s^{u_{k}}_{t-1}, \theta^{u_{k}}_{t-1}, w^{u_{k}}_{t-1}\right)|u_{k} \in U_{t-1}\},& \text{if } \text{alg} = \text{PF}\\
    (c_{t-1}, h_{t-1}),              & \text{if otherwise}
\end{cases}
\]
\FUNCTION{\textbf{BELIEF\_INFERENCE}($o_{t}, a^{i}_{t-1}, \text{alg}, \rho_{t-1}$)}
\IF{alg = PF}
\STATE  With $\mathbf{w}_{t-1} = \{w^{u_{k}}_{t-1}|u_{k}\in{}\rho_{t-1}\}$, sample $K$ particles from $\rho_{t-1}$, $$\bar{U}_{t-1} = \{u_{1}, u_{2}, ..., u_{K}\},$$ \text{with,} $$u_{1}, u_{2}, ..., u_{K} \overset{\mathrm{i.i.d.}}{\sim} \text{Categorical}(\text{Softmax}(\mathbf{w}_{t-1}))$$ 
\FORALL{$u_{k} \in \bar{U}_{t-1}$}
\STATE $a_{t-1}^{u_k} \sim q_{\alpha}(a^{u_{k}}_{t-1}|s^{u_{k}}_{t-1},\theta^{u_{k}}_{t-1}, a^{i}_{t-1},o_{t})$  \hfill\COMMENT{Action Inference}
\STATE  $s^{u_{k}}_t \sim q_{\beta}(s^{u_{k}}_{t}|s^{u_{k}}_{t-1},a^{u_{k}}_{t-1}, a^{i}_{t-1},o_{t})$  \hfill\COMMENT{State Inference}
\STATE $\theta_{t}^{u_k} = f_{\delta}(s^{u_{k}}_{t},\theta^{u_{k}}_{t-1},a^{u_{k}}_{t-1},a^{i}_{t-1},o_{t})$ \hfill\COMMENT{Type Update}
\STATE Compute $w_{t-1,\alpha}^{u_{k}}$ and $w_{t-1,\beta}^{u_{k}}$ following Equation~\ref{eq:w_action_update} and Equation~\ref{eq:w_state_update}.
\STATE $ w^{u_{k}}_{t} = \text{log}(q_{\zeta}(o_{t}|s^{u_{k}}_{t},a^{u_{k}}_{t-1})) + w_{t-1,\beta}^{u_{k}} + w_{t-1,\alpha}^{u_{k}}$ \hfill\COMMENT{Particle Weight Update}
\ENDFOR
\STATE $U_{t} = \{(a^{u_{k}}_{t-1},s^{u_{k}}_{t},\theta^{u_{k}}_{t}, w^{u_{k}}_{t})|u_{k}\in\bar{U}_{t-1}\}$

\STATE \textbf{Return:} $ U_{t} $, $U_{t}$
\ELSE
\IF{alg = VAE}
\STATE $\mu_{t},\Sigma_{t},(c_{t}, h_{t}) = \text{Encoder}_{\alpha}(a^{i}_{t-1},o_{t},\rho_{t-1})$ \hfill\COMMENT{Based on Appendix~\ref{Sec:VAEInputPreprocessingAndArchitecture}}
\STATE Sample $K$ representations based on the parameters outputed by the encoder,
$$Z_{t} = \{(z_{1}, p(z_{1}|\mu_{t}, \Sigma_{t})), (z_{2}, p(z_{2}|\mu_{t}, \Sigma_{t})), ..., (z_{K}, p(z_{K}|\mu_{t}, \Sigma_{t}))\},$$
such that,
$$z_{1}, z_{2}, ..., z_{K} \overset{\mathrm{i.i.d.}}{\sim}\mathcal{N}(\mu_{t}, \Sigma_{t}).$$
\STATE \textbf{Return:} $ Z_{t} $, $(c_{t}, h_{t})$
\ELSE
\STATE $z_{t}, (c_{t}, h_{t}) = \text{Encoder}_{\alpha}(a^{i}_{t-1},o_{t},\rho_{t-1})$  \hfill\COMMENT{Based on  Appendix~\ref{Sec:AEInputPreprocessingAndArchitecture}}
\STATE \textbf{Return:} $ z_{t} $, $(c_{t}, h_{t})$
\ENDIF
\ENDIF
\ENDFUNCTION

\end{algorithmic}
\end{algorithm*}

The second difference between action value computation under partial and fully observable scenarios is the way inferred representations are computed for action value computation. In Algorithm~\ref{alg:ActValComputation}, this process is illustrated by the lines following the calls to the LSTM. However, under partial observability the action value computation depends on the belief inference model being used. We illustrate the way different belief inference models use their outputted representations for decision-making in Algorithm~\ref{alg:ActValCompPartObs}. Note that regardless of the belief inference method, the way an action value function is computed for a single sampled representation is the same, which is indicated by Algorithm~\ref{alg:SingleParticleActValComputationPartObs}.

\begin{algorithm}[!htb]
   \caption{Single Sample Action Value Computation Under Partial Observability}
   \label{alg:SingleParticleActValComputationPartObs}
\begin{algorithmic}[1]
   \STATE {\bfseries Input:} Input representation $\rho$, \\ joint-action value model parameters ($P_{val}$),\\ agent model parameters ($P_{ag}$)
   \FUNCTION{\textbf{QV\_PART}($\rho, P_{val}, P_{ag}$)}
   \STATE $\forall{j}, \bar{n}_{j} \leftarrow (RFM_{P_{ag}}(\rho))_{j}$
   \STATE $\forall{j}, q_{P_{ag}}(.|\rho) \leftarrow \mathrm{Softmax}(\mathrm{MLP}_{P_{ag}}(\bar{n}_{j}))$
   \STATE $\forall{j, a^{j}}, Q^{j}_{P_{val}}(a^{j}|\rho) \leftarrow \textrm{MLP}_{P_{val}}(\rho^{j}, \rho^{i})(a^{j})$
   \STATE $\forall{j, a^{j}, a^{k}}$,
   $$
   Q^{j,k}_{P_{val}}(a^{j},a^{k}|\rho) \leftarrow \textrm{MLP}_{\delta}(\rho^{j},\rho^{k}, \rho^{i})(a^{j},a^{k})$$
   \STATE Compute $\bar{Q}(\rho,a^{i})$ using Equation~\eqref{SingleActionValue}
   \STATE $\bar{Q}(\rho,.) \leftarrow \mathrm{\textbf{MARGINALIZE}}($
   $$q_{P_{Ag}}(.|\rho), Q_{P_{Val}}(.|\rho)), Q_{P_{val}}(.,.|\rho)$$
   )
   \STATE \textbf{return} $\bar{Q}(\rho,.)$
   \ENDFUNCTION
\end{algorithmic}
\end{algorithm}

\begin{algorithm}[!htb]
\caption{Action Value Computation Under Partial Observability}\label{alg:ActValCompPartObs}
\begin{algorithmic}[1]
\STATE \textbf{Input:} \text{The belief inference algorithm}, $\text{alg}\in \{\text{PF, AE, VAE}\},$ \\
Joint-action value model parameters ($P_{val}$),\\ Agent model parameters ($P_{ag}$), \\
\text{Representations resulting from the belief inference model},
\[
    \rho_{t}= 
\begin{cases}
    \{\left(a^{u_{k}}_{t-1}, s^{u_{k}}_{t}, \theta^{u_{k}}_{t}, w^{u_{k}}_{t}\right)|u_{k} \in U_{t}\},& \text{if } \text{alg} = \text{PF}\\
    \{(z_{1}, p(z_{1}|\mu_{t}, \Sigma_{t})), ..., (z_{K}, , p(z_{K}|\mu_{t}, \Sigma_{t}))\},& \text{if } \text{alg} = \text{VAE} \\
    z_{t},              & \text{if otherwise}
\end{cases}
\]
\FUNCTION{\textbf{QV\_P\_OBS}($\text{alg},\rho, P_{val}, P_{ag}$)}
\IF{alg = AE}
    \STATE \textbf{return} $\textbf{QV\_PART}(\rho_{t}, P_{val}, P_{ag})$
\ELSE
\IF{alg = PF}
    \FORALL{$u_{k}\in{\rho_{t}}$}
        \STATE $x^{u_{k}} \leftarrow \textbf{CONCATENATE}(e_{t}^{u_{k}}, s_{t}^{u_{k}},\theta_{t}^{u_{k}})$
        \STATE $\bar{Q}(e_{t}^{u_{k}}, s_{t}^{u_{k}},\theta_{t}^{u_{k}},.) \leftarrow \textbf{QV\_PART}(x^{u_{k}}, P_{val}, P_{ag})$
    \ENDFOR
    \STATE  \textbf{return} $\sum_{u_{k}\in{U_{t}}}\left( \dfrac{\text{exp}(w^{u_{k}}_{t})}{\sum_{u_{j}\in{U_{t}}} \text{exp}(w^{u_{j}}_{t})}\right) \bar{Q}_{\pi^{i,*}}(e_{t}^{u_{k}}, s_{t}^{u_{k}},\theta_{t}^{u_{k}},.)$
\ELSE
\FORALL{$(z_{k}, p(z_{k}|\rho_{t}, \Sigma_{t}))\in{\rho_{t}}$}
        \STATE $\bar{Q}(z_{k},.) \leftarrow \textbf{QV\_PART}(z_{k}, P_{val}, P_{ag})$
\ENDFOR
\STATE \textbf{return} $\dfrac{\sum_{(z_{k}, p(z_{k}|\rho_{t}, \Sigma_{t}))\in{\rho_{t}}} \bar{Q}(z^{k}_{t},.) p(z_{k}|\rho_{t},\Sigma_{t})}{\sum_{(z_{k}, p(z_{k}|\rho_{t}, \Sigma_{t}))\in{\rho_{t}}}p(z_{k}|\rho_{t},\Sigma_{t})}$
\ENDIF
\ENDIF
\ENDFUNCTION
\end{algorithmic}
\end{algorithm}

Finally, the last difference between the pseudocode for training and decision making under full and partially observable scenarios is the loss function. Under full observability, we do not have loss functions associated to belief inference. However, we now incorporate this loss function for training the belief inference model according to the losses defined in Section~\ref{sec:PseudocodeAndLearningObj}. This results in an additional optimised term that we add to algorithm~\ref{alg:fullGPL} to train the belief inference model.
\section{GPL Overview }
\subsection{Input Preprocessing}
\label{sec:GPLInputPreprocessing}
GPL's input preprocessing step ensures that a type vector is computed solely based on relevant information associated with the teammate which it characterises. This preprocessing step starts by separating the observed state features associated with different agents into an agent feature input batch, $x$. All vectors in the agent feature input batch are subsequently concatenated with the remaining state features that are not associated with any agent, $u$, to create an input batch $B$. This preprocessing step is illustrated in Figure~\ref{fig:inp-preprocessing}.

To provide a concrete example of this first preprocessing step, consider a pickup soccer environment. Example agent features that are included in $x$ are the position and orientation features which values are different for each agent. In contrast, example features in $u$ is the location of the ball, which value is shared between the different agents in the environment. Using $B$ as input to the type inference model ensures that a player's type only depends on its own trajectory when moving around the pitch.

\subsection{Type Computation and Output Postprocessing}
\label{Sec:GPLTypeComputation}

\begin{figure*}[t]
    \centering
    \subfloat[Observation preprocessing.]{%
        \includegraphics[trim={3.5cm 14.5cm 4cm 5.5cm},clip, scale=0.40]{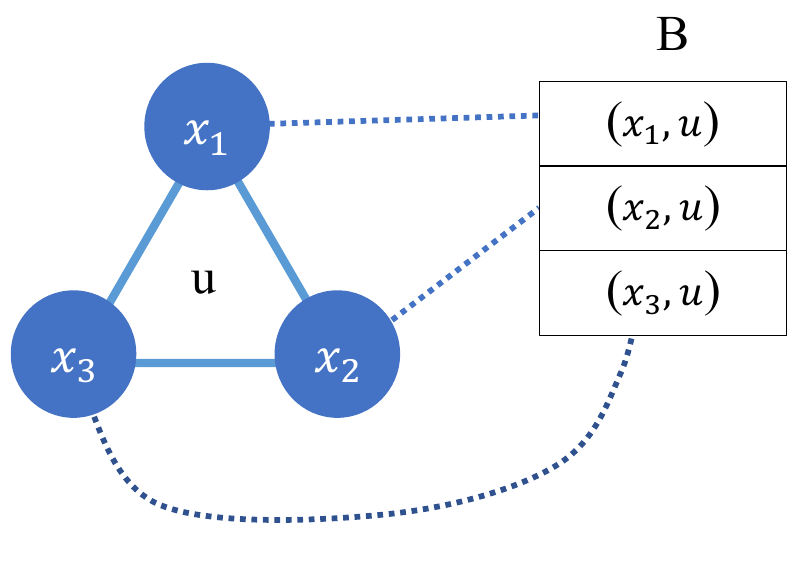}}
        \label{fig:inp-preprocessing}
    \subfloat[LSTM hidden vector preprocessing.]{%
        \includegraphics[trim={3.cm 4.5cm 4.2cm 2.5cm},clip, scale=0.35]{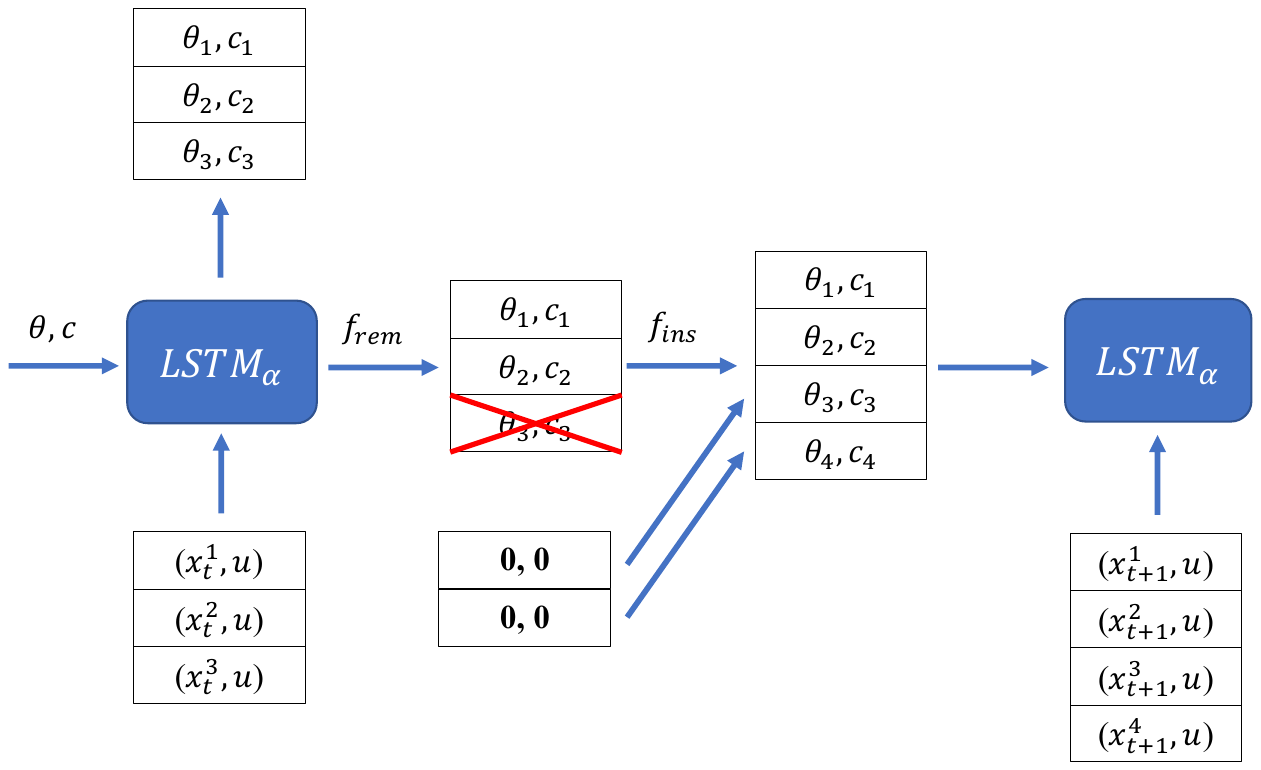}}
        \label{fig:h-postprocessing}
    \caption{The figure shows (a) the preprocessing of observation information into input for the GPL algorithm along with (b) the additional processing steps done to the agent embedding vectors to handle environment openness. Part (b) shows an example processing step where agent 3 is removed from the environment and two new agents join the environment.}  
    \label{fig:pre_post_processing}
\end{figure*}

The input batch $B$ resulting from the preprocessing step is presented into the RNN-based type inference model to update teammate type vectors from previous timesteps. In this work, we particularly use an LSTM as the type inference model. The LSTM-based type update is illustrated on the left side of Figure~\ref{fig:h-postprocessing}.

After the update process, additional processing steps are required to ensure only type vectors of existing agents are used in GPL's optimal action value estimation. Between subsequent timesteps, GPL removes the type vectors of teammates that are removed from the environment following environment openness. On the other hand, type vectors of teammates that are added to the environment are set to the default value of zero vectors. 

We formally define this additional LSTM output processing step with Equation~\ref{PostprocessingEq}. Assuming $i_{t}$ and $d_{t}$ correspond to the sets of added and removed agents at time $t$, $f_{rem}$ removes the states associated to agents leaving the environment while $f_{ins}$ inputs a zero vector for the states associated to agents joining the environment. An example of this preprocessing step is illustrated by the computational steps occurring between two LSTM blocks in Figure~\ref{fig:h-postprocessing}. 

\begin{equation}
    \label{PostprocessingEq}
    \text{Prep}(\theta_{t}, c_{t}) = f_{ins}(f_{rem}(\theta_{t}, c_{t}, d_{t}), i_{t})
\end{equation}

\begin{proof}[Proof of Equation \ref{SingleActionValue}] By substituting Equation~\ref{JointActionValueComputation} and~\ref{GNNequations} into Equation~\ref{Eq:Marginalisation}, we can derive the following expression:
\begin{align}
\begin{split}
\label{SubbedModels}
  \bar{Q}_{\pi^{i}}(s_{t},a^{i}_{t}) 
  &= \mathbb{E}_{a_{t}^{-i}\sim\boldsymbol{\pi}^{-i}(.|s_{t},\theta^{-i}_{t})}\bigg[Q_{\pi^{i}}(s_{t},a)\bigg|{}a^{i}=a_{t}^{i}\bigg] \\
  &= \sum_{a^{-i}\in{A^{-i}}} Q_{\pi^{i}}(s_{t}, a) \pi^{-i}(a^{-i}|s_{t},\theta_{t}^{-i}) \\
  &= \sum_{a^{-i}\in{A^{-i}}} (\sum_{ \mathclap{\substack{a^{j}\in{A_{j}}}}} Q^{j}_{\beta}(a^{j}|s_{t}) + \sum_{ \mathclap{\substack{a^{j}\in{A_{j}}, a^{k} \in{A_{k}}}}}Q^{j,k}_{\delta}(a^{j}, a^{k}|s_{t})) q_{\zeta,\eta}(a^{-i}|s_{t},a^{i}) \\
  &= Q^{i}_{\beta}(a^{i}_{t}|s_{t}) + \sum_{ \mathclap{\substack{a^{j}\in{A_{j}},  j\neq{i}}}}\big(Q^{j}_{\beta}(a^{j}|s_{t})+Q^{i,j}_{\delta}(a^{i}_{t}, a^{j}|s_{t})\big)q_{\zeta,\eta}(a^{j}|s_{t}) \\
&+ \sum_{\mathclap{\substack{a^{j}\in{A_{j}},a^{k}\in{A_{k}}, j, k\neq{i}}}}Q^{j,k}_{\delta}(a^{j}, a^{k}|s_{t})q_{\zeta,\eta}(a^{j}|s_{t})q_{\zeta,\eta}(a^{k}|s_{t}).
\end{split}
\end{align}
\end{proof}

\section{Input Preprocessing and Model Architecture for Methods Addressing Partial Observability}
This section details the preprocessing steps and latent variable inference model architectures designed for estimating the learner's optimal action-value function. We structure this section in terms of the different latent variable inference models defined in Section~\ref{sec:POGPLOverview}.

\subsection{Particle-based Belief Inference}
\label{Sec:ParticleBeliefModelArchitecture}

\begin{figure*}
    \centering
    \includegraphics[scale=0.5, trim={3.5cm 17.5cm 6cm 3cm},clip]{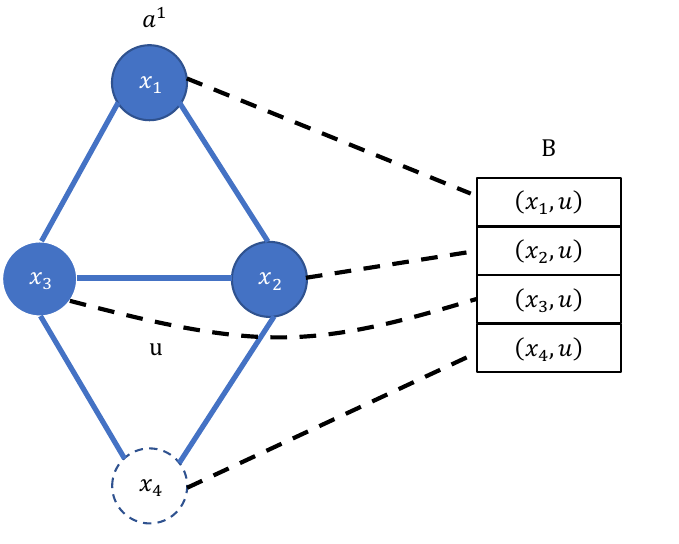}
    \caption{\textit{Preprocessing Under Partial Observability.} An illustration of the preprocessing step assuming that at most 4 agents exists in the environment. In this visualization, unobserved teammates are visualised by the dashed circles. Our preprocessing method assigns a zero vector as personal agent features to teammates that are unobserved by the learner.}
    \label{Fig:POPreprocess}
\end{figure*}
\textit{Input Preprocessing.} To preprocess the learner's observations in partially observable environments, we assume knowledge over the set of agents that exists in the environment, $N$. Based on the learner's observation, for each agent in $N$ we also assume access to their ID, personal features, and visibility in the observation. Finally, the learner also remembers the action which it has executed in the previous timestep $a^{i}_{t-1}$.

The aforementioned knowledge is subsequently preprocessed into a set of feature vectors for belief updates. For each $i\in{N}$, their ID, personal features, and visibility in the observation are concatenated as their personal agent features, $x_{i}$. The remaining global information not associated to any agent ($u$) is subsequently concatenated to the personal agent features to finally form an observation batch for learning, $B_{obs}$. This preprocessing step is illustrated in Figure~\ref{Fig:POPreprocess}.

\textit{Joint Action Inference Models.} The proposal and target action distribution for joint action inference are implemented as networks with similar architecture. The only difference is that the proposal action distribution includes $o_{t}$ and $a^{i}_{t-1}$ as its input. Formally, the input for the proposal network and for the target network are defined as follows:
\begin{equation}\label{Eq:JointActionInferenceModelInputParticle}
D_{in}= 
\begin{cases}
    \text{Concatenate}(e^{u_{k}}_{t-1},s^{u_{k}}_{t-1},\theta^{u_{k}}_{t-1}),& \text{if target distribution}\\
    \text{Concatenate}(e^{u_{k}}_{t-1},s^{u_{k}}_{t-1},\theta^{u_{k}}_{t-1}, B_{obs}, a^{i}_{t-1}),              & \text{otherwise.}
\end{cases}
\end{equation}
The network architecture to compute the proposal and target action distribution subsequently evaluates the joint action probability distribution as:

\begin{equation}
p_{\alpha}(a_{t}|D_{in}) = \prod_{j\in{N}} p_{\alpha}(a^{j}|D_{in}),
\end{equation}
with,
\begin{align}
\begin{split}
\bar{n}_{j} &= (\text{GNN}_{\alpha}(D_{in}))_{j},\\
p_{\alpha}(a^{j}|D_{in}) &= \textrm{Softmax}(\text{MLP}_{\alpha}(\bar{n}_{j}))(a^{j}).
\end{split}
\end{align}
Assuming that $\alpha^{p}$ and $\alpha^{t}$ are the parameters of the proposal and target action distribution respectively, our implementation uses separate neural networks for estimating these distributions, such that $\alpha = (\alpha^{p}, \alpha^{t})$.

\textit{Existence and State Inference Models.} As in the case with joint action inference models, input for the proposal and target distributions in existence and state inference is derived from concatenating all the necessary information as defined below:
\begin{equation}\label{Eq:StateInferenceModelInput}
D_{in}= 
\begin{cases}
    \text{Concatenate}(e^{u_{k}}_{t-1}, s^{u_{k}}_{t-1},a^{u_{k}}_{t-1}),& \text{if target distribution}\\
    \text{Concatenate}(e^{u_{k}}_{t-1}, s^{u_{k}}_{t-1},a^{u_{k}}_{t-1}, B_{obs}, a^{i}_{t-1}),              & \text{otherwise.}
\end{cases}
\end{equation}

For existence inference, we assume a unique integer index assigned to each agent in $N$, where the index assigned to $j\in{N}$ is denoted as $j_{id}$. Subsequently, both the target and proposal distribution are implemented as a neural network which computes agents' existence in the following manner:
\begin{equation}
    p_{\beta}(e_{t}^{u_k}|D_{in}) = \prod_{j\in{N}} p_{\beta}(e_{t}^{u_{k},j}|e_{t}^{<j_{id}}, D^{<j_{id}}_{in}),
\end{equation}
with,
\begin{align}
\begin{split}
\bar{n}(j)&= \sum_{\{k|k_{id}< j_{id}\}}\text{MLP}_{\alpha}(\text{Concatenate}(e^{k}_{t},D^{k}_{in})),\\
E_{in}^{j} &= \text{Concatenate}(e_{t-1}^{j},D_{in}^{j},\bar{n}(j)),\\
p_{\beta}(e^{u_{k},j}_{t}=1|e_{t}^{<j_{id}}, D^{<j_{id}}_{in}) &= \textrm{Sigmoid}(\text{MLP}_{\alpha}(E_{in}^{j})).
\end{split}
\end{align}
This autoregressive existence inference technique resembles GraphRNN~\citep{you2018graphrnn}, which is a generative model that generates graphs with varying numbers of nodes in an autoregressive fashion.

For state inference, both the proposal and target distributions are represented as multivariate normal distribution with a diagonal covariance matrix, $\mathcal{N}(\mu_{\beta}, \Sigma_{\beta})$, which parameters are evaluated by neural networks following this expression:
\begin{align}
    \mu_{\beta}(D_{in}) &= \text{MLP}_{\beta}^{\mu}(D_{in}),\\
    \Sigma_{\beta}(D_{in}) &= \text{Softplus}(\text{MLP}_{\beta}^{\Sigma}(D_{in}))
\end{align}
In our implementation, the target and proposal distribution for teammate existence and state inference are implemented as separate models which parameters are denoted as $\beta^{t}$ and $\beta^{p}$ respectively.

\textit{Type Update Network.} We implement the type update network as an LSTM which accounts for agents' previous types and recently inferred state representation and actions to compute their respective types. The type update process in the type update network is provided in the following expression:
\begin{align}
    c^{u_{k}}_{t}, h^{u_{k}}_{t} &= LSTM_{\delta}(D_{in}, c^{u_{k}}_{t-1}, h^{u_{k}}_{t-1}),\\
    \theta^{u_{k}}_{t} &= MLP_{\delta}(c^{u_{k}}_{t}),
\end{align}
with $c^{u_{k}}_{t-1}$ and $h^{u_{k}}_{t-1}$ being the cell and hidden state that represents the sequence of previously inferred state and type representations. The input to the LSTM model is subsequently defined below:
\begin{equation}
    D_{in} = \text{Concatenate}(s^{u_{k}}_{t},\theta^{u_{k}}_{t-1},a^{u_{k}}_{t-1}).
\end{equation}

\textit{Observation Likelihood Model.} We assume that the observation vector that we reconstruct is the preprocessed data vector, $B_{obs}$. Since $B_{obs}$ is a collection of continuous vectors, we use a multivariate normal distribution, $\mathcal{N}(\mu_{\zeta}, \Sigma_{\zeta})$, which parameters are computed as defined below:
\begin{align}
    \mu_{\zeta}(D_{in}) &= \text{MLP}_{\zeta}^{\mu}(D_{in}),\\
    \Sigma_{\zeta}(D_{in}) &= \text{Softplus}(\text{MLP}_{\zeta}^{\Sigma}(D_{in})),
\end{align}
with the input to this model defined as:
\begin{equation}
    D_{in} = \text{Concatenate}(s^{u_{k}}_{t},a^{u_{k}}_{t-1}).
\end{equation}

\subsection{Variational Autoencoder-based Belief Inference}
\label{Sec:VAEInputPreprocessingAndArchitecture}

Given the observations from the environment, $o_t$, we first preprocess them to obtain the vector $B_{obs}$. This preprocessing step is similar as the one done in particle-based methods which we detail in  Appendix~\ref{Sec:ParticleBeliefModelArchitecture}. 
Then, at every timestep, $B_{obs}$ is used as input to the encoder architecture to compute the distribution over latent variables $z_{t}$. Furthermore, $B_{obs}$ also acts as the information which will be reconstructed by the decoder. The architecture of the encoder and decoder is provided below.

\textit{Encoder Network} The encoder network is implemented as an LSTM which receives the learner's preprocessed observation, $B_{obs}$, as input. It subsequently produces the mean and covariance matrix for the variational parametric distribution following this expression:
\begin{align}
    \mu_{t} &= MLP_{\alpha^{\mu}}(c_{t}),\\
    \Sigma_{t} &= MLP_{\alpha^{\Sigma}}(c_{t}),
\end{align}
where,
\begin{align}
     c_{t}, h_{t} &= LSTM_{\alpha}(B_{obs}, c_{t-1}, h_{t-1}),
\end{align}
with $c_{t-1}$ and $h_{t-1}$ being the LSTM's cell and hidden state that represents the sequence of previous observations. 

\textit{Decoder Network.} Since the role of the decoder is to reconstruct the information observed by the learner, we train the decoder to reconstruct the learner's observations alongside predicting its observed teammates' actions. The decoder network subsequently outputs both the likelihood of $B_{obs}$ alongside the likelihood of observed teammates' actions. Since $B_{obs}$ is a collection of continuous vectors, we compute the likelihood of $B_{obs}$ based on a multivariate normal distribution, $\mathcal{N}(\mu_{\beta}, \Sigma_{\beta})$, which parameters are computed as defined below:
\begin{align}
    \mu_{\beta}(z_{t}) &= \text{MLP}_{\beta}^{\mu}(z_{t}),\\
    \Sigma_{\beta}(z_{t}) &= \text{Softplus}(\text{MLP}_{\beta}^{\Sigma}(z_{t})),
\end{align}
assuming $z_{t}$ is sampled from the variational parametric distribution outputted by the learner.

On the other hand, the part of the decoder that predicts the likelihood of teammates' actions has a similar implementation as the joint action inference model for the particle-based approach in Section~\ref{Sec:ParticleBeliefModelArchitecture}. Given $z_{t}$, the decoder computes the likelihood of observed agents' actions using a GNN following this equation:
\begin{equation}\label{Eq:JointActionInferenceModelVAEDecoderAct}
p_{\gamma}(a^{V}_{t}|z_{t}) = \prod_{j\in{V}} p_{\gamma}(a^{j}|z_{t}),
\end{equation}
assuming $V\subseteq{N}$ denotes the set of visible teammates and with,
\begin{align}\label{GNNequationDefinitionPartObs}
\begin{split}
\bar{n}_{j} &= (\text{GNN}_{\gamma}(z_{t}))_{j},\\
p_{\gamma}(a^{j}|z_{t}) &= \textrm{Softmax}(\text{MLP}_{\gamma}(\bar{n}_{j}))(a^{j}).
\end{split}
\end{align}

\subsection{Autoencoder-based Inference}
\label{Sec:AEInputPreprocessingAndArchitecture}
Given input data $B_{obs}$ which is obtained in a similar way as the VAE-based inference model, the encoder outputs a representation $\rho_{t}$ that is defined as the following:
\begin{align}
    \rho_{t} &= MLP_{\alpha}(c_{t}),
\end{align}
where,
\begin{align}
     c_{t}, h_{t} &= LSTM_{\alpha}(B_{obs}, c_{t-1}, h_{t-1}),
\end{align}
and $c_{t}$ and $h_{t}$ are the cell and hidden states of the LSTM. The interaction data gathered by the learner is then used to train the encoder's parameters to produce $\rho_{t}$ that contains important information regarding the learner's interaction experience.

The training process utilises a decoder network to produce $\rho_{t}$ that is representative of the learner's interaction experience. The decoder network receives $\rho_{t}$ as input and is trained to reconstruct the learner's current observation alongside predicting the observed teammates' previous joint actions. Given $\rho_{t}$, the reconstructed observation outputted by the decoder is denoted by:
\begin{equation}
    B_{pred}(\rho_{t}) = \text{MLP}_{\beta}(\rho_{t}).
\end{equation}
On the other hand, the part of the decoder that predicts teammates' joint actions is similar to the action prediction part of the VAE's decoders defined in Equation~\ref{Eq:JointActionInferenceModelVAEDecoderAct}. Assuming that the parameters of this model is denoted by $\gamma$, the only difference is that this model receives $\rho_{t}$ as input rather than $z_{t}$.

\section{Agent and Joint Action Value Modelling Learning Under Partial Observability}
\label{sec:AgentJointActionLoss}

In this section, we describe the loss functions to train GPL's agent and joint action value modelling components under partial observability. While the details of the loss functions depend on the latent variable inference being used, all loss functions are based on GPL's optimised loss functions described in Equation~\ref{ActionModelLoss} and~\ref{ValueLoss}. Details of the agent and joint action value modelling losses for each latent variable inference model are provided in the following sections.

\subsection{Particle-based Belief Inference}
\textit{Agent Modelling Loss Function.}  While other methods use separate models for agent modelling and latent variable inference, the target action distribution estimation model, $p_{\alpha^{t}}(a^{u_{k}}_{t}|e^{u_{k}}_{t},s^{u_{k}}_{t},\theta^{u_{k}}_{t})$, is reused for agent modelling when using particle-based belief models. This reuse is motivated by how GPL's agent modelling process introduced in Section~\ref{sec:GPLAgentModelling} aims to estimate the target action distribution in the first place. Therefore, the negative log likelihood loss is computed by assuming $P_{ag} = \{\alpha\}$. The negative log likelihood loss is then defined as:
\begin{equation}
    L^{NLL}_{P_{ag}}(D) = -\sum_{H_{n}\in{D}}\left(\sum_{u_{k}\in{U_{n}}} \text{log}\left(q_{P_{ag}}(a^{V,n}_{T_{n}}|e^{u_{k}}_{T_{n}}, s^{u_{k}}_{T_{n}},\theta^{u_{k}}_{T_{n}})\right)\right),
\end{equation}
where the joint action log likelihood is only evaluated over observed teammates' joint actions.

\textit{Reinforcement Learning Loss Function.} We assume that the CG-based model used in action value computation defined in Section~\ref{sec:ActionSelectionParticles} is parameterised by $\eta$ such that $P_{val}=\{\eta\}$. Assuming $\boldsymbol{A}^{-V}$ denotes the set of possible joint actions of unobserved agents, the CG-based joint action value model is then trained to estimate the optimal joint action value function by optimising:
\begin{equation}
    \label{Eq:RLLoss}
    L^{RL}_{P_{val}}(D) = \sum_{H_{n}\in{D}}\left(\sum_{u_{k}\in{U_{n}}}\left( \dfrac{\text{exp}(w^{u_{k}}_{T_{n}})}{2\sum_{u_{j}\in{U_{n}}} \text{exp}(w^{u_{j}}_{T_{n}})}\right) \left(y_{P_{val}}({u_k},a^{V,n}_{T_{n}}) - y(u'_{k})\right)^{2}\right),
\end{equation}
where,
\begin{equation}
    \label{Eq:JointActionMarginalized}
     y_{P_{val}}({u_k},a^{V,n}_{T_{n}}) = \sum_{\substack{a^{-V} \in \boldsymbol{A}^{-V}}} Q_{P_{val}}(e^{u_{k}}_{T_{n}}, s^{u_k}_{T_{n}},\theta_{T_{n}}^{u_k},a^{V}_{T_{n}},a^{-V}) p_{\alpha}(a^{-V}|e^{u_{k}}_{T_{n}},s^{u_{k}}_{T_{n}},\theta^{u_{k}}_{T_{n}}),
\end{equation}
is the estimated joint action value of agents that are visible to the learner at $T_{n}$ based on the contents of particle $u_{k}$. Specifically, $Q_{P_{val}}(e^{u_{k}}_{T_{n}}, s^{u_k}_{T_{n}},\theta_{T_{n}}^{u_k},a^{V}_{T_{n}},a^{-V})$ is computed via Equation~\ref{JointActionValueComputation} while $p_{\alpha}(a^{-V}|e^{u_{k}}_{T_{n}},s^{u_{k}}_{T_{n}},\theta^{u_{k}}_{T_{n}})$ is evaluated based on Equation~\ref{GNNequations}.

The target value for particle $u_{k}$ is then defined based on $u'_{k}$, which is the particle resulting from updating $u_{k}$ based on $o^{n}_{T_{n}+1}$ and $a^{i,n}_{T_{n}}$ according to Section~\ref{Sec:ParticleBeliefUpdate} excluding the particle sampling step. The target value is then defined as:
\begin{equation}
    \label{Eq:RLTarget}
    y(u'_{k}) = 
    r^{n}_{T_{n}}+\gamma\max\limits_{a'}\bar{Q}(s_{T_{n}}^{u'_{k}},\theta_{T_{n}}^{u'_{k}},a'),
\end{equation}
with $\bar{Q}(s_{T_{n}}^{u'_{k}},\theta_{T_{n}}^{u'_{k}},a')$ evaluated according to Equation~\ref{SubbedModels}. Note that unlike in the RL loss under full observability, we consider the particle weights in the loss computation to allow less likely particles to have higher temporal difference errors.

\subsection{Variational Autoencoder-based Belief Inference}

\textit{Agent Modelling Loss Function.} Agent modelling under the VAE-based model is done via the action prediction component of the decoder, which in our description at Section~\ref{Sec:VAEInputPreprocessingAndArchitecture} is parameterised by $\gamma$. This model is chosen for agent modelling since its purpose is also to predict teammates' joint actions. Assuming $P_{ag}=\{\gamma\}$, the loss function of this model is defined as:
\begin{equation}
    L^{NLL}_{P_{ag}}(D) = -\sum_{H_{n}\in{D}} \mathbb{E}_{z_{T_{n}}\sim{q(z_{T_{n}}|H_{n})}}\left[p_{P_{ag}}(a^{V}_{T_{n}}|z_{T_{n}})\right]
\end{equation}

\textit{Reinforcement Learning Loss Function.} As in GPL, we define a CG-based model to estimate the joint action values of the learner. Assuming that the parameters of this model is denoted as $\delta$, this model must be trained to estimate the joint action value given the variational parametric distribution, $q(z_{t}|H_{t})$. Since exactly computing $q(z_{t}|H_{t})$ is generally intractable, we use a Monte Carlo approach for training this model. Under this approach, we sample $m$ vectors from $q(z_{t}|H_{t})$ such that:
\begin{equation}
    z^{1}_{t}, z^{2}_{t}, ..., z^{m}_{t} \overset{\mathrm{iid}}{\sim} q(z_{t}|H_{t}).
\end{equation}
The sampled $z_{t}$ are subsequently used as input to the CG model, which loss function for joint action value modelling is subsequently computed as:
\begin{equation}
    \label{Eq:RLLossVAE}
    L^{RL}_{P_{val}}(D) = \sum_{H_{n}\in{D}}\left(\sum_{k=1}^{m}\left( \dfrac{p(z^{k}_{T_{n}}|H_{n})}{2\sum_{l=1}^{m} p(z^{l}_{T_{n}}|H_{n})}\right) \left(y_{P_{val}}({z^k_{T_{n}}},a^{V,n}_{T_{n}}) - y(z^{'k}_{T_{n}})\right)^{2}\right),
\end{equation}
assuming $P_{val}=\{\delta\}$. In Equation~\ref{Eq:RLLossVAE}, the predicted joint action value of observed teammates' joint actions is defined as:
\begin{equation}
    \label{Eq:JointActionMarginalizedVAE}
     y_{P_{val}}({u_k},a^{V,n}_{T_{n}}) = \sum_{\substack{a^{-V} \in \boldsymbol{A}^{-V}}} Q_{P_{val}}(z^{k}_{T_{n}},a^{V}_{T_{n}},a^{-V}) p_{\alpha}(a^{-V}|z^{k}_{T_{n}}),
\end{equation}
which is similar to the predicted value under particle-based approaches. Finally, the target value is defined as:
\begin{equation}
    \label{Eq:RLTargetVAE}
    y(u'_{k}) = 
    r^{n}_{T_{n}}+\gamma\max\limits_{a'}\bar{Q}(Z_{T_{n}}^{k},a'),
\end{equation}
with $\bar{Q}(Z_{T_{n}}^{k},a')$ computed according to Equation~\ref{Eq:IntegralApprox}.

\subsection{Autoencoder-based Inference}
\textit{Agent Modelling Loss Function.} Given $\rho_{T_{n}}$ produced by the encoder, GPL's agent and joint action model are trained to estimate the learner's action value function. We use the decoder's action prediction component for agent modelling since it is also designed to predict teammates' actions. Assuming $P_{ag} = \{\beta\}$, the agent model is subsequently trained to predict observed teammates' actions by minimising the following loss function:
\begin{equation}
    L^{NLL}_{P_{ag}}(D) = -\sum_{H_{n}\in{D}} \text{log}(p_{P_{ag}}(a^{V}_{T_{n}}|\rho_{T_{n}})).
\end{equation}
\textit{Reinforcement Learning Loss Function.} We train the joint action value model by optimising the temporal difference error defined below:
\begin{equation}
    \label{Eq:RLLossAE}
    L^{RL}_{P_{val}}(D) = \sum_{H_{n}\in{D}} \left(y_{P_{val}}({\rho^k_{T_{n}}},a^{V,n}_{T_{n}}) - y(\rho^{'k}_{T_{n}})\right)^{2}.
\end{equation}
Given the parameters of the joint action value model $P_{val}=\{\delta\}$, the predicted and target joint action value are evaluated following Equation~\ref{Eq:JointActionMarginalizedVAE} and~\ref{Eq:RLTargetVAE}.

\section{Partial observability results}
\label{annex:PO_results}
\subsection{Baselines}

To run the single agent baselines in environments with a changing number of teammates we assign a value of -1 to features associated with inactive agents. In our experiments we can have up to five agents in the environment, so we add these placeholder values to the input to match the size of the input vector when five agents are present. To prevent teammates' features from always being assigned a placeholder, which could hurt generalisation, we assign agents entering the environment an index number. We use this index to determine the location of their features in the input vector. This index remains the same while an agent is active in the environment.

For DVRL we utilised the code made available by the original authors. We utilised similar hyperparameters as the authors. We used 10 particles, with an action encoding of 16, the $z$ dimension is 100, and the $h$ dimension is 100. All neural networks utilise a hidden layer size of 100 unless stated. We used RMSProp with $\alpha = 0.99$, a gradient clipping of 0.5, learning rate $ = 1\times10^{-3}$, and gamma 0.99. For encoding the set particles into $\hat{h}$ we utilised a fully connected neural network. Actions are encoded by a fully connected neural network with two layers of 64 units. The policy is one fully connected layer whose size is determined by the action space. DVRL uses A2C, with 16 parallel environments, and a 5-step learning. 

For PPO we utilised the following hyperparameter: a fully connected network with two hidden layers of 128 neurons each, a learning rate of $3\times10^{-4}$, a batch size of 64, a number of epochs 10, and a number of steps of 2048.

\subsection{Unseen teammates}
\label{annex:unseen_teammates}

For the experiments with unseen teammates, we created a set of 8 different reinforcement learning agents all trained with different seeds, which were added to the pool of existing teammates.
The policies of the set of unseen teammates were obtained via reinforcement learning, particularly the PPO algorithm. The set of unseen agents is able to observe the full state of the system in order to make decisions. To encode the policies of the unseen teammates we utilised an MLP network with two hidden layers of 128 neurons each, a learning rate of $3\times10^{-4}$, a batch size of 64, a number of epochs 10, and a number of steps of 2048.